\documentclass{article}
\usepackage[utf8]{inputenc}
\usepackage[colorlinks=true,linkcolor=green,citecolor=blue,allcolors=green,hypertexnames=false]{hyperref}
\hypersetup{
   colorlinks=true,
   linkcolor=blue,
   filecolor=blue,
   urlcolor=blue,
   allcolors=blue
}
\usepackage{fullpage}
\usepackage{amsfonts,amsmath}
\usepackage{graphicx,color}
\usepackage{paralist}
\usepackage{enumitem}

\newcommand{\bitset}{\{0,1\}}
\newcommand{\eps}{\epsilon}

\newcommand{\poly}{\textrm{poly}}

\newcommand{\Ex}{\mathbb{E}}
\newcommand{\Var}{\mathbb{V}}
\newcommand{\eqdef}{\stackrel{\rm def}{=}}

\newcommand{\calA}{{\mathcal{A}}}
\newcommand{\mA}{{\mathcal{A}}}
\newcommand{\calB}{{\mathcal{B}}}
\newcommand{\calC}{{\mathcal{C}}}
\newcommand{\calD}{{\mathcal{D}}}
\newcommand{\calG}{{\mathcal{G}}}
\newcommand{\mG}{{\mathcal{G}}}
\newcommand{\calP}{{\mathcal{P}}}
\newcommand{\calT}{{\mathcal{T}}}

\newcommand{\trn}{{n'}}
\newcommand{\trG}{{G'}}
\newcommand{\trm}{{m'}}
\newcommand{\trA}{{\calA'}}
\newcommand{\treps}{{\eps'}}
\newcommand{\trR}{{R'}}
\newcommand{\tre}{e'}

\newcommand{\Sup}{\textrm{Sup}}

\newenvironment{proof}{\smallskip\noindent{\bf Proof:}}%
        {\hspace*{\fill}$\Box$\par}
\newenvironment{proofof}[1]{\smallskip\noindent{\bf Proof of #1:}}%
        {\hspace*{\fill}$\Box$\par}

\newtheorem{theorem}{Theorem}
\newtheorem{corollary}[theorem]{Corollary}
\newtheorem{lemma}{Lemma}
\newtheorem{claim}[lemma]{Claim}
\newtheorem{definition}{Definition}

\newcommand{\commentt}[1]{ {}}
\newcommand{\talya}[1]{{} }
\newcommand{\reut}[1]{ {}}
\newcommand{\tnew}[1]{{}}

\newcommand{\SodaReb}[1]{{ #1}}

\newcommand{\ICALPCR}[1]{{#1}}

\newcommand{\ForFuture}[1]{}

\makeatletter
\def\moverlay{\mathpalette\mov@rlay}
\def\mov@rlay#1#2{\leavevmode\vtop{%
   \baselineskip\z@skip \lineskiplimit-\maxdimen
   \ialign{\hfil$\m@th#1##$\hfil\cr#2\crcr}}}
\newcommand{\charfusion}[3][\mathord]{
    #1{\ifx#1\mathop\vphantom{#2}\fi
        \mathpalette\mov@rlay{#2\cr#3}
      }
    \ifx#1\mathop\expandafter\displaylimits\fi}
\makeatother

\newcommand{\cupdot}{\charfusion[\mathbin]{\cup}{\cdot}}

\newenvironment{packed_enum}{
\begin{enumerate}
  \setlength{\itemsep}{0pt}
  \setlength{\parskip}{0pt}
  \setlength{\parsep}{0pt}
}{\end{enumerate}}

 \newcommand{\SelectEdge}{\hyperref[alg:select-edge]{\color{black}{\sf Select-an-Edge}}}
  \newcommand{\TestCfour}{\hyperref[alg:C4]{\color{black}{\sf Test-$C_4$-freeness}}}
 
\newcommand{\TestCsix}{\hyperref[alg:C6]{\color{black}{\textsf{Test-$C_6$-freeness}}}}

\newcommand{\TestFfree}{\hyperref[alg:Ffree]{\color{black}{\textsf{Test-subgraph-freeness}}}}

\begin{document}

\title{Testing $C_k$-freeness in bounded-arboricity graphs}
	\author{
	Talya Eden\thanks{Department of Computer Science, Bar Ilan University, Israel. Email: {\tt talyaa01@gmail.com}}
		\and
		Reut Levi\thanks{Efi Arazi School of Computer Science,  Reichman University, Israel. Email: {\tt reut.levi1@runi.ac.il}.}
			\and
		Dana Ron\thanks{School of Electrical Engineering, Tel Aviv University, Israel. Email: {\tt danaron@tau.ac.il}.}
}

\maketitle

\date{}

\begin{abstract}
We study the problem of testing 
$C_k$-freeness ($k$-cycle-freeness) for fixed constant $k > 3$ in graphs with bounded arboricity (but unbounded degrees).
In particular, we are interested in one-sided error algorithms, so that they must detect a copy of $C_k$ with high constant probability when the graph is $\epsilon$-far from $C_k$-free.

We next state our results for constant arboricity and constant $\epsilon$ with a focus on the dependence on the number of graph vertices, $n$. The query complexity of all our algorithms grows polynomially with $1/\epsilon$. 
\begin{enumerate}
\item As opposed to the case of $k=3$, where the complexity of testing $C_3$-freeness grows with the arboricity of the graph but not with the size of the graph (\emph{Levi, ICALP 2021})\footnote{\label{foot:loglogn}\SodaReb{As presented in (\emph{Levi, ICALP 2021}), the complexity of the algorithm depends on $\log\log n$, but this dependence can be replaced with at most a polylogarithmic dependence on the arboricity.}} this is no longer the case already  for $k=4$.
We show that $\Omega(n^{1/4})$ queries are necessary for testing $C_4$-freeness, and that $\widetilde{O}(n^{1/4})$ are sufficient. The same bounds hold for $C_5$.
\item For every fixed $k \geq 6$, any one-sided error algorithm for testing $C_k$-freeness must perform $\Omega(n^{1/3})$ queries.
\item For $k=6$ we give a testing algorithm whose query complexity is $\widetilde{O}(n^{1/2})$.
\item For any fixed $k$, the query complexity of testing $C_k$-freeness is upper bounded by ${O}(n^{1-1/\lfloor k/2\rfloor})$.  
\end{enumerate}

\SodaReb{The latter upper bound builds on another result in which we show that for any fixed subgraph $F$, the query complexity of testing $F$-freeness is upper bounded by $O(n^{1-1/\ell(F)})$, where $\ell(F)$ is a parameter of $F$ that is always upper bounded by the number of vertices in $F$ (and in particular is $k/2$ in $C_k$ for even $k$).}

We extend some of our results  to bounded (non-constant) arboricity, where in particular, we obtain sublinear upper bounds
for all $k$.

Our $\Omega(n^{1/4})$ lower bound for testing $C_4$-freeness in constant arboricity graphs provides a negative answer to an open problem posed by (Goldreich, 2021). 
\end{abstract}


\section{Introduction}\label{sec:intro}
Detecting small subgraphs with  specific structures  (referred to as \emph{finding network motifs}) is a basic algorithmic task, with a variety of applications in biology, sociology and network science (see e.g.~\cite{HoLe70, Co88, portes2000social, EcMo02, burt2004structural, milo2002network, alon2007network, foucault2010friend, BerryHLP11, SeKoPi11, JRT12}). 
Of special  interest is the natural 
case of subgraphs that are cycles of a fixed size $k$, which we denote by $C_k$.
When the algorithm receives the entire graph as input, then by the well known 
result of Alon, Yuster and Zwick~\cite{AYZ}, this task can be solved in  time $\widetilde{O}(n^{\omega})$ 
where $n$ is the number of graph vertices and $\omega$ is the exponent of matrix multiplication.\footnote{The dependence on $k$ is exponential, but $k$ is considered a constant.}
But what if we seek a sublinear-time (randomized) algorithm that does not read the entire graph? Namely, the algorithm is given  query access to the graph\footnote{\label{foot:queries} The  types of queries typically considered are neighbor queries (``what is the $i$th neighbor of a vertex $v$?''), degree queries (``what is the degree of a vertex $v$?''), and pair queries (``is there an edge between a pair of vertices $v$ and $u$?'').} 
and should find a $C_k$ when the graph is not $C_k$-free.
This is clearly not possible if the graph contains only a single copy of $C_k$. 
However, is it possible to detect such a copy in sublinear-time when the graph is relatively far from being $C_k$-free?
By ``relatively far'' we mean that it is necessary to remove a non-negligible fraction, denoted $\eps$, of its edges in order to obtain an $C_k$-free graph. 
A closely related formulation of the question is whether we can design a one-sided error algorithm for testing $C_k$-freeness.\footnote{The problems are equivalent if the algorithm is not given access to degree queries, otherwise the algorithm might find evidence to the existence of a $C_k$ without actually detecting one. We note that all our algorithms do detect copies of $C_k$ when they reject.}

If the maximum degree in the graph is upper bounded by a parameter $d_{\max}$, then the $C_k$-freeness testing problem can easily be solved by performing a number of queries that grows polynomially with $d_{\max}$ and exponentially with $\Theta(k)$~\cite{GR-bound}. \commentt{Explain that this holds in general model or no need?}
In particular, when $d_{\max} = O(1)$, then there is no dependence on the size of the graph $G$. 
We are however interested in considering graphs with varying degrees, so that, in particular, the maximum degree may be   much larger than the average degree, and possibly as large as $\Theta(n)$.

For the special and interesting case where $k=3$, i.e., the cycle is a triangle,  Alon, Kaufman, Krivelevich and Ron~\cite{AKKR08} gave several upper and lower bounds on the query complexity of testing triangle-freeness as a function of the average degree $d$ of the graph (in addition to the dependence on $n$ and $\eps$).
While the upper and lower bounds are not tight  in general, they are tight for $d=O(1)$, where the complexity is $\Theta(\sqrt{n})$ (for constant $\epsilon$). 
The lower bound in this case is essentially based on ``hiding'' a small clique.

Since the aforementioned lower bound relies on the existence of a small dense subgraph, a natural question, studied by Levi~\cite{Reut}, is whether it is possible to obtain improved (and possibly tight) results when the arboricity of the graph, denoted $arb(G)$, is bounded.\footnote{The arboricity of a graph $G$ is the minimum number of forests required to cover its edges, and is equal (up to a factor of $2$) to the maximum average degree of any subgraph of $G$.} Focusing on the result under the assumption that $m\geq n$ (i.e., $d= \Omega(1)$) Levi showed that  
$\widetilde{O}(arb(G))$  queries are sufficient for testing triangle-freeness (the dependence on $1/\epsilon$ is polynomial), and that $\Omega(arb(G))$ queries are necessary.\footnote{To be precise, this lower bound holds for $m\geq (arb(G))^3$ -- if $m< (\arg(G))^3$ then the lower bound is $\Omega(m^{1/3})$. \SodaReb{See also Footnote~\ref{foot:loglogn} regarding the upper bound.}}
In particular, when $arb(G)$ is a constant, the complexity is polynomial in $1/\eps$ and does not depend on the size of the graph.

In this work we seek to understand the complexity of testing $C_k$-freeness, in particular with one-sided error, for fixed $k> 3$. Our main focus is on constant arboricity graphs and some of our results extend to bounded arboricity graphs, as well as to $F$-freeness for any general subgraph $F$ (of constant size). 
We note that the problem of testing cycle-freeness
without requiring the cycle to be of specific length, is  different from our problem. We further discuss this in Section~\ref{subsec:related}.
In the next subsection we state our findings.

\subsection{Our results}

Since our main focus is on graphs with constant arboricity, we first state our results in this setting, and later discuss our extensions to graphs with non-constant arboricity.
Throughout this paper we assume that $m = \Omega(n)$ since  
even obtaining a single edge in the graph requires $\Omega(n/m)$ queries. 
Our algorithms use degree and neighbor queries and our lower bounds also allow pair queries (see Footnote~\ref{foot:queries}).
For simplicity, we state our results for constant $\epsilon$. All our algorithms have a polynomial dependence on $1/\eps$, which is stated explicitly in the corresponding technical sections.

Our first finding is that, as opposed to the case of $k=3$, where the complexity of testing $C_3$-freeness grows with the arboricity of the graph but not with the size of the graph,\footnote{We note that this is true also for other $k$-cliques for $k > 3$.} this is no longer the case for $k=4$ (and larger $k$). In particular: 

\begin{theorem}\label{thm:C4-5}
The query complexity of one-sided error testing of $C_4$-freeness in constant-arboricity graphs over $n$ vertices is $\widetilde{\Theta}(n^{1/4})$.
The same bound holds for testing $C_5$-freeness.
\end{theorem}
 The more detailed statements for testing $C_4$ and $C_5$ freeness 
can be found in Theorems~\ref{thm:C4-ub}--\ref{thm:C5-const-alpha-lb}.

Theorem~\ref{thm:C4-5} \ICALPCR{(together with the upper bound in~\cite{GR-bound})}
answers negatively  the following open problem raised by Goldreich.

\smallskip\noindent\textsf{Open problem (number 3.2 in~\cite{goldreich2021open}): From bounded degree to bounded arboricity.} \newline
\textit{Suppose that property $\Pi$ is
testable within complexity $Q(n,\eps)$ in the bounded-degree graph model. Provide an upper
bound on the complexity of testing $\Pi$ in the general graph model under the promise that the tested graph has constant arboricity. For example, can the latter complexity be linear in $Q(n,\eps)$ while
permitting extra $poly(log n)$ or $1/\eps$ factors?}
\smallskip

The $\Omega(n^{1/4})$ lower bound for testing $C_4$-freeness, answers this question negatively. Indeed, 
testing $C_4$-freeness in $d$-bounded degree graphs  can be done with $poly(d,\eps)$ queries~\cite{GR-bound}, while our lower bound suggest that even in constant  arboricity graphs, a polynomial dependence on $n$ is necessary.

\smallskip
When $k \geq 6$, we show that it is no longer possible to obtain a complexity of $\widetilde{O}(n^{1/4})$ as is the case for $k=4,5$.

\newcommand{\GenLB}{
Let $k\geq 6$.
Any one-sided error tester for the property of $C_k$-freeness in graphs of constant arboricity over  $n$ vertices 
must perform $\Omega(n^{1/3})$ queries.
}

\begin{theorem}\label{thm:lb_ck_const_alpha}
\GenLB
\end{theorem}

While for $C_6$ we were not able to match the lower bound of $\Omega(n^{1/3})$, we were able to obtain a sublinear-time algorithm, as stated next.

\begin{theorem}\label{thm:C6-ub-intro}
There exists a one-sided error algorithm for testing $C_6$-freeness in graphs of constant arboricity  over $n$ vertices whose query complexity 
is $\widetilde{O}(n^{1/2})$.
\end{theorem}


\smallskip
For general (fixed) $k$ we prove the following upper bound.

\begin{theorem}\label{thm:Ck-ub-intro}
There exists a one-sided error algorithm for testing $C_k$-freeness in graphs of constant arboricity over $n$ vertices whose query complexity 
is $O(n^{1-1/\lfloor k/2 \rfloor })$. 
\end{theorem}

\SodaReb{
We also prove a more general result for testing $F$-freeness for any constant size subgraph $F$. Below, $\ell(F)$ is as defined in Definition~\ref{def:ell}, and is always upper bounded by the number of vertices in $F$.}

\begin{theorem}\label{thm:Gen-F-ub-intro}
\SodaReb{There exists  a one-sided error algorithm for testing  $F$-freeness in graphs of constant arboricity over $n$ vertices whose query complexity is $O(n^{1-1/\ell(F)}).$}
\end{theorem}

 \subsubsection{Extensions for general arboricity}

We state our results for general arboricity graphs assuming that the algorithm is given an 
upper bound $\alpha$ on the arboricity of the graph (in  the lower bounds the algorithm may be assumed to know the arboricity).
Alternatively,
if the algorithm receives as an input the number of edges, $m$, (as in previous results for $C_k$-freeness~\cite{AKKR08, Reut}) instead of an upper bound on the arboricity, then we can estimate a notion known~\cite{Reut} as the ``effective arboricity'' of the graph, and depend on it instead of $\alpha$. This is potentially beneficial since the effective arboricity can be much smaller than the actual arboricity of the graph, \SodaReb{and it does not affect the asymptotic running times of our algorithms in terms of the dependence on the size of the graph and $\alpha$}. For further details see Section~\ref{sec:prel}.

For $C_4$-freeness we give both an upper bound and a lower bound for general arboricity graphs. In particular, we show that a linear dependence on $\alpha$ is sufficient and a $\sqrt{\alpha}$-dependence is necessary (both for one-sided error algorithms) as stated next.

\ICALPCR{
\begin{theorem}\label{thm:C4-ub}
There exists a one-sided error algorithm for testing $C_4$-freeness in graphs of arboricity at most $\alpha$ over $n$ vertices whose query complexity  is $\tilde{O}\left(\min\{n^{1/4}\alpha, \alpha+n^{3/4}\}/\eps^3\right)$.\footnote{More precisely, for values $\alpha<\log n$, the complexity is $O(n^{1/4}\alpha^{1/2}\log^{1/2}n/\eps^3)$, for values $\log n < \alpha<\sqrt n$, the complexity is $O(n^{1/4}\alpha/\eps^3)$, and for values $\alpha>n^{1/2}$, it is $O( (\alpha+n^{3/4})/\eps^3)$.}
\end{theorem}

\begin{theorem}\label{thm:C4-gen-alpha-lb}
Testing $C_4$-freeness with one-sided error in graphs over $n$ vertices with arboricity $c_1\log n < \alpha < n^{1/2}/c_1'$ for sufficiently large constants $c_1$ and $c_1'$ requires $\Omega(n^{1/4}\alpha^{1/2})$ queries for constant $\epsilon$.\footnote{Note that the two-sided error lower bound of $\Omega(n^{1/4})$ for constant arboricity graphs (as stated in Theorem~\ref{thm:C4-5}) also holds for graphs with higher arboricity $\alpha$, and in particular, $\alpha = O(\log n)$. This is the case since we can simply add a small subgraph with arboricity $\alpha$ and no $C_4$s to the lower bound construction.} 

\end{theorem}
}

For general
\SodaReb{constant size subgraphs $F$ (and in particular $C_k$)}
our upper bound also has at most a linear dependence on $\alpha$
--
see Theorem~\ref{thm:Gen-F-ub} and Corollary~\ref{cor:Ck-ub} for a precise statement.

We comment that our lower bound of $\Omega(n^{1/3})$  for one-sided error algorithms, $k \geq 6$ and constant arboricity (stated in Theorem~\ref{thm:lb_ck_const_alpha}) also applies to graphs with non-constant arboricity (by adding a $C_k$-free subgraph with  higher arboricity).\footnote{For an odd $k$, it suffices to add a dense bipartite graph, and for even $k$, by the Erd\H{o}s girth conjecture~\cite{E63}, one can add a subgraph with arboricity $n^{2/k}$.} 
 
We also note that it is possible to extend our algorithms for $C_5$ and $C_6$ freeness so as to get a polynomial (but not linear) dependence on $\alpha$. However,  these extensions do not introduce new techniques (and are most probably not optimal), so we do not present them here.

\subsection{A high-level discussion of our algorithms and lower bounds}\label{subsec:high-level-intro}
\ForFuture{reconsider epsilon in this part}
Before discussing each of our results in more detail, we highlight some common themes. The starting point of all our algorithms is that if a graph is $\eps$-far from being $C_k$-free  (for a constant $k$), then it contains $\Omega(\eps m)$ edge-disjoint cycles.\footnote{\ICALPCR{To verify this, let $G$ be a graph that is $\epsilon$-far from being $C_k$-free for a fixed constant $k$. Consider any maximal set $S$ of edge-disjoint $k$-cycles. Since by removing all $k\cdot |S|$ edges on these cycles, the graph can be made cycle-free, $|S| \geq \epsilon m /k = \Omega(\epsilon m)$.}}
We next use the bounded arboricity of the graph. Specifically, if a graph has arboricity at most $\alpha$, then the number of edges between pairs of vertices that both have degree greater than $\theta_0 = \Theta(\alpha/\eps) $, is at most $O(\eps m)$. 

Hence, there is a set of edge-disjoint $C_k$s, which we denote by $\calC$, such that $|\calC| =\Omega(\eps m)$, and no $C_k$ in $\calC$ contains any edge between two vertices with degree greater than $\theta_0$. In other words, for every $k$-cycle $\rho$ in $\calC$, and for every vertex $v$ with degree greater than $\theta_0$ in $\rho$, the two neighbors of $v$ in $\rho$ have degree at most $\theta_0$.
In particular, when $\alpha$ is a constant, the two neighbors have degree $O(1/\eps)$.

At this point our algorithms diverge, but there are two common aspects in the case of $k=4,5,6$ which we would like to highlight. The first is that for the sake of ``catching'' one of the $C_k$s in $\calC$, it will be useful to  consider a subset, $\calC'$, in which every vertex $v$ that participates in one of the edge-disjoint $C_k$s in $\calC'$ actually participate in $\Omega(\eps\cdot d(v))$  $C_k$s in $\calC'$. The existence of such a subset follows by applying (as a mental experiment) a simple iterative process that removes $C_k$s with vertices that do not obey this constraint.

To illustrate why it is useful to have such a set $\calC'$, consider the case of $k=4$, and assume that a relatively large fraction of the $C_4$s in $\calC'$ contain, in addition to the two vertices of degree at most $\theta_0 = O(\alpha/\eps)$, at least one other vertex  that has degree at most $\theta_1=O(n^{1/2}/\eps)$. 
\SodaReb{
In this case we can obtain such a vertex $v$ with high probability (as discussed below), and then sample 
roughly $\sqrt{d(v)/\eps} = O(\sqrt{\theta_1/\eps})=O(n^{1/4}/\eps)$ of its neighbors, so that the following holds.}
By (a slight variant of) the birthday paradox, with high constant probability we hit two of its neighbors, $u$ and $u'$, that reside on the same $C_4$ in $\calC'$ (and hence have degree at most $\theta_0$). By querying all the neighbors of $u$ and $u'$, we obtain this $C_4$.

However, what if for most of the $C_4$'s in $\calC'$ there are two vertices with degree significantly larger than $\sqrt{n}$ (that are ``one opposite the other'' on the $C_4$s)? Roughly speaking, in this case we exploit the fact that the number of such high-degree vertices is bounded, and we show how to detect a $C_4$ by performing random walks of length $2$. A related issue arises in the case of $k=6$, when there are three very high degree vertices on most $C_6$s in $\calC'$. In this case we show how to essentially reduce the problem to testing triangle-freeness in a certain auxiliary graph. 
More precisely, the auxiliary graph is a multi-graph to which we have access only to certain types of queries, so that we cannot apply the algorithm of~\cite{AKKR08}. However, we can still show how to obtain a triangle in this graph, and hence a $C_6$ in the original graph. Interestingly, our general lower bound of $\Omega(n^{1/3})$ for $C_k$-freeness, $k\geq 6$ builds on the lower bound for testing triangle-freeness of~\cite{AKKR08}.


\smallskip
In the following subsections we assume for the sake of the exposition that $\eps$ is a constant.

\subsubsection{The  results for $C_4$-freeness (and $C_5$-freeness)}
We discuss our results for $C_4$-freeness in graphs with general arboricity. 
The results for $C_5$-freeness in constant arboricity graphs are obtained using very similar techniques. 

\paragraph{The algorithm.}
Our algorithm for testing $C_4$-freeness,  \TestCfour, \ICALPCR{which has query complexity $\widetilde{O}(n^{1/4}\alpha)$,}  
is governed by two thresholds: $\theta_0=\Theta(\alpha)$, and $\theta_1=\Theta(n^{1/2})$. 
For the sake of the current high-level presentation, 
we 
assume that\footnote{Indeed, graphs with arboricity greater than $n^{1/2}$ necessarily contain at least one $C_4$, but since we are interested in a one-sided error algorithm, and $\alpha$ is only known to be an upper bound on $\alpha$, the algorithm cannot reject if it is provided with $\alpha > n^{1/2}$.} $\alpha \leq n^{1/2}$, 
so that $\theta_0\leq \theta_1$.

The algorithm first samples $O(1)$ edges approximately uniformly by invoking a procedure
\SelectEdge,\footnote{\SodaReb{This is a fairly standard and simple procedure, where we use the fact that graph has bounded arboricity, so that most of its edges have at least one endpoint with degree $\theta_0$.}}
and then randomly selects one of their endpoints. For each vertex $v$ selected, it queries its degree, $d(v)$. If $d(v)\leq \theta_1$, then the algorithm selects $O(\sqrt{d(v)})$ random neighbors of $v$, and for each selected neighbor $u$ such that $d(u)\leq \theta_0$,  it queries all the neighbors of $u$. If $d(v)>\theta_1$, then the algorithm
performs $\widetilde{O}(n^{1/4}\alpha^{1/2})$ random walks of length two from $v$. If a $C_4$ is observed in any one of these steps, then the algorithm rejects, otherwise it accepts.

\paragraph{The analysis of the algorithm.}
By the above description, the algorithm will only reject a graph if it detects a $C_4$, implying that it never errs on $C_4$-free graphs. Hence, consider a graph $G$ that is far from being $C_4$-free. As discussed at the start of Section~\ref{subsec:high-level-intro}, the setting of $\theta_0=\Theta(\alpha)$ (together with the fact that $G$ is $\Omega(1)$-far from being $C_4$-free) implies the following. There exists a set, denoted $\calC$, of $\Omega(m)$ \emph{edge-disjoint} $C_4$s in $G$, such that
 no $C_4$ in $\calC$ contains an edge
between two vertices that both have degree greater than $\theta_0$. 
Thus, for each $C_4 $ in $\calC$, there are at most two vertices with degree greater than $\theta_0$, and they do not neighbor each other. 

Considering the second aforementioned  degree threshold $\theta_1$
(and recalling that 
$\theta_1 \geq \theta_0$), we partition $\calC$ into two subsets. The first, $\calC_1$, consists of those $C_4$s in $\calC$ that contain at most one vertex with degree greater than $\theta_1$, and the second, $\calC_2$, of the remaining $C_4$s in $\calC$, which contain exactly two vertices with degree greater than $\theta_1$.
Since $\calC = \calC_1 \cupdot \calC_2$, at least one of these subsets is of size $\Omega(m)$.

\begin{figure}[htb!]
\centerline{\mbox{\includegraphics[width=0.5\textwidth]{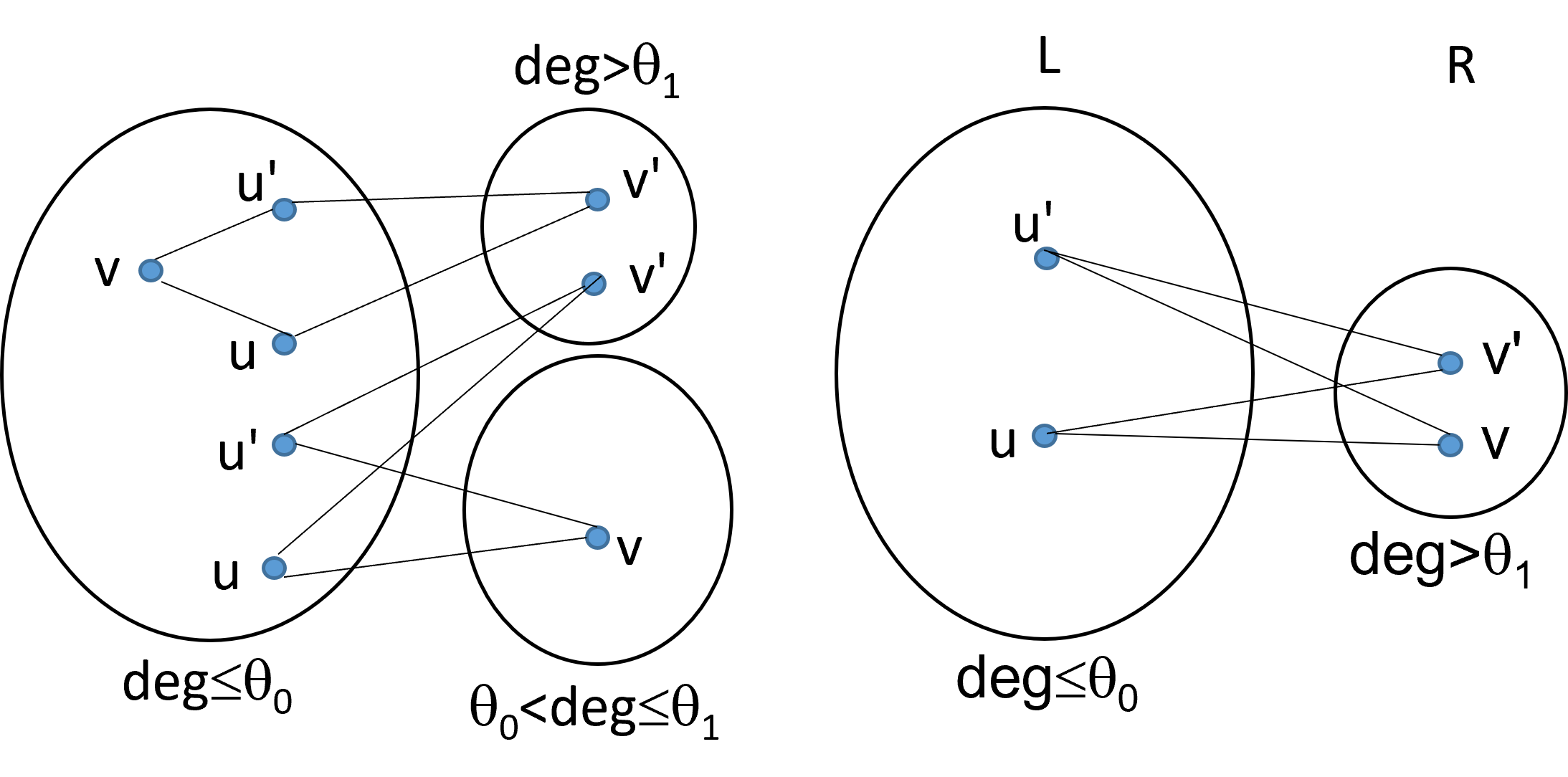}}}
\caption{An illustration for some of the cases considered in the analysis of the algorithm for $C_4$-freeness. On the left side are two examples in which there is a single vertex $v'$ with degree greater than $\theta_1$, so that there is a vertex $v$ with degree at most $\theta_1$ with two neighbors whose degree is at most $\theta_0$.
On the right is an illustration when there are two such vertices with degree greater than $\theta_1$.}
\label{fig:C4-ub}
\end{figure}

\smallskip\noindent\textsf{$C_4$s with at most one high-degree vertex.}
Suppose first that $|\calC_1|=\Omega(m)$. Observe  that since each $4$-cycle $\rho\in \calC_1$ contains at least two vertices with degree at most $\theta_0$ and at most one vertex with degree greater than $\theta_1$, it must contain at least one vertex, with degree at most $\theta_1$ whose neighbors on the $C_4$ both have degree at most $\theta_0$. For an illustration, see the LHS of Figure~\ref{fig:C4-ub}. 
Furthermore, we show that there exists a subset of $\calC_1$, which we denote by $\calC_1'$, such that $|\calC_1'|=\Omega(m)$, and every vertex $v$ that participates in one of the $C_4$s in $\calC_1'$, actually participates in $\Omega(d(v))$ edges-disjoint $C_4$s in $\calC_1$. 
It follows that in this case, when the algorithm selects a random edge (almost uniformly), with high constant probability it will obtain an edge with (at least) one endpoint $v$ having the above properties. Conditioned on the selection of such a vertex $v$,
the algorithm selects $\Theta(\sqrt{d(v)})$ random neighbors of $v$. By the birthday paradox, with high constant probability, among these neighbors there will be a pair of vertices that reside, together with $v$, on a common $C_4$ in $\calC_1'$. Once their (at most $\theta_0$) neighbors are queried, this $C_4$ is revealed.

\smallskip\noindent\textsf{$C_4$s with two high-degree vertex.}
We now turn to the case in which $|\calC_2| = \Omega(m)$.
Here too we can show that there exists a subset of $\calC_2$, denoted $\calC_2'$, such that $|\calC_2'|=\Omega(m)$, and every vertex $v$ that participates in one of the $C_4$s in $\calC_2'$ actually participates in $\Omega(d(v))$ edges-disjoint $C_4$s in $\calC_2$. 

Recall that by the definitions of $\calC$ and $\calC_2$ and since $\calC_2'\subseteq \calC_2\subseteq \calC$, the following holds. For each $4$-cycle $\rho$ in $\calC_2'$, since it is in $\calC_2$, there are two vertices whose degree is greater than $\theta_1$. Therefore, by the definition of $\calC$, they are both adjacent on $\rho$ to two vertices whose degree is at most $\theta_0$. Hence, if we consider the subgraph induced by the vertices and edges of the $C_4$s in $\calC_2'$, it is a bipartite graph, where on one side, denoted $L$, all vertices have degree at most $\theta_0$, and on the other side, denoted $R$, all vertices have degree greater than $\theta_1$. Furthermore, by the definition of $\calC'_2$, for each vertex in $R$, a constant fraction of its neighbors (in the original graph $G$) belong to $L$, and for each vertex in $L$, a constant fraction of its neighbors belong to $R$. For an illustration, see the RHS of Figure~\ref{fig:C4-ub}. 

Hence, if we select an edge almost uniformly and pick one of its endpoints with equal probability, with high constant probability we obtain a vertex $v\in R$. Conditioned on this event, since $d(v)> \theta_1$, the algorithm will perform $\widetilde{O}(n^{1/4}\alpha^{1/2})$ random walks of length two from $v$, and with high constant probability, a constant fraction of these walks will be of the form $(v,u,v')$ where $u\in L$ and $v' \in R$.
If for some $v'$ we get two walks, $(v,u,v')$ and $(v,u',v')$ for $u \neq u'$, then a $C_4$ is detected.

Observe that since all vertices in $R$ have degree greater than $\theta_1 = \Theta(n^{1/2})$, we have that $|R| \leq 2m/\theta_1 = O(n^{1/2}\alpha)$. This can be used to show that the expected number of pairs of walks that induce a $C_4$ is greater than $1$. In order to show that we actually get such a pair with high constant probability, we perform a more careful analysis to bound the variance.

\paragraph{A (two-sided error) lower bound for testing $C_4$-freeness in constant arboricity graphs.}
To obtain 
this lower bound \ICALPCR{of $\Omega(n^{1/4})$},
we define two distributions over graphs.
In the support of the first distribution, $\calD_0$, all graphs are $C_4$-free, and in the support of the second distribution, $\calD_1$, all graphs are $\Omega(1)$-far from being $C_4$-free. Furthermore, $\calD_0$ is uniform over all graphs isomorphic to a specific  graph $G_0$, and $\calD_1$ is uniform over all graphs isomorphic to a specific graph $G_1$.

\begin{figure}[htb!]
\centerline{\mbox{\includegraphics[width=0.5\textwidth]{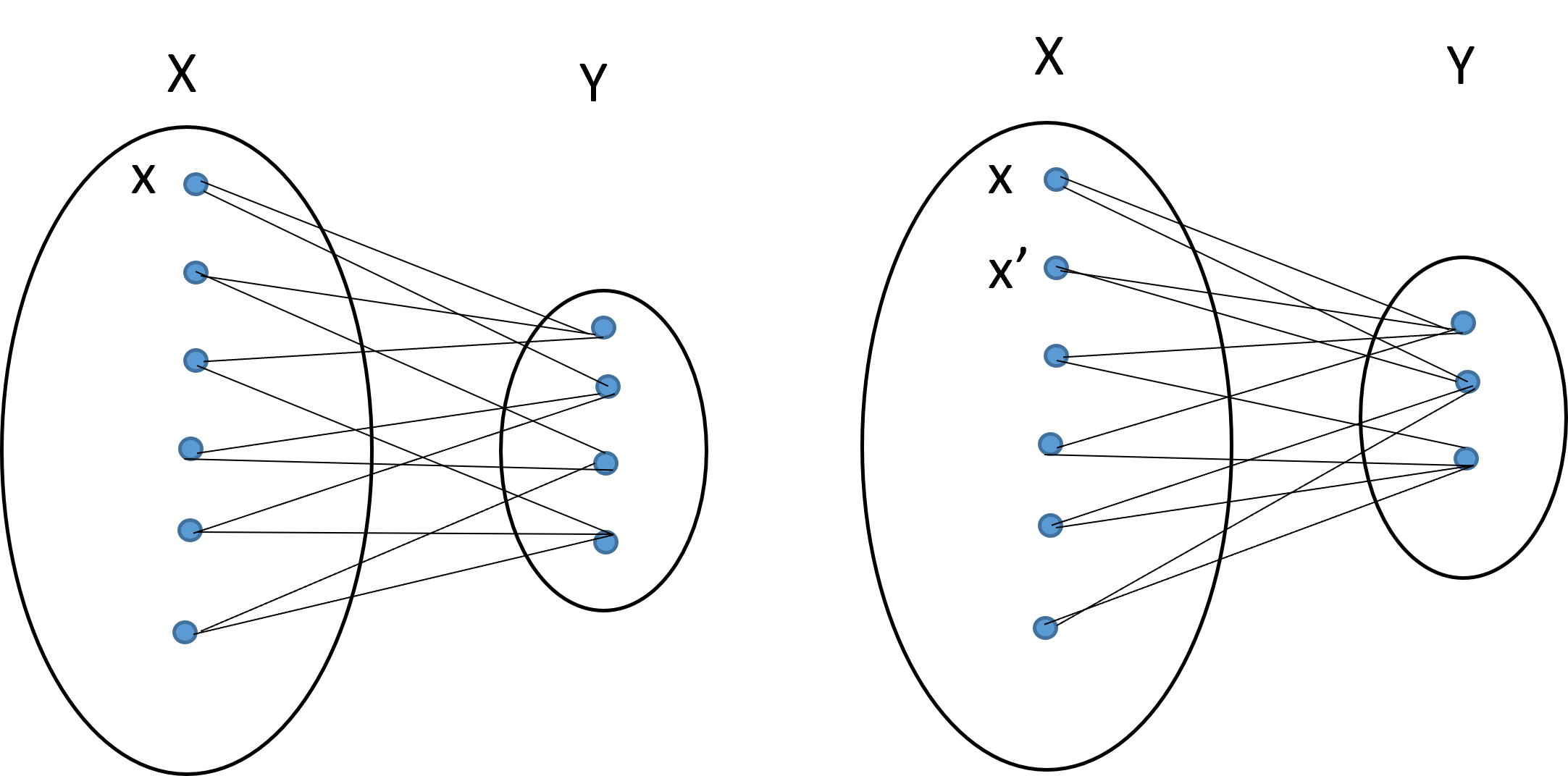}}}
\caption{An illustration for the lower bound construction. The graph on the left is $C_4$-free while the graph on the right contains $\Omega(m)$ edge-disjoint $C_4$s and is hence $\Omega(1)$-far from being $C_4$-free.}
\label{fig:C4-lb}
\end{figure}

We next describe a slightly simplified version of the two graphs (which cannot be used to prove the lower bound, but gives the essence of the proof). Both graphs are bipartite graphs, where one side, $Y$, contains $\Theta(\sqrt{n})$ vertices, and the other side, $X$, contains $\Theta(n)$ vertices, In $G_0$, each vertex in $X$ has a unique pair of neighbors in $Y$ (so there are no $C_4$s). On the other hand, in $G_1$, each vertex $x$ in $X$ has a ``twin'', $x'$, where $x$ and $x'$ have the same pair of neighbors in $Y$ (thus creating $\Omega(n)$ edge-disjoint $C_4$. See Figure~\ref{fig:C4-lb}.
\ICALPCR{Observe that the arboricity of both graphs is $2$ as for any subset of vertices $S$, the number of edges within $S$ is at most $|S\cap X| \cdot 2$ so the average degree in the subgraph induced by $S$ is at most $2$.}

In order to prove that no (possibly adaptive) algorithm can distinguish between a graph selected according to $\calD_0$ and a graph selected according to $\calD_1$, we define two processes, $\calP_0$ and $\calP_1$, which answer the queries of a testing algorithm while selecting a graph from $\calD_0$ (respectively, $\calD_1$) ``on the fly''.
The lower bound of $\Omega(n^{1/4})$ follows from the fact that when performing 
\SodaReb{fewer than $n^{1/4}/c$ queries (where $c$ is a sufficiently large constant),}
for both distributions, with high constant probability, each new neighbor query is answered by a uniformly selected vertex id.

\paragraph{A one-sided error lower bound for testing $C_4$-freeness in graphs with arboricity $\mathbf{\alpha}$.}
We next  discuss the lower bound of $\Omega(n^{1/4}\alpha^{1/2})$ for graphs with arboricity $\alpha = \Omega(\log n)$ and one-sided error algorithms. 

Here we define a single distribution $\calD$ which is uniform over a family of graphs with arboricity $\alpha$ such that almost all graphs in this family are $\Omega(1)$-far from $C_4$-free.

Roughly speaking, the graphs in the support of $\calD$ are random bipartite graphs, where one side, $Y$, is of size $\Theta(\sqrt{n}\alpha)$ and the other side, $X$, is of size $\Theta(n)$. Every vertex in $X$ has $\alpha$ neighbors in $Y$, and every vertex in $Y$ has $\Theta(\sqrt{n})$ neighbors in $X$. We need to show that if we select such a graph randomly, then on one hand it will be $\Omega(1)$-far from $C_4$-free, and on the other hand, in order to detect a $C_4$, any algorithm must perform $\Omega(n^{1/4}\alpha^{1/2})$ queries.

We next discuss the high-level idea as to why the resulting graphs are (with high constant probability) far from being $C_4$-free. Consider a fixed edge $(x,y)$ in the bipartite graph, where $x\in X, y\in Y$. The number of $C_4$s this edge participates in is determined by the number of edges between the sets of neighbors of $x$ and $y$, respectively $\Gamma(x)$ and $\Gamma(y)$. Recall that $x$ has $\Theta(\alpha)$ neighbors and $y$ has $\Theta(\sqrt n)$ neighbors. Since overall there are $|X|\cdot |Y|=\Theta(n^{3/2}\alpha)$ potential pairs in the bipartite graph, and $\Theta(n\alpha)$ edges, each pair in $X \times Y$ is an edge with probability $\Theta(1/\sqrt n)$. Hence, the expected number of edges between $\Gamma(x)$ and $\Gamma(y)$ is $|\Gamma(x)| \cdot |\Gamma(y)| \cdot (1/\sqrt n)=\Theta(\alpha)$. By analyzing the variance between pairs of edges, we  furthermore show that with high constant probability, most edges do not participate  in too many $C_4$s. Combining the two insights, it follows that with high constant probability, the graph is indeed far form being $C_4$-free.

In order to prove that any algorithm that performs at most  $n^{1/4}\alpha^{1/2}/c$   queries (for a sufficiently large constant $c$), will not detect a $C_4$ with high constant probability, we actually prove that it will not detect \emph{any} cycle.  Roughly speaking, we show that by the randomness of the construction, 
since $|Y|=\Theta(\sqrt n \alpha)$, and the algorithm performs $O(\sqrt{|Y|})$ queries,
each new neighbor query is answered by a uniformly   distributed vertex that has not yet been observed. Therefore, the algorithm essentially views a forest.

A central challenge that we need to overcome is that we do not want to allow parallel edges, where the above construction might lead to their existence. One possibility is to first define the distribution over graphs with parallel edges and then to remove them. The benefit is that due to the higher degree of independence in the construction, it is somewhat easier to formally prove that the graphs obtained (with parallel edges) are with high probability $\Omega(1)$-far from $C_4$-free, and this remains the case when we remove parallel edges.

However, this creates a 
difficulty when we turn to argue that no (one-sided error) algorithm can detect a $C_4$ unless it makes $\Omega(n^{1/4}\alpha^{1/2})$ queries. The difficulty is due to the fact that 
in the formal proof we need to deal with dependencies that arise due to varying degrees (which occur because parallel edges are removed). While intuitively, varying degree should not actually ``help'' the algorithm, this intuition is difficult to formalize.
Hence, we have chosen to define the distribution, from the start, over graphs that do not have parallel edges.
This choice creates some technical challenges of its own (in particular in the argument that the graphs obtained are $\Omega(1)$-far from $C_4$-free), but we are able to overcome them. For the full details see Section~\ref{sec:C4-lb}.


\subsubsection{The algorithm for $C_6$-freeness}
\ICALPCR{Recall that for $C_6$ we have a (one-sided error) testing algorithm whose query complexity is $\widetilde{O}(n^{1/2})$.}
In addition to assuming (for the sake of the exposition) that $\eps$ is a constant, we also 
ignore polylogarithmic factors in $n$. Similarly to the algorithm for testing $C_4$-freeness, the algorithm for testing $C_6$-freeness, \TestCsix,  in constant arboricity graphs is governed by two thresholds. The first, $\theta_0$, is of the order of the arboricity, so that it is a constant (recall that we assume that $\eps$ is a constant). The second, $\theta_2$, is of the order of $\sqrt{n}$.

The algorithm repeats the following process several times. It selects a vertex $v$ uniformly at random, and if $d(v) \leq \theta_0$, it performs a \emph{restricted} BFS starting from $v$ to depth $4$. Specifically:
\begin{packed_enum}
           \item Whenever a vertex $u$ is reached such that $d(u) \leq \theta_0$, all its neighbors are queried.
           \item Whenever a vertex $u$ is reached such that $d(u) > \theta_0$ and $u$ is reached from a vertex $u'$ such that $d(u') \leq \theta_0$, there are two sub-cases. If $d(u) \leq \theta_1$, then  all of $u$'s neighbors are queried. Otherwise,  $\theta_1$ neighbors of $u$ are selected uniformly at random. \label{step:C6_rand_nbr}
           \item Whenever a vertex $u$ is reached from a vertex $u'$ such that both $d(u)>\theta_0$ and $d(u') > \theta_0$, the BFS does not continue from $u$.
\end{packed_enum}
         
The algorithm rejects if and only if it observes a $C_6$.

Consider a graph that is far from being $C_6$-free, so that it contains a set  of $\Omega(m)=\Omega(n)$ edge-disjoint $C_6$s. Furthermore, it contains such a set, denoted $\calC$ for which every $C_6$ in $\calC$ contains at most three vertices with degree greater than $\theta_0$, and furthermore, these vertices are  \emph{not} adjacent on the $C_6$. We partition $\calC$ into three subsets: $\calC_1$, $\calC_2$, and $\calC_3$, depending on the number of vertices with degree greater than $\theta_0$ that it contains.

If either $|\calC_1| = \Omega(m)$, or $|\calC_2| = \Omega(m)$, then it is not hard to show that the algorithm will detect a $C_6$ with high constant probability. The more interesting part of the proof is handling the case in which only $|\calC_3| = \Omega(m)$.

In this case we define an auxiliary multi-graph, denoted $G'$, over the set of vertices that participate in $C_6$s belonging to $\calC_3$, and have degree greater than $\theta_0$ (in $G$). We denote this set of vertices by $M$, and the set of vertices with degree at most $\theta_0$ that participate in these $C_6$s, by $L$. 

\begin{figure}[htb!]
\centerline{\mbox{\includegraphics[width=0.5\textwidth]{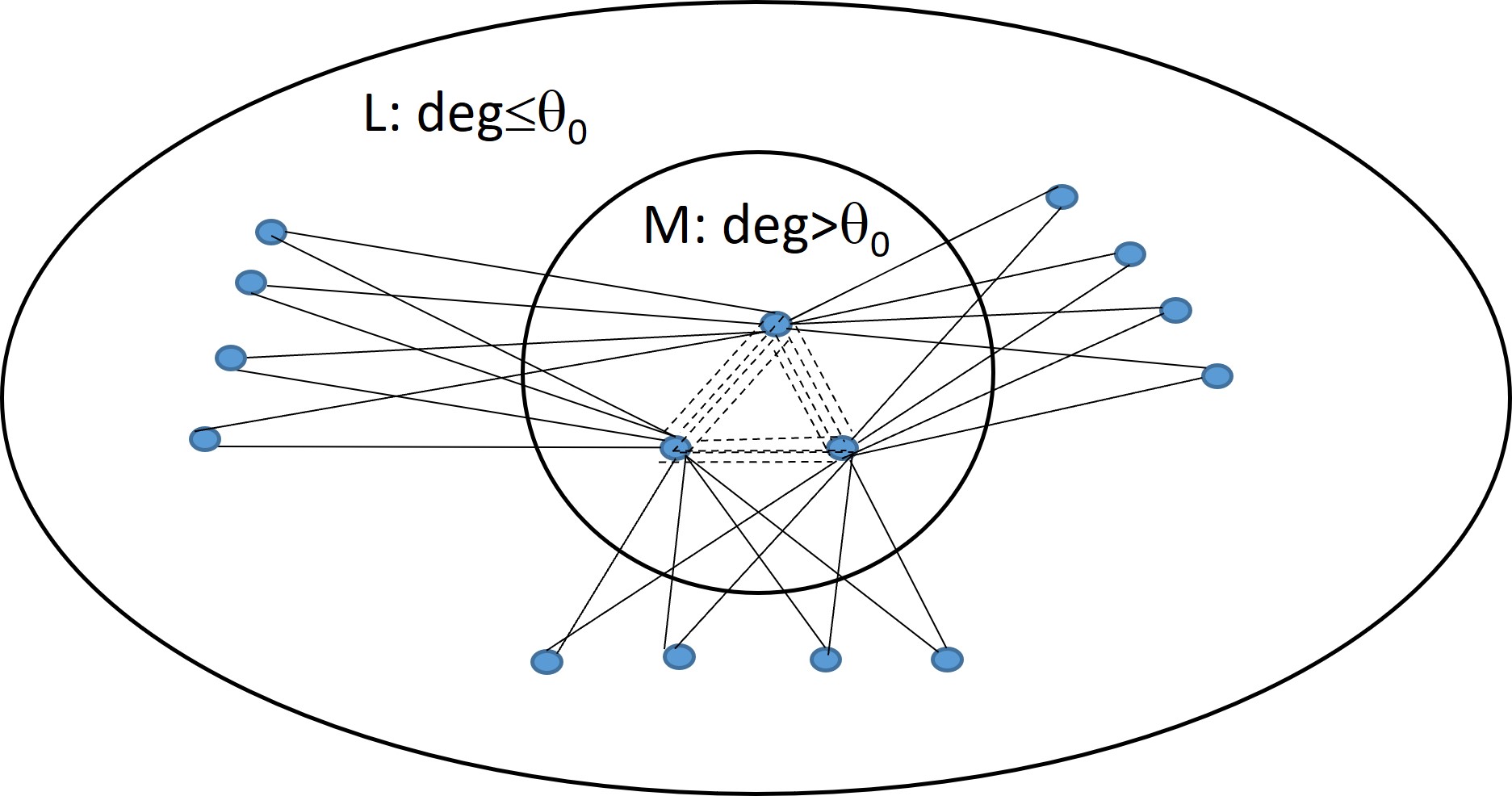}}}
\caption{An illustration of the auxiliary (multi-)graph $G'$ in the $C_6$-freeness testing algorithm. The dashed lines represent edges in $G'$, each one corresponding to a length-2 path in $G$ that passes through a vertex with degree at most $\theta_0$.
}
\label{fig:C6}
\end{figure}

Assume for simplicity that each vertex in $L$ has degree exactly $2$ (i.e., it participates in a single $C_6$).
For each pair of vertices in $M$, we put in $G'$ a set of parallel edges, whose size equals the number of  length-2 paths between them in $G$ that pass through vertices in $L$.
Hence, for each $C_6$ in $\calC$, we have a triangle in $G'$, where these triangles are edge-disjoint, and we denote their set by $\calT$. See Figure~\ref{fig:C6}.

Observe that selecting a vertex uniformly at random from $L$ 
and querying its two neighbors in $M$ 
corresponds to selecting an edge uniformly at random in $G'$. If we add an additional simplifying assumption by which (in $G$), vertices belonging to $M$ only neighbor vertices belonging to $L$, then our algorithm on $G$ essentially translates to picking a random edge in $G'$. Then depending on the degree of the endpoints, either querying all their neighbors in $G'$ or $\theta_1$ random neighbors. 

Let $H$ denote the subset of vertices in $M$ whose degree in $G$ is greater than $\theta_1$.
If relatively many triangles in $\calT$ contain at most
one vertex in $H$, then we are done, since these triangles contain an edge for which  both endpoints have degree at most $\theta_1$.
Hence, it remains to address the case in which almost all triangles in $\calT$ have two or three vertices in $H$.

Roughly speaking, in this case we show that the existence of many edge-disjoint, but not vertex-disjoint, 
 triangles in $G'$ that contain such high-degree vertices  implies the existence of ``many more'' triangles that may be caught by our algorithm. As an illustrative extreme (but easy) special case, assume that in $G'$ there are only three vertices. Then the existence of some number $t$ of edge-disjoint triangles between them, actually implies the existence of $t^3$ (non edge-disjoint) triangles.

\subsubsection{The general lower bound 
   for $C_k$-freeness, $k\geq 6$}
We establish our general lower bound \ICALPCR{of $\Omega(n^{1/3})$}
for one-sided error testing of $C_k$-freeness when $k\geq 6$ by 
building on a lower bound for testing triangle-freeness that appears in~\cite[Lemma 2]{AKKR08}. This lower bound on testing triangle-freeness is based on the difficulty of detecting a triangle in graphs selected uniformly from a family $\mG_{\trn}$ of graphs 
in which almost all graphs are $\Omega(1)$-far from being triangle-free. All graphs in the family are $d$-regular tri-partite graphs over $\trn$ vertices and the lower bound on the number of queries necessary to detect a triangle (with constant probability), is $\Omega(\min\{d,\trn/d\})$. By setting $d= \sqrt{\trn}$, the lower bound is $\Omega(\sqrt{\trn})$.

We show that, for any constant $k\geq 6$, if we had a one-sided error testing algorithm $\mA$ for  $C_k$-freeness of graphs with $n$ vertices and constant arboricity using at most $n^{1/3}/c$ queries (for a constant $c$), then we would be able to detect triangles in graphs selected uniformly from $\mG_{\trn}$ using at most $\sqrt{\trn}/c'$ queries (for a constant $c'$).

\begin{figure}[h!]
\centerline{\mbox{\includegraphics[width=0.4\textwidth]{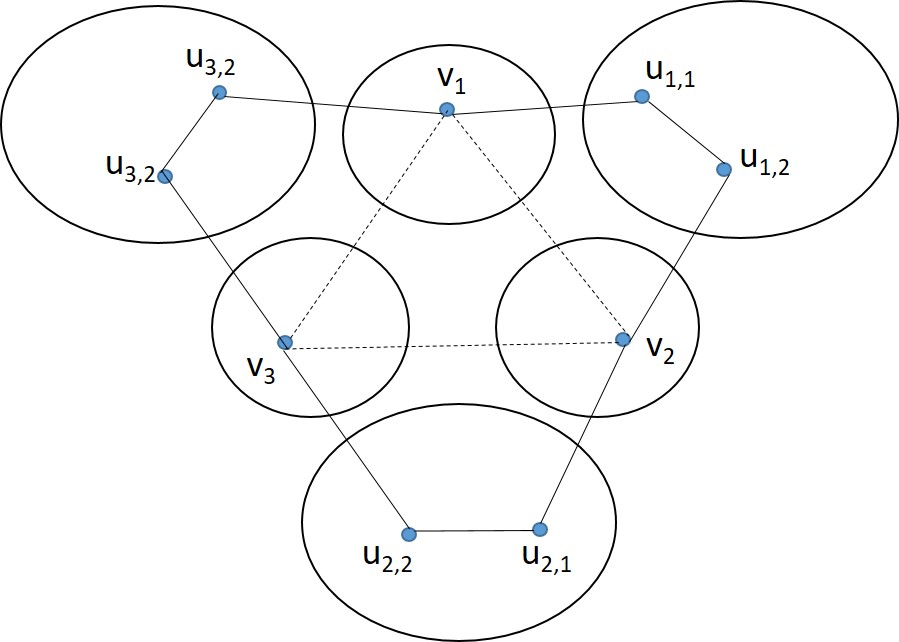}}}
\caption{An illustration for the lower bound construction for $C_k$-freeness in constant arboricity graphs when $k=9$.
The three circles in the middle and the dashed lines represent a graph $G'\in \calG_{n'}$. The outer circles represent the additional vertices in $G$. Since $k=9$ in this example, each edge in $G'$ is replaced by a path of length 3 in $G$.
}
\label{fig:gen-lb}
\end{figure}

To this end we define an algorithm $\mA$ that, given query access to a graph $\trG \in \mG_{\trn}$, implicitly defines a graph $G$ for which the following holds. First, the number of vertices in $G$ is $n=\Theta({(\trn)}^{3/2})$, and the number of edges is $m = \Theta(\trm)$, where $\trm$ is the number of edges in $\trG$ (so that $\trm = \Theta((\trn)^{3/2})$). Second, $G$ has arboricity $2$. Third, the distance of $G$ to $C_k$-freeness is of the same order as the distance of $\trG$ to triangle-freeness. Fourth, there is a one-to-one correspondence between triangles in $\trG$ and $C_k$s in $G$. The basic idea is to replace edges in the tri-partite graph $\trG$ with paths of length $k/3$. See Figure~\ref{fig:gen-lb}

Assuming there existed a testing algorithm $\mA$ as stated above, the algorithm $\trA$ would use it to try and find a $C_k$ in $G$ (and hence a triangle in $\trG$). In order to be able to run $\mA$ on $G$, the algorithm $\trA$ must be able to answer queries of $\mA$ to $G$ by performing queries to $\trG$. We show how this can be done with  a constant multiplicative overhead.
Hence, the lower bound of $\Omega(\sqrt{\trn})$ for testing triangle-freeness (when the degree is $\Theta(\sqrt{\trn})$) translated into a lower bound of $\Omega(n^{1/3})$ for testing $C_k$-freeness.

\subsubsection{The general upper bound for  $C_k$-freeness}
Recall that our starting point is that if $G$ is $\Omega(1)$-far from being $C_k$-free, then it contains a set $\calC$ of $\Omega(m)$ edge-disjoint $C_k$'s that do not contain any edge between vertices that both have degree greater than $\theta_0 = \Theta(\alpha)$. We refer to vertices with degree at most $\theta_0$ as \emph{light} vertices, and to those with degree greater than $\theta_0$ as \emph{heavy}.
Hence, each $C_k$ in $\calC$ has at least $\lceil k/2\rceil$ light vertices, and each heavy vertex
on it neighbors two light vertices.

We present two different algorithms, where each of them is suitable for a different setting. The basic idea of both algorithms is to take a large enough sample of vertices and edges so that the  subgraph determined by 
the sampled light vertices and their incident edges, as well as the sampled edges, contains a copy of $C_k$. \ICALPCR{The query complexity of each algorithm is stated following its high-level description.}

\paragraph{The first algorithm.}
Our first algorithm simply samples vertices uniformly, independently at random, and then performs queries that reveal the neighbors of all light vertices in the sample.
To analyze what is the sufficient sample size for this algorithm, consider the following generalization of the birthday paradox for $k$-way collisions. Assume we sample elements under the uniform distribution over $[n]$. Then we obtain a $k$-way collision after taking $\Theta(n^{1-1/k})$ samples.
Similarly, suppose  we sample vertices uniformly from a graph that is composed only of $n/k$  vertex-disjoint copies of $C_k$. Then, after sampling $\Theta(n^{1-1/k})$ vertices, we will hit all the vertices of at least one of the copies (with high constant probability).
Conditioned on this event, if we reveal the neighborhood of all the vertices in the sample, then  we obtain a $C_k$.

The next observation is that, in fact, we only need to hit a \emph{vertex cover} of a copy of a $C_k$ (as opposed to all its vertices). In particular we would like to hit such a cover that contains only light vertices, which we refer to as a \emph{light vertex cover}. 
For constant $\alpha$, this yields an improved dependence on $k$ in the exponent, i.e., $O(n^{1-1/\lfloor k/2\rfloor})$ sampled vertices suffice.
%

When taking into account the dependence on $\alpha$ (so that it is not necessarily true that $m=O(n)$) and incorporating this in the analysis, we prove that the query complexity is upper bounded by $O(m \cdot (\alpha/m)^{2/k})$ for even $k$ and $O(m \cdot (\alpha/m)^{2/(k+1)})$ for odd $k$ 
(up to a polynomial dependence on $k$).
Since $\alpha \leq \sqrt{m}$ it follows that the above bounds are at most $O(m^{1-1/k})$ and $O(m^{1-1/(k+1)})$, respectively. 

\paragraph{The second algorithm.}
Our second algorithm is designed for the case in which $k$ is odd and $m = \Omega(\alpha^{(k+3)/2})$. In particular it is preferable when $\alpha$ is constant. 
We observe that when $k$ is odd, for each $C_k$ in $\calC$, there is an edge in 
 which both endpoints are light vertices.
 \SodaReb{Therefore, if we sample edges (almost) uniformly from the graph (using a variant of the procedure \SelectEdge), then} we are likely to hit one of these edges. This additional step reduces the number of vertices we need to hit in each copy by $2$, which  results in improved complexity for some range of the parameters. In particular, the query complexity of this algorithm is $O(m \cdot (\alpha^2/m)^{2/(k-1)})$.
 Specifically, when $\alpha$ is a constant, the query complexity of this algorithm (which works for odd $k$) is $O(m^{1-2/(k-1)})$ (instead of $O(m^{1-2/(k+1)})$).

 \paragraph{General subgraph $F$.}
 Our first algorithm also works for any constant-size subgraph, $F$, where the upper bound on the sample size is of the form $m^{1-1/\ell(F)}$ where $\ell(F)$ depends on the structure of $F$, and is at most $k=|V(F)|$.
 The idea is that if we want to find a copy of $F$, it suffices to hit a light vertex cover of this copy and then query all neighbors of the sampled light vertices. 


\subsection{Related work}\label{subsec:related}
In this subsection we shortly discuss several related works, in addition to the two aforementioned works regarding testing $C_3$-freeness~\cite{AKKR08,Reut}.

Testing subgraph-freeness for fixed, constant size subgraphs in the dense-graphs model can be done using a number of queries that depends only on $1/\eps$ (where the dependence is a tower of height $\poly(1/\eps)$), as shown by Alon, Fischer, Krivelevich and Szegedy~\cite{AFKS00}.  Alon~\cite{Alo02} proved that a super polynomial dependence on $1/\eps$ is necessary, unless the subgraph $F$ is bipartite. Goldreich and Ron addressed the problem in the bounded-degree model~\cite{GR-bound}, and gave a simple algorithm that depends polynomially on $1/\eps$ and the maximum degree in the graph, and exponentially on the diameter of $F$.

A special case of graphs that have bounded arboricity is the family of graphs that exclude a fixed minor (a.k.a. minor-free graphs).
Newman and Sohler~\cite{NS13} showed that 
for this family of graphs, in the bounded-degree model, all properties 
can be tested with no dependence on the size of the graph $G$. Moreover, it was recently shown~\cite{SS21,LS21} that any property which is monotone and additive\footnote{A property is monotone if it closed under removal of edges and vertices. A property is 
additive if it is closed under the disjoint union of graphs.} (and in particular $F$-freeness where $F$ is a connected graph) can be tested using a number of queries that is only polynomial in $1/\eps$ and $d$, where $d$ is the degree bound (and $O(d^{\rho(\eps)})$ in general $(\eps, \rho(\eps))$-hyperfinite graphs\footnote{Let $\rho$ be a function from $\mathbb{R}_+$ to $\mathbb{R}_+$. A graph $G = (V, E)$ is $(\eps, \rho(\eps))$-hyperfinite if for every $\epsilon > 0$ it is possible to
remove $\eps |V|$ edges of the graph such that the remaining graph has connected components of size at most $\rho(\eps)$.}).
For minor-free graphs with unbounded degrees, Czumaj and Sohler~\cite{CS19} showed that a property is testable 
with one sided error and a number of queries that does not depend on the size of the graph if and only if it can be reduced to testing for a finite family of finite forbidden subgraphs.\footnote{They consider a model in which they can perform only random neighbor queries.} The correctness of their algorithm relies on the fact that the arboricity of minor-free graphs remains constant even after contractions of edges (which is not the case for general constant-arboricity graphs). 

In general graphs, it was shown that $k$-path freeness~\cite{IY18} and more generally $T$-freeness where $T$ is a tree of order $k$~\cite{EFFGLMMOORT17}, can be tested with time and query complexity that depend only on $k$, assuming the edges of the graph can be accessed uniformly at random.
Testing cycle-freeness (where a no instance is a graph that is far from being a forest) was studied in the bounded-degree model in~\cite{GR-bound}, where a two-sided error algorithm was given whose query complexity is polynomial in $1/\eps$ and the degree bound. Czumaj  et. al~\cite{CGRSSS14} showed that the complexity of this problem for one-sided error algorithms in the bounded-degree model is $\widetilde{\Theta}(\sqrt{n})$ (for constant $\eps$ -- their algorithm has a polynomial dependence on $1/\eps$), and the algorithm can be adapted to the general-graphs model.

Other sublinear-time graph algorithms 
for counting and sampling (rather than detecting) subgraphs 
that give improved results when the graph $G$ has bounded arboricity include~\cite{ERS19,ERR19,ERS20, ERR22}. 

\subsection{Organization}
We start in Sections~\ref{sec:prel} with some preliminaries. In 
Section~\ref{sec:C4} we give the upper bound for testing $C_4$-freeness. Few details are deferred to Appendix~\ref{sec:4.1}, and the lower bound for testing $C_4$-freeness is given in 
Appendix~\ref{sec:C4-lb}.
In Appendix~\ref{sec:C5} we provide the adaptation of the algorithm for testing $C_4$-freeness, to testing $C_5$-freeness (in graphs with constant arboricity). 
Our algorithm for testing $C_6$-freeness (in graphs with constant arboricity) is given in Appendix~\ref{sec:C6-ub}. Our lower bound for one-sided error testing of $C_k$-freeness is proved in Appendix~\ref{sec:Ck-lb}, and the upper bound is given in Appendix~\ref{sec:Ck}.

\section{Preliminaries}\label{sec:prel}

Unless stated explicitly otherwise, the graphs we consider are simple, so that in particular they do not contain any parallel edges. We denote the number of vertices in the graph by $n$ and the number of edges by $m$. Every vertex $v$ in the graph has a unique id, denoted $id(v)$, and its degree is denoted by $d(v)$.

We work in what is known as the \emph{general graph model}~\cite{ParnasR02,KKR04}. In particular, under this model,
the \emph{distance} of a graph $G$ to $C_k$-freeness, denoted $dist(G,C_k\text{-free})$, is the minimum fraction of edges that should be  removed from $G$ in order to obtain a $C_k$-free graph.
As for the allowed queries, 
a neighbor query to the $i$th neighbor of a vertex $v$ is denoted by $nbr(v,i)$, and to its degree by $deg(v)$. A pair query between two vertices $v_1$ and $v_2$ is denoted by $pair(v_1,v_2)$.
Given query access to a graph $G$ and a parameter $\eps$, a one-sided error testing algorithm for $C_k$-freeness should accept $G$ if it is $C_k$-free, and should reject $G$ with probability at least $2/3$ if $dist(G,C_k\text{-free}) > \eps$.
If the algorithm may also reject $C_k$-free graphs with probability at most $1/3$, then it has two-sided error.

As noted in the introduction, we assume our algorithms for graphs whose arboricity is not promised to be constant,  are given an upper bound $\alpha$ on the arboricity $arb(G)$ of the tested  graph $G$, and their complexity depends on this upper bound. Alternatively, if the algorithm is provided with the number of edges, $m$, then it may run a procedure from~\cite{Reut} to obtain a value $\alpha^*$ that with high constant probability satisfies the following: (1) $\alpha^* \leq 2arb(G)$; (2) The number of edges between vertices whose degree is at least $\alpha^*/(c\eps)$ for a constant $c$ is at most $(1-\eps/c')m$ (for another, sufficiently large, constant $c'$).
Up to polylogarithmic factors in $n$, the query complexity and running time of the procedure are $O(arb(G)/\eps^3)$ with high probability (assuming the average degree is $\Omega(1)$). 


Throughout this work we assume, whenever needed, that $\epsilon$ is upper bounded by some sufficiently small constant (or else it can be set to that constant).


\section{An upper bound of $\widetilde{O}(n^{1/4} \alpha)$ for testing $C_4$-freeness}\label{subsec:C4-ub}
\label{sec:C4}

\ICALPCR{In this subsection we prove the more general (arboricity-dependent) form of the upper bound for testing $C_4$-freeness which is stated as Theorem~\ref{thm:C4-ub} in the introduction.}

Recall that the assumption on $\alpha$ is that it is an upper bound on the arboricity $arb(G)$ - hence, while it is known that for graphs with $arb(G)>\sqrt{n}$ there exists a $C_4$,  we cannot simply reject if we get $\alpha>n^{1/2}$ since it might be that $arb(G)<\sqrt{n}$ (and we want one-sided error). 
The above expression shows that in the  case that $\alpha>n^{1/2}$, the $n^{1/4}\alpha$ term is replaced by $n^{3/4}$ (and the additive $\alpha$ term is due to the edge sampling).


The algorithm referred to in Theorem~\ref{thm:C4-ub} is described next. 

\medskip\noindent
\textsf{Test-$C_4$-freeness}($n,\epsilon,\alpha$)\label{alg:C4}
\begin{enumerate}
\item Let $\theta_0 = 4\alpha/\epsilon$,
$\theta_1 = c_1\cdot \sqrt{n}/\eps$ (where 
$c_1$ will be determined subsequently) and $\theta_{min}=\min\{\theta_0, \theta_1\}$  (it is useful to read the algorithm while having in mind that $\theta_0\leq \theta_1$ (i.e., $\alpha = O(\sqrt{n})$) so that $\theta_{min}=\theta_0$).
\label{step:set_params_C4}
\item Repeat the following $t=\Theta(1/\epsilon)$ times:
  \begin{enumerate}
     \item Select an edge $e$ by calling the procedure \SelectEdge($\alpha,\epsilon$), which is provided in Appendix~\ref{append:sample-edge}.  If it
     does not return an edge, then continue to the next iteration.
     \item Select an endpoint $v$ of $e$ by flipping a fair coin.
     \item \label{step:random-neighbors} If $d(v) \leq \theta_1$, then select $s_1=\Theta( \sqrt{d(v)/\eps})$ ~($=O(n^{1/4}/\eps)$) random neighbors of $v$, and for each neighbor $u$ such that  
     $d(u) \leq \theta_{min}$
     query all the neighbors of $u$. 
     \item \label{step:random-walk} Otherwise ($d(v) > \theta_1$), perform $s_2=\Theta(\sqrt{  (n\alpha/\theta_1)\log n}/\eps^2)$ 
      \;$(=\tilde{O}(n^{1/4}\alpha^{1/2}/\eps^2))$ 
     random walks of length $2$ starting from $v$. 
     \item If a $C_4$ is detected, then return it, `Reject' and terminate.
  \end{enumerate}
  \item Return `Accept'.
\end{enumerate}
We note that the algorithm can be unified/simplified so that it only performs random walks of length-2 
where the number of walks is $\Theta(n^{1/4}\alpha/\eps^2)$, 
but then the analysis becomes slightly more complicated. 

We start by stating a claim concerning the procedure \SelectEdge\ -- where both the procedure and the proof of the claim are deferred to Appendix~\ref{append:sample-edge}. We then state and prove two additional claims that will be used in the proof of Theorem~\ref{thm:C4-ub}.

\begin{claim}\label{clm:select-edge} 
With probability at least $2/3$ the procedure \SelectEdge\ returns an edge. Conditioned on it returning an edge, each edge incident to a vertex with degree at most $\theta_0$ is returned with probability at least $1/(2m')$ and at most $1/m'$, where $m'$ is the number of edges incident to vertices with degree at most $\theta_0$.
\end{claim}

\begin{claim}\label{clm:C4-ub1}
Let $v$ be a vertex and let $C(v,\theta_{min})$ be a set of edge-disjoint $C_4$'s 
containing $v$ such that the neighbors of $v$ on these $C_4$s all have degree at most $\theta_{min}$, where $\theta_{min}$ is as defined in the algorithm.\footnote{Actually, we do not rely on the setting of $\theta_{min}$, so this claim holds for any threshold value.} Suppose that $|C(v,\theta_{min})|\geq 1$ and let $\eps' = |C(v,\theta_{min})|/d(v)$.
If we select $s = 16\sqrt{d(v)/\eps'}$ random neighbors of $v$, and for each selected neighbor $u$ such that {$d(u)\leq \theta_{min}$} we query all the neighbors of $u$,
then the probability that we obtain a $C_4$ is at least $9/10$.
\end{claim}

\begin{proof}
Let $E'(v)$ denote the set of edges incident to $v$ that participate in the set $C(v,\theta_{min})$. By the premise of the claim, $|E'(v)|/d(v)= 2|C(v,\theta_{min})|/d(v)= 2\eps'$. Let $s'$ be the number of neighbors  of $v$ that are incident to edges in $E'(v)$ among the $s$ selected random neighbors of $v$.
It holds that $\Ex[s']=2\eps' \cdot s$, and  by the multiplicative Chernoff bound, $s' \geq \eps' \cdot s$
with probability at least $1-e^{-\eps'\cdot s/4}$. 
We first show that this probability is at least $19/20$, and then condition on this event.
By the setting of $s=16\sqrt{d(v)/\eps'}$, it holds that $  \eps'\cdot s = 16\sqrt{\eps'\cdot d(v)},$ and by the setting of  $\eps'=|C(v,\theta)|/d(v)$, we get $\eps'\cdot s \geq 16\sqrt{|C(v, \theta_{min})|}\geq 16$. 
Therefore, with probability at least $19/20$, $s'>\eps'\cdot s= 16\sqrt{\eps' \cdot d(v)}$.
We condition on this event and consider only those $s'$ 
selected neighbors of $v$ that are endpoints of $E'(v)$.

For each $4$-cycle $\rho\in C(v,\theta_{min})$, let $u_1(\rho)$ and $u_2(\rho)$ be the two neighbors of $v$ on this $C_4$ (so that they are endpoints of edges in $E'(v)$). Since the $C_4$s in $C(v,\theta_{min})$ are edge-disjoint, these vertices are distinct. 
Observe that the $s'$ selected neighbors of $v$   are uniformly distributed in $\bigcup_{\rho\in C(v,\theta_{min})}\{u_1(\rho),u_2(\rho)\}$, and that $s' \geq 16\cdot\sqrt{|C(v,\theta_{min})|}$. Hence, by the ``birthday paradox'',  with high constant probability, the sample of neighbors of $v$ contains two vertices, $u_1(\rho)$, and $u_2(\rho)$
for some $\rho \in C(v,\theta_{min})$. 
Conditioned on this event, once the (at most $\theta_{min}$) neighbors of
$u_1(\rho)$ and $u_2(\rho)$ are queried, the four-cycle $\rho$ is observed.
\end{proof}

\begin{claim}\label{clm:C4-ub2}
Let $G$ be a graph over $n$ vertices and $m$ edges, and let $\theta_1, \eps', \eps''$ be parameters.
Suppose that $G$ contains a bipartite subgraph $G' = (L,R,E(G'))$ such that every vertex in $R$ has degree at least $\theta_1$ in $G$. 
Let $v$ be a vertex in $R$ such that $v$ has at least $\eps' \cdot d(v)$ neighbors in $L$ where each  of these neighbors, $u$, has at least $\eps'' \cdot \max\{d(u),\frac{m}{n}\}$ neighbors in $R$. 
If $\theta_1\geq 2\sqrt{n/(\eps'\cdot \eps'')}$ and we take $s_2 \geq \frac{32}{\eps'\cdot \eps''}\cdot \sqrt{2\log n\cdot |R|}$ random walks of length 2 from $v$ for a sufficiently large constant $c'$, then with probability at least $9/10$ we shall detect a $C_4$ in $G$.
\end{claim}

\begin{proof}
For a pair of vertices $v$ and $v'\neq v$ in $R$, let $\ell_2(v,v')$ be the number of length-2 paths between $v$ and $v'$, and let $\ell_2(v,R) = \sum_{v'\in R} \ell_2(v,v')$.  Consider taking two random length-2 walks from $v$, and let $\mathcal{E}_1$ be the event that both of them end at vertices in $R$. Let $\mathcal{E}_2$ be the event that these two paths are distinct and end at the same vertex. Then for each single vertex $v'\in R$, conditioned on $\mathcal{E}_1$, the probability that the two walks end at $v'$ is exactly $\frac{\ell_2(v,v')}{\ell_2(v,R)}\cdot   \frac{\ell_2(v,v')-1}{\ell_2(v,R)}$.
Therefore,
\begin{equation} \label{eq:col}
\Pr[\mathcal{E}_2\,| \, \mathcal{E}_1] = \sum_{v'\in R} \frac{\ell_2(v,v')}{\ell_2(v,R)}\cdot   \frac{\ell_2(v,v')-1}{\ell_2(v,R)}
=  \frac{1}{(\ell_2(v,R))^2} \cdot \sum_{v'\in R} (\ell_2(v,v'))^2 - \frac{1}{\ell_2(v,R)}\;.
\end{equation}
We would like to lower bound the above probability.
For the first term on the right-hand-side, by applying the Cauchy-Schwartz inequality we get that
\begin{equation}\label{eq:col1}
\frac{1}{(\ell_2(v,R))^2} \cdot \sum_{v'\in R} (\ell_2(v,v'))^2 \geq  \frac{1}{(\ell_2(v,R))^2} \cdot |R| \cdot \left( \frac{\ell_2(v,R))}{|R|}\right)^2 = \frac{1}{|R|}\;.
\end{equation}
By combining Equations~\eqref{eq:col} and~\eqref{eq:col1} we get that
$\Pr[\mathcal{E}_2\,| \, \mathcal{E}_1] \geq \frac{1}{|R|} - \frac{1}{\ell_2(v,R)}\;.
$
Since each vertex in $R$ has degree at least $\theta_1$, we have that $|R| \leq \frac{2m}{\theta_1}$. 
By the premise of the claim regarding $v$ and its neighbors, $v$ has $\eps' d(v)\geq \eps'\cdot \theta_1$ neighbors in $L$, and each of them has 
at least $\eps''\cdot (m/n)$ neighbors in $R$. Therefore,
\begin{equation}
\ell_2(v,R) \geq \eps'\cdot \theta_1\cdot \eps''\cdot \frac{m}{n} 
 \geq \frac{\eps'\cdot \eps''\cdot \theta_1^2 \cdot |R|}{2n}\geq 2|R|,
\end{equation}
where the last inequality is by the premise $\theta_1 \geq 2\sqrt{n/(\eps'\cdot \eps'')}$. Therefore,
$\Pr[\mathcal{E}_2\,| \, \mathcal{E}_1] \geq \frac{1}{2|R|} \;.$
So far we have shown that when taking two distinct random walks from $v$, and conditioned on them both ending at $R$ (the event $\mathcal{E}_1$), the two paths collide on the end vertex (and hence result in a $C_4$) with probability at least $1/2|R|$. We shall now prove, 
that when taking $s$ length-2 random walks from $v$, sufficiently many of them indeed end at $R$, and that with high probability, at least two of them collide, resulting in a $C_4$.

Consider first the event $\mathcal{E}_1$.
By the premise of the claim, $v$ has at least $\eps'\cdot d(v)$ neighbors in $L$, and each $u$ of them has at least $\eps'' \max\{d(u), m/n\}\geq \eps'' d(u)$ neighbors in $R$. Therefore, 
the probability that a single random walk from $v$ ends at $R$ is at least $\eps'\cdot \eps''$. 
Hence, if we take $s\geq  \frac{32}{\eps' \cdot \eps''}\sqrt{2\log n \cdot |R|}$ length-2 random walks from $v$, and let $s'$ denote the number of  walks that end at a vertex in $R$,  we have that $\Ex[s']=32\cdot \sqrt{2\log n \cdot |R|}$, and that with probability at least $9/10$, we have $s' \geq 16\cdot \sqrt{2\log n \cdot |R|}$. We henceforth condition on this event. 

 Let $\chi'_{i,j}$ denote the event that the $i$th and $j$th random walks among the ones that end at $R$ collide on the ending vertex (and thus result in a $C_4$). By the above discussion, we have that for a specific pair $i\neq j$, $\Pr[\chi'_{i,j}=1]\geq 1/2|R|$. We now  lower bound the probability that at least one pair of random walks from the $s'$ that end in $R$ detects a $C_4$, i.e. lower bound $\sum_{i,j\in[s']}\chi_{i,j}$,  using Corollary~\ref{cor:cheb1}. For that end  we also need to upper bound the variance of the sum. 
 
 Partition the vertices in $R$ according to $\ell_2(v,v')$, where $R_x(v) = \{v': 2^{x-1} < \ell_2(v,v') \leq 2^x\}$
for $x = 0, \dots \log L \leq \log n$.
Since $\sum_{v'\in R} \frac{\ell_2(v,v')}{\ell_2(v,R)}\cdot   \frac{\ell_2(v,v')-1}{\ell_2(v,R)}>\frac{1}{2|R|}$,  there exists at least one setting of $x$ for which
$\sum_{v'\in R_x} \frac{\ell_2(v,v')}{\ell_2(v,R)}\cdot   \frac{\ell_2(v,v')-1}{\ell_2(v,R)} \geq \frac{1}{2|R|\log n}$. We denote this setting by $x^*$ and observe that $x^* > 0$ (since for every $v'\in R_0$, $\ell_2(v,v')-1 =0$).

For every $i,j \in [s']$, $i<j$, we define a Bernoulli random variable $\chi_{i,j}$ that is 1 if and only if the $i$th and the $j$th random walks from $v$ (among the $s'$ considered) end at the same $v'\in R_{x^*}$ and pass through a different vertex in $L$. We next show that we can apply Corollary~\ref{cor:cheb1} (with $s$ in that claim set to $s'$) to get an upper bound on the probability that $\sum_{i,j \in [s'], i<j} \chi_{i,j} = 0$ (which is an upper bound on the probability that we do not detect a $C_4$).

 By the definition of the random variables, for every $i_1\neq i_2, j_1\neq j_2$, it holds that $\chi_{i_1,j_1}, \chi_{i_2, j_2}$ are independent,  so that 
the first condition in Corollary~\ref{cor:cheb1} is satisfied. Next, for any pair $i,j\in [s']$, $i<j$ we have that
\begin{equation} \label{eq:mu_clm6}
\mu = \Pr[\chi_{i,j}=1] = \sum_{v'\in R_{x^*}} \frac{\ell_2(v,v')}{\ell_2(v,R)}\cdot   \frac{\ell_2(v,v')-1}{\ell_2(v,R)} \geq \frac{1}{2|R|\log n}\;.
\end{equation}
Therefore, we have that  $s' \geq 16\cdot\sqrt{2|R|\log n}=16/\sqrt{\mu}$, and so  
the third condition in Corollary~\ref{cor:cheb1}
is satisfied (for $c_2=16$, where $s'$ serves as the parameter $s$ in the corollary).  

It remains to verify that the second condition holds. For any four indices $i_1,j_1,i_2,j_2\in [s']$, $i_1 < j_1$, $i_2 < j_2$ such that exactly two of the four indices are the same,
\begin{equation}
\Pr[\chi_{i_1,j_1}=\chi_{i_2,j_2} = 1] = \sum_{v'\in R_{x^*}} \frac{\ell_2(v,v')}{\ell_2(v,R)}\cdot   \left(\frac{\ell_2(v,v')-1}{\ell_2(v,R)}\right)^2 \leq \mu\cdot \frac{2^{x^*}-1}{\ell_2(v,R)} \;.
\end{equation}
Since by Equation~\eqref{eq:mu_clm6}, $\mu = \sum_{v'\in R_{x^*}} \frac{\ell_2(v,v')}{\ell_2(v,R)}\cdot   \frac{\ell_2(v,v')-1}{\ell_2(v,R)} \geq \frac{2^{2(x^*-1)}}{2(\ell_2(v,R))^2}$ (as $\ell_2(v,v') \geq 2^{x^*-1}$ for every $v'\in R_{x^*}$  and $\ell_2(v,v') - 1 \geq \ell_2(v,v')/2$), 
we get that $\Pr[\chi_{i_1,j_1}=\chi_{i_2,j_2} = 1]<\sqrt 2 \cdot \mu^{3/2}$, and so
the second condition in Corollary~\ref{cor:cheb1} holds as well (for $c_1=\sqrt 2$).
Thus, the current claim follows. 
\end{proof}

\medskip
We are now ready to prove Theorem~\ref{thm:C4-ub}.

\begin{proofof}{Theorem~\ref{thm:C4-ub}}
Since the algorithm only rejects a graph $G$ if it detects a $C_4$, it will always accept graphs that are $C_4$-free.
Hence, we focus on  the case that $G$ is $\eps$-far from being $C_4$-free.

Recall that $\theta_0 = 4\alpha/\epsilon$
and let $E_{>\theta_0}$ be the subset of edges in $G$ where both endpoints have degree greater than $\theta_0$. 
Since the arboricity of $G$ is at most $\alpha$,
there are at most $2m/\theta_0$ vertices with degree greater than $\theta_0$, so that $|E_{>\theta_0}| \leq (2m/\theta_0)\cdot \alpha=\eps m/2$ edges.

Since $G$ is $\eps$-far from $C_4$-free, if we remove all edges in $E_{>\theta_0}$, then we get a graph that is at least $(\eps/2)$-far from $C_4$-free. It follows that there exists a set of edge-disjoint $C_4$s, denoted $\calC$, such that
no $C_4$ in $\calC$ contains an edge in $E_{>\theta_0}$, and $|\calC| \geq \eps m/8$.

 We next  partition $\calC$ into two disjoint subsets: $\calC_1$ contains those $C_4$s that have at most one vertex with degree at least $\theta_1$ in them, and $\calC_2$ contains those that have at least two such vertices (where in the case $\theta_0\leq \theta_1$ there will be exactly two).  Since $\calC = \calC_1 \cup \calC_2$, either $|\calC_1| \geq \eps m/16$ or $|\calC_2| \geq \eps m/16$ (possibly both).

\paragraph{The case $\mathbf{|\calC_1| \geq \eps m/16}$.}
Consider first the case that $|\calC_1| \geq \eps m/16$. In order to analyze this case, we apply a process of ``coloring'' vertices and edges. Initially, all vertices and edges that participate in $C_4$s that belong to $\calC_1$ are colored \emph{green}, and all other vertices and edges are colored \emph{red}. We next apply the following iterative process. As long as there is a green vertex $v$ whose number of incident green edges is less than $\eps d(v)/64$, color $v$ and its green incident edges by red. Observe that the total number of edges colored red by this process is at most $\eps m/32$.
Furthermore, at the end of this process, every green vertex $v$ has at least $\eps d(v)/64$ incident green edges (and if a vertex is red, then all its incident edges are red).
Let $\calC_1'$ be the subset of $\calC_1$ that consists of those $C_4$s in $\calC_1$ whose edges all remain green after the process (and hence they are green), so that $|\calC'_1| \geq \eps m/32$.

By the definition of $\calC_1$, and hence also $\calC'_1$, in each $C_4$ in $\calC'_1$ there is at most  one vertex with degree greater than $\theta_1$, and no edges such that both endpoints have degree greater than $\theta_0$. Assume without loss of generality that for each four-cycle $\rho\in \calC'_1$, 
where $\rho=(v_0(\rho), v_1(\rho), v_2(\rho), v_3(\rho))$,  $v_2(\rho)$ is the highest degree vertex (where $d(v_2(\rho))$ could be any value between $1$ to $n$).  
Let $V_0(\calC'_1) =\bigcup_{\rho \in \calC'_1} \{v_0(\rho)\}$ denote this set of vertices (i.e., the ones that are across from the highest degree vertex in a (green) four-cycle in $\calC_1'$).

\medskip\noindent\textsf{Observation: }
\emph{For every $\rho\in\calC'_1$,
\begin{inparaenum}[(1)]
    \item $d(v_0(\rho)) \leq \theta_1$, and
    \item  $v_1(\rho)$ and $v_3(\rho)$ are of degree at most $\theta_{min}=\min\{\theta_0, \theta_1\}$.
\end{inparaenum}
}

\smallskip
To verify this observation, note that
    by the definition of $\calC'_1$, for every $\rho\in \calC'_1$, there is at most one vertex with degree greater than $\theta_1$, and since $v_2(\rho)$ is the highest degree vertex in $\rho$, it follows  that all three other vertices in $\rho$ are of degree at most $\theta_1$.
    
    We now show that $d(v_1(\rho))\leq \theta_0$, and the proof for $v_3(\rho)$ is identical. If $d(v_2(\rho))>\theta_0$, then it 
    must be the case that $d(v_1(\rho))<\theta_0$, as otherwise both have degree greater than $\theta_0$  and so they cannot be connected, which is a contradiction to them both being incident on the four-cycle $\rho$.  
    If  $d(v_2(\rho))\leq \theta_0$, then since  $v_2(\rho)$ is the highest degree vertex in $\rho$, $d(v_1(\rho))\leq d(v_2(\rho))\leq \theta_0$.

\medskip
Therefore, for every   $v\in V_0(\calC'_1)$, it has at least $\eps d(v)/64$ neighbors $u$ such that $(v,u)$ is green and  $d(u) \leq \theta_{min}$. Hence, overall in the graph,   the set of vertices $V_0(\calC'_1)$ has  at least $\eps m/32$ green edges that are incident to it and their second endpoint is  of degree at most $\theta_{min}\leq \theta_0$. It follows that     conditioned on an edge being returned by  procedure \SelectEdge, by Claim~\ref{clm:select-edge}, it returns an edge incident to a vertex $v\in V_0(\calC'_1)$ with probability at least $(\eps m/32)/2m'>\eps/128$ (since $m'>\frac{1}{2} m$).
So the probability that in some iteration of \TestCfour\ a vertex $v_0\in V_0(\calC'_1)$ is selected, is at least $1-(1-\frac{\eps}{128})^t >9/10$ (recall that $t = \Theta(1/\eps)$ so that it suffices to set $t= 500/\eps$).

 Conditioning on this event, we apply Claim~\ref{clm:C4-ub1}. 
 Specifically:
 \begin{itemize}
     \item $\theta_0=4\alpha/\eps$ 
      (as defined in Step~\ref{step:set_params_C4} in Algorithm~\TestCfour); 
     \item $C(v_0,\theta_{min})$ is the set of $C_4$s in $\calC'_1$ that are incident to $v_0$;
     \item $\eps' = |C(v_0,\theta_{min})|/d(v) \geq \eps/128$ (since $v_0$ has at least $\eps d(v)/64$ incident green edges, and they can be partitioned into pairs such that each pair belongs to exactly one $C_4$ in $C(v_0,\theta_{min})$);
     \item $d(v_0) \leq \theta_1$ (by 
     the above observation);
 \end{itemize}
 In order to apply the claim, we must ensure that $s>16\sqrt{d(v_0)/\eps'}$. By the above, it is sufficient to set $s_1$ in Step~\ref{step:random-neighbors}, to be $s_1=512\sqrt{d(v_0)/\eps}$.
 
 Hence, by Claim~\ref{clm:C4-ub1}, if Step~\ref{step:random-neighbors} is applied to $v_0$, then a $C_4$ is observed with probability at least $9/10$.

The analysis for the case that $\mathbf{|\calC_2| \geq \eps m/16}$ is similar, and due to space constraints, it is deferred to Appendix~\ref{append:case-C2}. 

We next turn to analyze the query complexity. By the settings of $\theta_0$, $\theta_1$, $t$, $s_1$ and $s_2$ in the algorithm,
the query complexity of the algorithm is upper bounded as follows.
\begin{equation}
O\left(\frac{1}{\eps}\cdot \left(\frac{\alpha}{\eps}+\max\{s_1, s_2\}\right) \right) =
O\left(\frac{1}{\eps}\left( \frac{\alpha}{\eps}+\max\left\{\sqrt{\frac{\theta_1}{\eps}}\cdot \theta_{min}, \frac{1}{\eps^2}\cdot \sqrt{\frac{n\alpha}{\theta_1}\cdot\log n}\right\} \right) \right) 
\end{equation}
For the case that $\alpha\leq(c_1/4)\sqrt n$, we have that $\theta_{min}=\theta_0=\Theta(\alpha/\eps)$ and that $\theta_1=\Theta(\sqrt n/\eps)$, and so we get a complexity of
\begin{equation}
     O\left(
\eps^{-3} \cdot n^{1/4}\alpha^{1/2} \cdot \max\{\alpha^{1/2},\log^{1/2} n\}
 \right) = O(\eps^{-3}\cdot n^{1/4}\alpha)\;,
\end{equation}
where the last inequality is for $\alpha>\log n$, and otherwise the complexity is $O(\eps^{-3}\cdot n^{1/4}\alpha^{1/2}\log^{1/2} n)$. 

For the case that $\alpha>(c_1/4)\sqrt{n}$, we have that $\theta_{min}=\theta_1=\Theta(\sqrt n/\eps)$. Therefore, the complexity is
\begin{equation}
     O(\eps^{-3}\cdot (\alpha+n^{3/4}))\;.
\end{equation}
Thus, the proof is complete.
\end{proofof}

\bibliographystyle{plain}
\bibliography{lit}

\appendix

\section{Missing details for the testing algorithm for $C_4$}\label{sec:4.1}
 
\subsection{The procedure \textsf{Select-an-Edge} and Proof of Claim~\ref{clm:select-edge}} \label{append:sample-edge}
\textsf{Select-an-edge}($\epsilon,\alpha$)\label{alg:select-edge}
\begin{enumerate}
\item Repeat the following $\Theta(\alpha/\eps)$ times:
\begin{enumerate}
  \item Select a vertex $u$ uniformly at random.
  \item If $d(u) \leq \theta_0$ for $\theta_0=4\alpha/\eps$, then with probability $d(u)/\theta_0$ select an edge incident to $u$ uniformly at random and return it. 
\end{enumerate}
\item If no edge was selected, then return `Fail'.
\end{enumerate}

\begin{proofof}{Claim~\ref{clm:select-edge}}
Let $L$ denote the set of vertices with degree at most $\theta_0$.
Consider any single iteration of the algorithm.
Since $m\leq n\alpha$, the fraction of vertices not in $L$
is at most $\eps$.  Hence, the probability of selecting a vertex in $L$ is at least $1-\eps > 1/2$.
Conditioned on this event, the probability that an edge is returned is at least $1/\theta_0$.
Since the procedure performs $\Theta(\theta_0)$ iterations, for an appropriate constant in the $\Theta$ notation, an edge is returned with probability at least $2/3$.

Turning to the second part of the claim,
let $E_1$ denote the set of edges that have a single endpoint in $L$ and let $E_2$ denote the set of edges that have both endpoints in $L$.
Conditioned on an edge being selected, an edge that has a single endpoint in $L$ is selected with probability 
$\frac{1}{|E_1| + 2|E_2|} \geq \frac{1}{2m'}$ and an edge with both endpoints in $L$ is selected with probability
$\frac{2}{|E_1| + 2|E_2|} \leq \frac{1}{m'}$.
\end{proofof}

\subsection{Missing details in the proof of Theorem~\ref{thm:C4-ub}}\label{append:case-C2}
\paragraph{The case $\mathbf{|\calC_2| \geq \eps m/16}$.}
We now turn to the case that $|\calC_2| \geq \eps m/16$, 
where recall that $\calC_2$ is the set of green $C_4$s with at least two vertices with degree greater than $\theta_1$. 
We again consider two sub-cases, depending on whether $\theta_0\leq \theta_1$ or not. 
We start with the former case $\theta_0\leq \theta_1$ (i.e., $\alpha \leq (c_1/4)\sqrt n$). 

\smallskip\noindent
\textsf{The sub-case $\mathbf{\alpha\leq (c_1/4)\sqrt n}$.}
Recall that by the definition of $\calC_2$, for every 4-cycle  $\rho\in \calC_2$, it has no edges with two endpoints greater than $\theta_0$. Since $\theta_0\leq \theta_1$, this also implies that $\rho$ has exactly two vertices with degree greater than $\theta_1$. 
Hence for each $\rho\in \calC_2$, there are two vertices $v,v'$ with degree greater than $\theta_1$ that do not neighbor one another on the $C_4$, and there are two vertices $u,u'$ with degree at most $\theta_0$ that each neighbors both $v$ and $v'$. Thus the vertices and edges that participate in $C_4$s belonging to $\calC_2$ induce a bipartite graph between vertices with degree greater than $\theta_1$ and vertices with degree at most $\theta_0$. Here too we run a coloring process, where initially all vertices and edges that participate in $C_4$s contained in $\calC_2$ are colored green, but the process is slightly modified as explained next. Let $\bar{d} = m/n$.  In addition to the (re)coloring rule (green to red) defined above,
we also color red a green vertex $u$ and its incident green edges, if the number of these edges is less than 
$\eps\bar{d}/128$. This process removes at most $\eps m/32+\eps m/64$ edges, and therefore at the end of the process we are still left with $\eps m/64$ many $C_4$'s.

Let $\calC_2'$  be the subset of  $\calC_2$ that consists of those $C_4$s in $\calC_2$ whose edges all remain green, and observe that by the above process, each vertex that belongs to a $C_4$ in $\calC_2'$ has at least $\max\{\eps \bar{d}/128,\eps d(v)/64\}\geq \eps''\cdot \max\{\bar{d},\eps d(v)\}$ many incident green edges (for $\eps''=\eps/128$). 
We have that $|\calC'_2| \geq \eps m/32$, and if we let $R$ be the subset of vertices with degree at least $\theta_1$ that are colored green, and $L$ be the subset of vertices with degree at most $\theta_0$ that are colored green, then the premises of Claim~\ref{clm:C4-ub2} hold for every $v\in R$.
Specifically, for
\begin{inparaenum}[(1)]
    \item $\eps'=\eps/16$ and $\eps''=\eps/128$,
    \item $\theta_1=c_1\sqrt n/\eps$ (where here $c_1 = 100$ is sufficient so that indeed $\theta_1>2\sqrt{n/(\eps'\cdot \eps'')}$ as required)
    \item $|R|\leq 2m/\theta_1\leq 2n\alpha/\theta_1 $,
    so that for an appropriate setting of the constant in the $\Theta(\cdot)$ notation for $s_2=\Theta(\sqrt{  (n\alpha/\theta_1)\log n}/\eps^2)$ in Step~\ref{step:random-neighbors} of the algorithm, we have that
    $s_2 \geq \frac{32}{\eps' \eps''}\sqrt{2|R|\log n}$, as required. 
\end{inparaenum}

The argument for why the algorithm will select a vertex $v\in R$ with  probability at least $9/10$ is as in the case that $|\calC_1| \geq \eps m/16$. Conditioned on this event, by Claim~\ref{clm:C4-ub2} and the setting of $\theta_1$, when Step~\ref{step:random-walk} is applied to $v$, a $C_4$ is observed with probability at least $9/10$.

\smallskip\noindent
\textsf{The sub-case $\mathbf{\alpha> (c_1/4)\sqrt n}$.}
Recall that the set $\calC_2$ consists of all $C_4$s that have at least two vertices with degree at least $\theta_1$, and no two incident vertices where both have degree greater than $\theta_0$.
In the case where $\alpha>(c_1/4)\sqrt n$, so that $\theta_0>\theta_1$, it is no longer true that all $C_4$s  in $\calC_2$ have exactly two vertices with degree greater than $\theta_1$. Indeed, as $\theta_1<\theta_0$, it can occur that the four-cycles in $\calC_2$ have edges with both endpoints of degrees greater than $\theta_1$. Therefore, we need a different argument in order to obtain the bipartite graph $(L,R)$ for which we would apply Claim~\ref{clm:C4-ub2}.

Consider further partitioning $\calC_2$ into two subsets: $\calC_{2,2}$ consists of those $C_4$s in $\calC_2$ that contain exactly two vertices with degree greater than $\theta_0$, and $\calC_{2,1}$ of those that contain at most one vertex with degree greater than $\theta_0$. 

If $|\calC_{2,2}| \geq |\calC_2|/2$, then 
the argument  proceeds essentially the same as for the sub-case  $\alpha \leq (c_1/4)\sqrt  n$ with $\calC_2$ replaced by $\calC_{2,2}$. Here we use the fact that $\theta_0 > \theta_1$, so that after running the ``coloring'' process on the vertices and edges of $C_4$s in $\calC_{2,2}$, if we let $R$ be the subset of vertices with degree greater than $\theta_0$ that are colored green, and $L$ be the subset of vertices with degree at most $\theta_0$ that are colored green, then the premises of Claim~\ref{clm:C4-ub2} hold for every $v\in R$.

 If $|\calC_{2,1}|>|\calC_2|$/2, then
we do the following. First, we put in 
a  subset $\widetilde{L}$ all vertices with degree at most $\theta_1$, and in a subset $\widetilde{R}$ all vertices with degree greater than $\theta_0$. As for the vertices with degree between $\theta_1$ and $\theta_0$, consider  randomly partitioning  them into the two sides. Observe that for each fixed $C_4$ in $\calC_{2,1}$, the probability that all its edges cross between $\widetilde{L}$ and $\widetilde{R}$, which we refer to as ``surviving'', is at least $1/8$, and so the expected fraction that survive is at least $1/8$. Therefore, there exists a partition 
$(\widetilde{L}^*,\widetilde{R}^*)$ 
for which at least this fraction survives. We denote the set of surviving $C_4$s with respect to $(\widetilde{L}^*,\widetilde{R}^*)$ by $\widetilde{\calC}_{2,1}$. 

The important thing to observe is that all vertices in $\widetilde{L}^*$ have degree at most $\theta_0$, all vertices in $\widetilde{R}^*$ have degree at least $\theta_1$, and all edges participating in $C_4$s that belong to $\widetilde{\calC}_{2,1}$ have two vertices in each side, with the edges crossing between them.
The argument continues essentially as in the subcase that $\alpha \leq (c_1/4)\sqrt{n}$, where $\calC_2$ is replaced with $\widetilde{\calC}_{2,1}$.

\section{A lower bound of $\Omega(n^{1/4}\alpha^{1/2})$ for testing $C_4$-freeness}\label{sec:C4-lb}

\subsection{A two-sided error lower bound for constant-arboricity graphs}
We start by proving the lower bound for constant-arboricity graphs, as stated in Theorem~\ref{thm:C4-5} in the introduction.

\begin{theorem} \label{thm:C4-const-alpha-lb}
Testing $C_4$-freeness in constant-arboricity graphs over $n$ vertices requires $\Omega(n^{1/4})$ queries for constant $\epsilon$.
\end{theorem}
We note that Theorem~\ref{thm:C4-const-alpha-lb} holds for two-sided error algorithms and not only one-sided error ones.

\begin{proof}
For any (sufficiently large) $n$ we shall define two bipartite graphs, $G_0$ and $G_1$, where both have constant arboricity. The graph $G_0$ is $C_4$-free and  the graph $G_1$ is $\Omega(1)$-far from being $C_4$ free.
Based on each $G_b$, $b\in \bitset$ we have a distribution, $\calD_b$, whose support consists of graphs isomorphic to $G_b$ (that differ only in vertex and edge labelings), and each distribution is uniform over its support.
We shall show that for a sufficiently large constant $c$, an algorithm that performs less than $n^{1/4}/c$ queries, cannot distinguish with success probability at least $2/3$ between a graph selected according to $\calD_0$ and a graph selected according to $\calD_2$. The lower bound stated in the theorem follows.

We start by defining $G_0$. The vertices in $G_0$ are partitioned into three sets: $X_0,Y_0,Z_0$, \SodaReb{whose sizes are $x_0,y_0$ and $z_0$, respectively. These sizes satisfy the following. First,}
 $x_0 = y_0(y_0-1)/2$. We set $y_0$ to be the largest odd integer satisfying 
 \ICALPCR{$x_0 + y_0 \leq n$}
 and $z_0 = n-(x_0+y_0)$.
For each pair of vertices in $Y_0$ there is a unique vertex in $X_0$ that neighbors exactly these two vertices.
The vertices in $Z_0$ have no incident edges.
By construction, $G_0$ is $C_4$-free, and we have $y_0 = \Theta(n^{1/2})$, 
\ICALPCR{$z_0 = O(n^{1/2})$,}
and $x_0 = \Theta(n)$.

Turning to $G_1$, its vertices are also partitioned into three sets: $X_1,Y_1,Z_1$, of sizes $x_1$, $y_1$ and $z_1$, respectively, where $X_1$ is further partitioned into two equal-size sets $X_1^1$ and $X_1^2$, and $Y_1$ is partitioned into two equal-size sets $Y_1^1$ and $Y_1^2$ (so that $x_1$ and $y_1$ are both even).
Here $x_1/2 = (y_1/2)^2$ so that $x_1 = y_1^2/2$, and we set 
\ICALPCR{
$y_1 = y_0 - 1$
}
and $z_1 = n- (x_1+y_1)$. For each pair of vertices, $u^1 \in Y_1^1$ and $u^2 \in Y_1^2$ there is one vertex in $X_1^1$ that neighbors exactly these two vertices, and one vertex in $X_1^2$ that neighbors them. The vertices in $Z_1$ have
no incident edges. By construction, since there are $x_1/2 = \Theta(n)$ edge-disjoint $C_4$s in $G_1$, it is $\Omega(1)$-far from being $C_4$-free.

Comparing the two graphs we observe that the vertices in
$Z_1$ have degree 0, which is the same as those in $Z_0$, The vertices in $X_1$ have degree 2, which is the same as those in $X_0$, and  the vertices in $Y_1$ have degree 
\ICALPCR{$2\cdot (y_1/2) = y_1 = y_0-1$,}
which is the same as those in $Y_0$.
Comparing the sizes of the sets of vertices, since 
\ICALPCR{
$y_1 = y_0 - 1$,
}
we get that
$$x_1 = \frac{y_1^2}{2} = \frac{(y_0-1)^2}{2} = \frac{y_0(y_0-1)}{2} - \frac{y_0-1}{2} = x_0 - \Theta(n^{1/2})\;,$$
and
\ICALPCR{
$$z_1 = n-(x_1+y_1) = n - (x_0 + (1/2)(y_0-1)) = z_0 + y_0/2 + 1/2 = O(n^{1/2})\;.$$
}
\ForFuture{D: maybe explain a bit more}

We next define two processes, $\calP_0$ and $\calP_1$, where $\calP_b$, for $b\in \bitset$ constructs ``on the fly'' a graph selected  according to $\calD_b$ while answering the queries of a testing algorithm. We shall show that unless one of several events occurs, the distributions induced on the sequences of queries and answers are identical. Let $q^t$ denote the $t$th query of the algorithm and $a^t$ the answer it gets, so that both are random variables. The query $q^t$ may depend on the \emph{query-answer history} $(q^1,a^1),\dots,(q^{t-1},a^{t-1})$ and possibly randomness of the algorithm. Both processes maintain a \emph{Knowledge graph}, which contains the vertices and edges that appear in these questions and answers, where $K^t=(V^t,E^t)$ is the knowledge graph after the $t$th query is answered. Let $X^t$, $Y^t$ and $Z^t$ denote the subsets of vertices with degree $2$, $y_0-1$ and $0$, respectively.
(In principle, the knowledge graph should also include information about the degrees of vertices and pairs of vertices that are known not to have edges between them, but, as we discuss below, this will be implicit from the graph.)
To determine $a^t$ given $q^t$,
the process $\calP_b$ simply considers all graphs in the support of $\calD_b$ that are consistent with $K^{t-1}$, and selects an answer with probability proportional to the number of such graphs that are also consistent with this answer.

We assume that before asking a query concerning a vertex $v$ that is not in the knowledge graph, the algorithm performs a degree query $q^t$ on $v$ (which in particular determines whether $v$ belongs to $X^t$, $Y^t$ or $Z^t$).
If the answer to this query is $2$, then the algorithm performs two neighbor queries to get $v$s two neighbors.
Also, if the algorithm gets a new vertex $v$ as an answer to a neighbor query from a vertex
$u \in Y^{t-1}$, then it performs a neighbor query on $v$ so as to gets its second neighbor.
A lower bound of $T$ on the query complexity of the algorithm under theses restrictions translates into a lower bound of $\Omega(T)$ without these restrictions. Note that the algorithm need not ask any further degree queries, as their answers are determined by the knowledge graph (which is known to be isomorphic to a subgraph of $G_0$ or $G_1$). Also note that because for each degree-2 vertex that is introduced into the graph, the algorithm obtains both its neighbors, we may assume (without loss of generality) that the algorithm does not perform any vertex-pair queries (as their answers would always be negative).

We observe that for both processes, if the algorithm performs $o(n^{1/2})$ queries, then the probability that any answer to a degree query (on a new vertex) is not $2$, is $o(1)$ (since $|Y_b|+|Z_b| = O(n^{1/2})$ for both $b=0$ and $b=1$). Let's refer to this ``bad'' event as $E_1$.
In addition, by the second restriction on the algorithm (i.e., that each degree-2 vertex essentially ``arrives with both its neighbors''), we have that for any neighbor query $q^t$ on a vertex in $Y^{t-1}$, the process $\calP_b$ answers with a vertex (in $X^t$) whose label is uniformly selected among all $n-|V^{t-1}|$ ``free'' labels. That is, the distribution on these answers is identical under both processes.


It remains to analyze the answers to neighbor queries for vertices $v$ in $X^{t-1}$.
If $v$ is the answer to a neighbor query from a vertex $u \in Y^{t-1}$, then each of the processes needs to select a single (additional) neighbor. Conditioned on this neighbor not belonging to $V^{t-1}$, for both processes, its label is uniformly distributed among all $n-|V^{t-1}|$ ``free'' labels. The same is true if $v$ is introduced following a degree query and both its neighbors are determined not to belong to $V^{t-1}$. We denote the event that we get at least one neighbor that already belongs to the knowledge graph, by $E_2$. Observe that conditioned on $E_1$ and $E_2$ not occurring,  the distributions on query-answer histories induced by the two processes, are identical (since, conditioned on these events, every degree query is answered by $2$, and every neighbor query is answered by a uniformly selected new label).

It hence remains to upper bound the probability of the event $E_2$, under both processes, when the algorithm performs at most $n^{1/4}/c$ queries, for some sufficiently large constant $c$. Consider any prefix $(q^1,a^1),\dots,(q^{t-1},a^{t-1})$ of the query-answer history (where neither event $E_1$ or $E_2$ has yet occurred), and the corresponding knowledge graph $K^{t-1}$.
Suppose that $q^t$ is a degree query for a vertex $v$ (that does not belong to $V^{t-1}$, and such that the answer to this query is $2$). Consider first the process $\calP_0$. 
 Since for each pair of vertices in $Y_0$ there is a single vertex in $X_0$ that neighbors them, and there are at most $2(t-1)$ vertices in $Y^{t-1}$ \ICALPCR{and at most $(t-1)$ in $X^{t-1}$}, the probability (over the distribution induced by $\calP_0$) that at least one of the neighbors of $v$ belongs to $Y^{t-1}$ is upper bounded by
\ICALPCR{
$$\frac{2(t-1)\cdot (y_0-1)}{x_0 - (t-1)} = \frac{2(t-1)\cdot (y_0-1)}{(y_0\cdot(y_0-1)/2) - (t-1)} 
    \leq \frac{c' t}{n^{1/2}}$$
    }
 for a constant $c'$ (since $y_0 = \Theta(n^{1/2})$ and $t = O(n^{1/4})$). A similar (slightly lower) upper bound is obtained for $\calP_0$ when $v$ is the answer to a neighbor query from a vertex $u \in Y^{t-1}$ (since one of $v$'s neighbors is already determined to be $u$).

Turning to $\calP_1$, we get a similar upper bound. Namely, for a new vertex $v$ (with degree 2), the probability that at least one of its neighbors belongs to $Y^{t-1}$ is upper bounded by
\ICALPCR{
$$\frac{2(t-1)\cdot (y_1/2)}{x_1-(t-1)} =\frac{2(t-1)\cdot (y_1/2)}{(y_1/2)^2-(t-1)} 
\leq \frac{c'' t}{n^{1/2}}\;,$$
}
for a constant $c''$,
 and a similar upper bound holds when $v$ is an answer to a neighbor query from $u\in Y^{t-1}$.
  Summing over $t=1,\dots,T= n^{1/4}/c$, we get that the probability that $E_2$ occurs (for either processes) is a small constant, as required.
\end{proof}

\subsection{A one-sided error lower bound for graphs with arboricity $\alpha$}
We now turn to a lower bound on non-constant arboricity graphs for one-sided error algorithms \ICALPCR{as stated in Theorem~\ref{thm:C4-gen-alpha-lb} in the introduction}. We assume that $\alpha =O(n^{1/2})$ (recall that every graph with  arboricity greater than $n^{1/2}$ must contain $C_4$s). 
The lower bound holds for $\alpha = \Omega(\log n)$. Note that the lower bound of Theorem~\ref{thm:C4-const-alpha-lb} also holds for $\alpha = O(\log n)$ that is non-constant, since, as noted in the introduction  we can add to the graphs in the lower-bound construction a subgraph with arboricity $\alpha$ that is $C_4$-free. \ForFuture{D: reference to Fox? or something else}

In order to prove Theorem~\ref{thm:C4-gen-alpha-lb}, we define
a distribution 
over $n$-vertex graphs for which the following holds. First, with high probability, a graph selected according to 
this distribution is $\Omega(1)$-far from being $C_4$-free. Second, the number of queries necessary to observe a $C_4$ with sufficiently high constant probability when the queries are answered by graph selected according to
the distribution 
is $\Omega(n^{1/4}\alpha^{1/2})$.
Furthermore, if a one-sided error algorithm performs less than 
$n^{1/4}\alpha^{1/2}/c$ queries for a sufficiently large $c$ and does not observe a $C_4$, then it must accept, since the subgraph observed (including the degrees of the vertices) can be extended to a $C_4$-free graph.

The vertex set of each graph in the support of $\calD$ is partitioned into five parts: $X_1$, $X_2$, $Y_1$, $Y_2$, and $Z$, where $|X_1| = |X_2|=n/4$, and for $d = n^{1/2}/c_2$,  $|Y_1| = |Y_2| = (n\alpha)/(4 d) = \Theta(n^{1/2}\alpha)$ and $|Z| = n - (|X_1|+|X_2|+|Y_1|+|Y_2|)$. (We assume for simplicity that $n$ is divisible by $2d$
and that $\alpha$ and $d$ are even integer -- the more general case can be handled similarly). 

The distribution $\calD$ is uniform over all graphs for which the following holds:
\begin{itemize}
\item There is a bipartite graph, denoted $G_{b,b'}$ between $X_b$ and $Y_{b'}$ for $b,b' \in \{1,2\}$, where vertices in $X_b$ have  $\alpha/2$ neighbors in $Y_{b'}$, vertices in $Y_{b'}$ have $d/2$ neighbors in $X_b$ and vertices in $Z$ have degree $0$.
\item There are no multiple edges.
\end{itemize}
 Observe that there exists at least one such graph: Let each $X_b$  consist of $\frac{n/4}{d/2}$ subsets, each of size $d/2$, and let each $Y_b$ consist of $\frac{n\alpha/(4d)}{\alpha/2}$ subsets, each of size $\alpha/2$. For the edges, let there be a complete bipartite graph between the $i$th subset in each $X_b$ to the $i$th subset of both $Y_1$ and $Y_2$. One can verify that indeed both conditions above hold. We stress that this graph is not a good candidate for a lower bound, as it contains many $C_4$s. However, as we prove below, when choosing uniformly over all graphs in the (non-empty) support of $\calD$, detecting a $C_4$ requires $\Omega(n^{1/4}\alpha^{1/2})$ queries.


 \begin{lemma}\label{lem:D-C4-far}
With probability at least $9/10$, a graph selected according to $\calD$ is $\Omega(1)$-far from being $C_4$-free.
\end{lemma}

 In order to establish Lemma~\ref{lem:D-C4-far},
 we define the following ``bad events'' (taken over the random choice of a graph according to $\mathcal{D}$). Let $\alpha' = \Theta(\alpha)$ (where the constant in the $\Theta$ notation will be determined subsequently).
First, $E^0$ is the event that the fraction of edges between $X_1$ and $Y_1$ that participate in less than $\alpha'/2$ $C_4$s is larger than $1/2$.
Second, for each $b,b' \in \{1,2\}$ and $\ell \geq 1$, let $D_{b,b'}^\ell$ be the set (\emph{bucket}) of edges between $X_b$ and $Y_{b'}$ that participate in a number of $C_4$s that is in the range $[(2^\ell+1)\cdot \alpha', (2^{\ell+1}+1)\cdot \alpha')$.
The event $E_{b,b'}^\ell$ is the event that $|D_{b,b'}^\ell| > (2^{-2\ell}/c_3)\cdot |X_b|\cdot (\alpha/2)$ for some constant $c_3$.

\begin{claim}~\label{clm:Pr-Es}
For a sufficiently large $\alpha = \Omega(\log n)$,
the probability that either $E^0$ occurs or some $E_{b,b'}^\ell$ occurs (for $b,b' \in \{1,2\}$ and $\ell \geq 1$) is at most $1/10$.
\end{claim}
We prove Claim~\ref{clm:Pr-Es} momentarily, and first show how it implies Lemma~\ref{lem:D-C4-far}.

\medskip
\begin{proofof}{Lemma~\ref{lem:D-C4-far}}
Assume that neither $E^0$ nor any $E_{b,b'}^\ell$ occur, which by Claim~\ref{clm:Pr-Es} holds with probability at least  $9/10$. Conditioned on $E^0$ not occurring, there are at least  $|X_1|\cdot (\alpha/2)\cdot (\alpha'/2)$ ($ = \Omega(n\alpha^2)$) $C_4$s in a graph selected according to $\mathcal{D}$.
Suppose we mark the edges that participate in any one of the buckets $D_{b,b'}^\ell$ for $b,b'\in \{1,2\}$ and $\ell\geq 1$ by ``red'' and all other edges by ``green''.
By our assumption on the sizes of the buckets (the events $E_{b,b'}^\ell$ not occurring), the total number of $C_4$s that contain at least one red edge is upper bounded by
\ForFuture{ D: verify constants etc.}
\begin{equation}
\sum_{b,b'}\sum_\ell |D_{b,b'}^\ell|\cdot (2^{\ell+1}+1)\cdot \alpha' \leq
  4 \cdot (2^{-2\ell}/c_3)\cdot |X_b|\cdot (\alpha/2) \cdot (2^{\ell+1} +1)\cdot \alpha' \leq   |X_1|\cdot(\alpha/2)\cdot (\alpha'/c_3') 
  \;.
\end{equation}
By ensuring the $c_3' \geq 4$,
the number of $C_4$s residing  on green edges (only) is at least
\begin{equation}
\frac{|X_1|\cdot\alpha\cdot\alpha'}{4}
- \frac{|X_1|\cdot\alpha\cdot\alpha'}{8} =
\Omega(n \alpha^2)\;.
\end{equation}

Hence, with probability at least $9/10$ over the choice of a graph according to $\calD$, the following holds. There exists a set of green edges,  such that the number of $C_4$s that reside  on green edges is $\Omega(n\alpha^2)$, and each green edge participates in  $O(\alpha)$ $C_4$s. This implies that there are $\Omega(n\alpha)$ edge-disjoint $C_4$s. To verify this consider running an iterative procedure, starting from the graph restricted to green edges. In each iteration, until there are no remaining $C_4$s, a $C_4$ is selected from the current graph and put in the set of edge-disjoint $C_4$s. Then all $O(\alpha)$ $C_4$s that include one of the edges in the selected $C_4$ are removed from the graph (i.e., their other edges are removed from the graph). Since we start with $\Omega(n\alpha^2)$ $C_4$s and in each iteration remove $O(\alpha)$ $C_4$s, the final set of edge-disjoint $C_4$s we get is of size $\Omega(n\alpha)$, implying that the graph is $\Omega(1)$-far from $C_4$-free.
\end{proofof}

\smallskip
\noindent
 We next prove Claim~\ref{clm:Pr-Es}.

 \smallskip
\begin{proofof}{Claim~\ref{clm:Pr-Es}}
We start by noting that in the selection of a random graph according to $\mathcal{D}$, the four graphs $G_{b,b'}$, $b,b' \in \{1,2\}$ are generated independently from each other (recall that $G_{b,b'}$ is a bipartite graph between $X_b$ and $Y_{b'}$, where each vertex in $X_b$ has $\alpha/2$ neighbors in $Y_{b'}$, and each vertex in $Y_{b'}$ has $d/2$ neighbors in $X_b$).

\medskip
For simplicity of notations, let $b=1$, $b'=1$.
The analysis below can be applied to any pair $b,b' \in \{1,2\}$.
Fix any choice of bipartite graph $G_{2,2}$.
We shall show that for any fixed edge $(x,y)$ between $x\in X_1$ and $y \in Y_1$, the following holds over the random choice of $G_{1,2}$ and $G_{2,1}$.
The expected number of $C_4$s that reside on $(x,y)$ is $\alpha'$, and furthermore, with sufficiently high probability, the actual number does not deviate by too much from its expectation.

For a fixed edge $(x,y)\in X_1\times Y_1$,
Let $\Gamma_2(x)$ denote  the set of $\alpha/2$ neighbors of $x$ in $Y_2$ (as determined by the choice of $G_{1,2}$), and let $\Gamma_2(y)$ denote the set of $d/2$ neighbors of $y$ in $X_2$ (as determined by the choice of $G_{2,1}$). Since we are considering a single vertex $x\in X_1$ and a single vertex $y\in Y_1$, these sets are uniformly distributed in $Y_2$ and $X_1$, respectively.

For $i\in [d/2]$  and $j \in [\alpha/2]$, let $\chi_{i,j}$ be a Bernoulli random variable that is 1 if there is an edge between the $i$th neighbor of $y$ and the $j$th neighbor of $x$ (and is 0 otherwise). Let  $\mu = \Pr[\chi_{i,j}]=1]$, so that
\begin{equation}
\mu = 
   \frac{\alpha/2}{|Y_2|} 
   = \Theta\left(\frac{n^{1/2}}{n}\right) = \Theta\left(\frac{1}{n^{1/2}}\right)\;.
\end{equation}
Therefore, if we let $\alpha' = (d/2)\cdot(\alpha/2)\cdot \mu$ (so that $\alpha' = \Theta(\alpha)$), then
\begin{equation}
\Ex\left[\sum_{\substack{i\in [d/2] \\ j\in [\alpha/2]}}\chi_{i,j}\right] = \frac{d}{2}\cdot\frac{\alpha}{2}\cdot \mu = \alpha'\;.
\end{equation}
We turn to analyze the variance of the sum.
\begin{align}
\Var\left[\sum_{\substack{i\in [d/2] \\ j\in [\alpha/2]}}\chi_{i,j}\right] & =
\Ex\left[\left(\sum_{\substack{i\in [d/2] \\ j\in [\alpha/2]}}\chi_{i,j}\right)^2\right] - \left(\Ex\left[\sum_{\substack{i\in [d/2] \\ j\in [\alpha/2]}]}\chi_{i,j}\right]\right)^2 \\
& =
\Ex\left[\sum_{\substack{i\in [d/2] \\ j\in [\alpha/2]}} (\chi_{i,j})^2\right] +
\Ex\left[\sum_{\substack{i\in [d/2] \\ j_1\neq j_2 \in [\alpha/2]}}\chi_{i,j_1}\chi_{i,j_2}\right]
+ \Ex\left[\sum_{\substack{i_1\neq i_2\in [d/2] \\ j\in  [\alpha/2]}}\chi_{i_1,j}\chi_{i_2,j}\right] \nonumber \\
&\;\;\;\;\;\;\;\;\;\;+ 
\Ex\left[\sum_{\substack{i_1\neq i_2\in [d/2] \\ j_1\neq j_2\in  [\alpha/2]}}\chi_{i_1,j_1}\chi_{i_2,j_2}\right] - \left(\frac{d}{2}\cdot\frac{\alpha}{2}\cdot \mu\right)^2 \;.
\label{eq:var-lb}
\end{align}
We analyze each of the terms separately.
Since the $\chi_{i,j}$s are Bernoulli random variables,
$(\chi_{i,j})^2 = \chi_{i,j}$ so that
\begin{equation} \label{eq:cheb-C4-lb-case1}
\Pr[(\chi_{i,j})^2 = 1] = \Pr[\chi_{i,j}=1] = \mu\;.
\end{equation}
 For every $i\in [d/2]$ and $j_1,j_2 \in [\alpha/2]$ such that $j_1 \neq j_2$,
\begin{equation}\label{eq:cheb-C4-lb-case2}
\Pr[\chi_{i,j_1}=\chi_{i,j_2}=1]
  \leq \left( \frac{\alpha/2}{|Y_2|}\right)^2 = \mu^2
  \;,
\end{equation}
and for every $i_1,i_2\in [d/2]$ and $j\in [\alpha/2]$ such that $i_1\neq i_2$,
\begin{equation}\label{eq:cheb-C4-lb-case3}
 \Pr[\chi_{i_1,j}=\chi_{i_2,j}=1]
\leq 
\left( \frac{d/2}{|X_2|}\right)^2 = \left( \frac{\alpha/2}{|Y_2|}\right)^2 = \mu^2 
\;.
\end{equation}
Finally,
for $i_1\neq i_2\in [d/2]$ and $j_1\neq j_2\in [\alpha/2]$, we have that
\begin{equation}\label{eq:cheb-C4-lb-case4}
\Pr[\chi_{i_1,j_1}=\chi_{i_2,j_2}=1]
=  \Pr[\chi_{i_1,j_1}=1] \cdot
\Pr[\chi_{i_1,j_1}=1\,|\,\chi_{i_2,j_1}=1]
  =\frac{\alpha/2}{|Y_2|}\cdot \frac{|X_2|(\alpha/2)-1}{(|X_2|-1)(|Y_2|-1)}\;,
\end{equation}
where the second term on the right-hand-side of Equation~\eqref{eq:cheb-C4-lb-case4} is the fraction of pairs in $X_2\times Y_2$ that have an edge between them, among all pairs with the exception of one pair that is an edge in $G_{2,2}$.
It is not hard to verify that
\begin{equation}\label{eq:cheb-C4-lb-5}
 \frac{|X_2|(\alpha/2)-1}{(|X_2|-1)(|Y_2|-1)} \leq \frac{\alpha/2}{|Y_2|} \cdot \left(1+\frac{2}{|Y_2|}\right) = \mu \cdot \left(1+\frac{2}{|Y_2|}\right)
\end{equation}
By combining Equations~\eqref{eq:var-lb}--\eqref{eq:cheb-C4-lb-5} (and recalling that $\alpha' = (d/2)(\alpha/2)\mu$), we get that
\begin{eqnarray}
\Var\left[\sum_{\substack{i\in [d/2] \\ j\in [\alpha/2]}}\chi_{i,j}\right]
&\leq& \frac{d}{2}\cdot\frac{\alpha}{2}\cdot\mu
+ \frac{d}{2}\cdot \left(\frac{\alpha}{2}\right)^2 \cdot \mu^2 + \left(\frac{d}{2}\right)^2\cdot \frac{\alpha}{2}\cdot \mu^2 \nonumber \\
&\;\;\;\;\;\;\;\;\;\;\;\;\;\;\;\;\;\;\;+& \left(\frac{d}{2}\right)^2\cdot \left(\frac{\alpha}{2}\right)^2 \cdot \mu^2\cdot \left(1+\frac{1}{|Y_2}\right) - \left(\frac{d}{2}\cdot\frac{\alpha}{2}\cdot \mu\right)^2 \nonumber \\
&=& \alpha'\cdot
\left( 1+ \frac{\alpha}{2}\cdot\mu + \frac{d}{2}\cdot \mu + \frac{d}{2}\cdot \frac{\alpha}{2}\cdot \frac{2}{|Y_2|}\cdot \mu\right)\nonumber \\
&\leq& 4\alpha' \;,
\end{eqnarray}
where in the last inequality we used the following facts: $\alpha \leq d$, $|Y_2|> \alpha$, and
$(d/2)\mu \leq 1$.

By Chebishev's inequality,
for any choice of $t>0$,
\begin{eqnarray}
\Pr\left[\left|\sum_{i\in [d/2],j\in [\alpha/2]}\chi_{i,j} - \alpha'\right|< t\cdot \alpha' \right] &\leq& \frac{\Var\left[\sum_{\i\in [d/2],j\in [\alpha/2]}\chi_{i,j}\right]}{t^2\cdot \left(\Ex\left[\sum_{i\in [d],j\in [\alpha]} \chi_{i,j}\right]\right)^2} \\
&\leq& \frac{4\alpha'}{t^2 \cdot (\alpha')^2}
\;\;= \;\;\frac{4}{t^2\cdot \alpha'}\;
\label{eq:C4-lb-cheb2}
\end{eqnarray}
By setting $t=1/2$ we get that for any fixed edge $(x,y)\in X_1\times Y_1$, the probability that $(x,y)$ participates in less than $\alpha'/2$ $C_4$s is at most $16/\alpha'$. By Markov's inequality and the definition of the event $E^0$, we have that
$\Pr[E^0] \leq 32/\alpha'$.

By applying the same argument to any other $b,b' \in \{1,2\}$ (i.e., not necessary $b=b'=1$), we get that
for any fixed edge $(x,y)\in X_b\times Y_{b'}$ and for
any choice of $t = 2^{\ell}$, $\ell \geq 0$,
 the probability that $(x,y)$ participates in more  than $(2^\ell +1)\cdot \alpha'$ $C_4$s is at most $\frac{4}{2^{2\ell}\cdot \alpha'}$.
Therefore, the expected number of such edges is at most
$|X_b|\cdot (\alpha/2)\cdot \frac{4}{2^{2\ell}\cdot \alpha'}$. By Markov's inequality, the probability that there are more than
$ |X_b|\cdot (\alpha/2)\cdot \frac{1}{c_3\cdot 2^{2\ell}}$ such edges is at most $\frac{4c_3}{\alpha'}$.

By the definition of the bucket $D_{b,b'}^\ell$ (which contains a subset of the aforementioned edges) and the event $E^\ell_{b,b'}$, we have that $\Pr[E_{b,b'}^\ell] \leq \frac{4c_3}{\alpha'}$.
Since $\alpha' = \Theta(\alpha)$ and $\alpha \geq c_1\log n$, for a sufficiently large constant $c_1$, if we take a union bound over all bad events, we get a total probability of at most $1/10$, and the claim follows.
\end{proofof}

\begin{lemma}\label{lem:D-C4-lb}
Let $\calA$ be an algorithm  that performs less than $n^{1/4}\alpha^{1/2}/c$ queries for a sufficiently large constant $c$. The probability that $\calA$ detects a $C_4$ in a graph selected according to $\calD$
is at most $1/10$.
\end{lemma}

\begin{proof}
Similarly to the proof of Theorem~\ref{thm:C4-const-alpha-lb}, consider a process $\calP$ that answers the query of a testing algorithm while constructing a graph from $\Sup(\calD)$ (the support of $\calD$) on the fly. Here too we define a knowledge graph $K^t$ following the query-answer history $(q^1,a^1),\dots,(q^t,a^t)$ and we require the algorithm to perform a degree query on each vertex $v$ that is not yet in the knowledge graph before performing any other query involving it. The algorithm also receives the subset to which the new vertex belongs (as this is not be implied by the degree of the vertex). Let $X_1^t$, $X_2^t$, $Y_1^t$, $Y_2^t$ and $Z^t$ denote the corresponding subsets in the knowledge graph $K^t$. 

The set of edges in $K^t$ is denoted by $E^t$, where for each $e\in E^t$ there are two associated labels,
$\phi^t_X(e)$ for its endpoint in $X^t = X_1^t \cup X_2^t$, and $\phi^t_Y(e)$ for its endpoint in $Y^t = Y_1^t \cup Y_2^t$.
 We assume that a neighbor query $(x_{b,i},j)$ is answered by a pair $(y_{b',i'},j')$ where
$y_{b',i'}$ is the $j$th neighbor of $x_{b,i}$ and $x_{b,i}$ is the $j'$th neighbor of $y_{b',i'}$. Similarly, if a vertex-pair query $(x_{b,i},y_{b',i'})$ is answered positively, then the algorithm also gets the corresponding edge labels $\phi_X$ and $\phi_Y$ for this edge. The knowledge graph also includes a set $\overline{E}^t$ of non-edges (corresponding to negative answers to vertex-pair queries).

Similarly to
the proof of Theorem~\ref{thm:C4-const-alpha-lb}, in order to
determine $a^t$ given $q^t$,
the process $\calP$ considers all graphs  $G\in \Sup(\calD)$
that are consistent with $K^{t-1}$,
It then selects an answer with probability proportional to the number of such graphs $G$ for which 
$G$ is also consistent with this answer.

Consider any knowledge graph $K^{t-1}$ for $t < n^{1/4}\alpha^{1/2}/c$.
We would like to bound the probability that either a neighbor query or a vertex-pair query adds to the knowledge graph an edge $(x,y)$ that creates a $C_4$ in the knowledge graph. We shall actually bound the probability that a neighbor query is answered with \emph{some} vertex that already belongs to the knowledge graph or a vertex-pair query is answered positively (even if this does not reveal a $C_4$).
We shall refer to such an answer as a \emph{collision}. 

Let $\calG(K^{t-1})$ denote the subset of graphs  $G\in \Sup(\calD)$ such that 
$G$ is consistent with $K^{t-1}$. For a query $q^t$ (either a neighbor query or a pair query), let $\calG_c(K^{t-1},q^t)$ denote the subset of graphs  $G\in \calG(K^{t-1})$ for which the answer to $q^t$ according to 
$G$ results in a collision, and similarly define $\calG_{nc}(K^{t-1},q^t)$ for the case in which the answer is not a collision. 

\smallskip\noindent
\textsf{Subclaim}~
\emph{
Let $K^{t-1}$ be any knowledge graph where $t < n^{1/4}\alpha^{1/2}/c$, and let $q^t$ be a neighbor or pair query.
Then
$$ \frac{|\calG_{nc}(K^{t-1},q^t)|}{|\calG_{c}(K^{t-1},q^t)|} = \Omega(n^{1/4}\alpha^{1/2}) \;.$$
}

\smallskip
\noindent
\textsf{Proof:}~
In order to prove the subclaim we define an auxiliary bipartite graph $\calB(K^{t-1},q^t)$. Each node in $\calB(K^{t-1},q^t)$ corresponds to a graph $G \in \calG(K^{t-1})$.
Nodes on the left side correspond to graphs in $\calG_{c}(K^{t-1},q^t)$, and nodes on the right side to graphs in $\calG_{nc}(K^{t-1},q^t)$.
 There is an edge between two nodes corresponding to graphs $G_1$ and $G_2$, if $G_2$ can be obtained from $G_1$ by a certain ``swap'' modification (which will be explained precisely below). We shall show that the number of nodes on the left is smaller by a factor of $\Omega(n^{1/4}\alpha^{1/2})$ than the number of nodes on the left by appropriately lower bounding the degree of the former and upper bounding the degree of the latter.
 From this point on we use the shorthand notations $\calB$, $\calG$, $\calG_c$ and $\calG_{nc}$ for $\calB(K^{t-1},q^t)$, $\calG(K^{t-1})$, $\calG_c(K^{t-1},q^t)$ and $\calG_{nc}(K^{t-1},q^t)$, respectively.

For a graph $G\in \calG$ and an edge $e$ in $G$ we shall use the notation $\phi^{G}_X(e)$
and $\phi^G_Y(e)$ for its label with respect to its endpoint in $X$ and in $Y$, respectively.
Let $q^t$ be a neighbor query $nbr(v,i)$ where 
$v\in X^{t-1}$, 
and let $G_{c}$ be a graph in $\calG$ for which the answer to $q^t$ results in a collision. That is,  the answer $a^t$ is $(u,j)$ for some $u\in Y^{t-1}$.
The neighbors of node $G_{c}$ in $\calG_{nc}$ are nodes representing graphs that can be obtained from $G_{c}$ by swapping the edge $(v,u)$ with a different edge $(v, u')$ where $u'\notin Y^{t-1}$, and  
such that the resulting graph belongs to $\calG$.
We prove that there are $\Omega(n^{1/2}\alpha)$ such graphs.

In order to modify the graph $G_c \in \calG_c$ to obtain a graph $G_{nc} \in \calG_{nc}$, we proceed as follows.
    We choose a vertex 
    $u'\in Y\setminus Y^{t-1}$ (where the constraints on this choice are explained momentarily), and let $v'$ denote the $j$th neighbor of $u'$ in the graph $G_{c}$. Note that $v'$ is in $X$. Let $i'=\phi^{G_c}_X((v',u'))$. Now we swap the pair of  edges $(v,u)$
    and $(v',u')$   with the pair  $(v,u'), (v',u)$  while maintaining the original edge-labels. That is, $\phi^{G_{nc}}_X((v,u'))=i, \phi^{G_{nc}}_Y((v,u'))=j$ and $\phi^{G_{nc}}_X((v',u))=i', \phi^{G_{nc}}_Y((v,u'))=j'$.
    Since $u' \notin Y^{t-1}$, if $G_{nc}$ is in $\calG$, then it is in $\calG_{nc}$. In order to ensure that $G_{nc}\in\calG$, it is sufficient (and necessary), that there is no edge in $G_c$ between $v$ and $u'$ and no edge between $v'$ and $u$ and that
    $(v',u)\notin \overline{E}^{t-1}$.

    Observe that the selection of $u'$ uniquely determines $v'$.
    There are $|Y^{t-1}| \leq t = O(n^{1/4}\alpha^{1/2})$ ``illegal'' choices of $u' \in Y^{t-1}$ and $2\alpha = O(n^{1/4}\alpha^{1/2})$ illegal choices of $u'$ that are neighbors of $v$. There are at most $t$ choices of $v'$ such that $(v',u)\in \overline{E}^{t-1}$, and $d$ choices of neighbors of $u$. For each of these $d+t$ choices of $v'$ there are at most $2\alpha$ neighbors $u'$ such that $v'$ may be the $j$th neighbor of $u'$. This rules out at most $2(d+t)\cdot\alpha$ choices of $u'$.
    Since $|Y|=|X|\alpha/d = n\alpha/(2d)$  \ForFuture{D: Be precise} and $d = n^{1/2}/c_2$, for a sufficiently large constant $c_2$, the number of choices of $u'$ for which $G_{nc} \in \calG$ is $\Omega(n^{1/2}\alpha)$.

    We now upper bound the number of neighbors of each node  in $\calB$ that corresponds to a graph $G_{nc} \in \calG_{nc}$.
    Assume that the answer to $q^t=nbr(v,i)$ in $G_{nc}$ is $a^t=(u',j')$. That is, $u'$ is the $i$th neighbor of $v$ in $G_{nc}$, and $v$ is the $j'$th neighbor of $u'$.
      By the definition of $\calB$, every neighbor of $G_{nc}$ in $\calB$ is a graph $G_{c}$ where the $i$th neighbor of $v$ is a vertex $u''\in Y^{t-1}$. Furthermore, by the definition of the swap process, if the edge $(v,u'')\in G_c$ was replaced by a swap process and resulted in an edge $(v,u')$ where $\phi^{G_{nc}}_Y((v,u'))=j'$, it must be the case that $\phi^{G_c}_{Y}((v,u''))=j'$ (since the swapping process  involves only two edges such that both have the same $\phi_Y$ value). Hence, for each choice of $u''\in Y^{t-1}$, the graph $G_c$ from which $G_{nc}$ was obtained by a swap process is uniquely defined:
      Let $v''$ denote the $j'$th neighbor of $u''$ in $G_c$.
      The graph $G_c$ is a graph where
      $(v,u'')$ is an edge with $\phi^{G_c}_{X}((v,u''))=i, \phi^{G_c}_Y((v,u''))=j'$ and $(v'', u')$ is an edge with $\phi^{G_c}_Y((v'', u'))=j'$ (and $\phi^{G_c}_X((v'', u'))=i''$ for some $i''$). Since there are at most $|Y^{t-1}|=O(n^{1/4}\alpha^{1/2})$ options to choose $u''$, it holds that $G_{nc}$ has $O(n^{1/4}\alpha^{1/2})$ neighbors in $\calB$ (it might be the case that some of the resulting graphs $G_c$ are not in
      $\calG$
      but the upper bound still holds). \ForFuture{D: doublecheck superscripts to $\phi$ (and other indices}

Since each node  corresponding to a graph in $\calG_c$ has $\Omega(n^{1/2}\alpha)$ neighbors and each node corresponding to a graph in $\calG_{nc}$ has $O(n^{1/4}\alpha^{1/2})$ neighbors, the subgraph follows for any neighbor query $nbr(v,i)$ for $v \in X^{t-1}$.

We next consider the case in which $q^t = nbr(u,j)$ for $u \in Y^{t-1}$. The swap operation and analysis are very similar to the case in which $q^t = nbr(v,i)$ for $v \in X^{t-1}$, and so we emphasize only the differences. Let  $G_c \in \calG_c$ be a graph for which $q^t$ is answered by $(v,i)$ where $v \in X^{t-1}$. To perform a swap operation and obtain a graph in $\calG_{nc}$ we select a vertex $v'\notin X^{t-1}$ such that $v'$ is not a neighbor of $u$ in $G_c$, and for the $i$th neighbor of $v'$, denoted $u'$ we have that
$u'$ is not a neighbor of $v$ in $G_c$, but
$(u',v)\notin \overline{E}^{t-1}$.  For each such legal choice of $v'$ we can perform a swap operation between the edges $(u,v)$ and $(u',v')$ while maintaining the edge labels as in the case that $q^t = nbr(v,i)$ for $v\in X^{t-1}$. The  constraints on $v'$ rule out at most $t+2d + (t+\alpha)\cdot 2d $ vertices, and so the number of legal choices of $v'$ is $\Omega(|X|) = \Omega(n)$. On the other hand, for any graph $G_{nc}\in \calG_{nc}$, for which the answer to $q^t = nbr(v,j)$ is $(v',i')$ for $v'\notin X^{t-1}$ there are at most $t$ choices of $v''\in X^{t-1}$ such that a swap operation between the edges $(v'',u)$ and $(v',u')$ resulted in the graph $G_{nc}$.
Hence, here the ratio between $|\calG_{nc}|$ and $|\calG_c|$ is $\Omega(n/t) = \Omega(n^{3/4}/\alpha^{1/2}) = \Omega(n^{1/4}\alpha^{1/2})$, as required.

 Finally we turn to the case in which $q^t$ is a pair query between two vertices $v\in X^{t-1}$ and $u\in Y^{t-1}$. Consider a graph $G_c \in \calG_c$ for which the answer is positive.
 We can perform a swap with any edge $(v',u')$ in $G_c$ such that
 $(v',u')\notin E(K^{t-1})$ and  in addition, $(v',u)\notin E(G_c)$, $(v,u')\notin E(G_c)$ and $(v',u), (v,u')\notin \overline{E}^{t-1}$. The number of edges in $G_c$ on which a swap cannot performed with $(u,v)$ is hence upper bounded by $t+ (d+t)\cdot \alpha + d\cdot (t+\alpha) = O(d\alpha)$. This leaves $\Omega(n\alpha)$ edges with which a swap can be performed (each resulting in a different graph in $\calG_{nc}$). On the other hand, for a graph $G_{nc}\in \calG_{nc}$, for which there is no edge between $u$ and $v$, the number of graphs in $\calG_c$ for which a swap operation resulted in $G_{nc}$ is at most $d\alpha$ (the number of pairs of vertices $(u',v')$ such that $(u,v'),(u',v)\in E(G_{nc})$). Hence the ratio between $|\calG_{nc}|$ and $|\calG_c|$ is $\Omega((n\alpha)/(d\alpha) = \Omega(n/d) = \Omega(n^{1/2})= \Omega(n^{1/4}\alpha^{1/2})$.
  {\hspace*{\fill}$\Diamond$\par}

\smallskip
By the above, the probability that the algorithm detects a $C_4$ following any one of its at most
$n^{1/4}\alpha^{1/2}/c$ queries is upper bounded by a small constant (e.g., $1/10$).
\end{proof}

\bigskip
Theorem~\ref{thm:C4-gen-alpha-lb} follows by combining Lemmas~\ref{lem:D-C4-far} and~\ref{lem:D-C4-lb}.

\section{Results for $C_5$}\label{sec:C5}

The results for $C_5$ when the graph has constant arboricity are very similar to those for $C_4$ (when the arboricity $\alpha$ is a constant), and hence we only provide a sketch in this section.

We start with the upper bound (as stated in Theorem~\ref{thm:C4-5} in the introduction).
\begin{theorem}\label{thm:C5-ub}
There exists a one-sided error algorithm for testing $C_5$-freeness in graphs with  arboricity at most $\alpha = O(1)$ over $n$ vertices whose query complexity is $\widetilde{O}(n^{1/4}/\eps^{4})$. 
\end{theorem}

The algorithm is an adaptation of the algorithm for testing $C_4$-freeness.
 While the algorithm for testing $C_4$-freeness attempted to find a $C_4$ by obtaining two length-$2$ edge-disjoint paths from a vertex $v$ to a vertex $v'$, the algorithm described next
 attempts to find a $C_5$ by obtaining two edge-disjoint paths from a vertex $v$ to a vertex $v'$ such that one of these paths is of length $2$ and the other of length $3$.

\medskip\noindent
\textsf{Test-$C_5$-freeness}($n,\epsilon,\alpha$)\label{alg:C5}
\begin{enumerate}
\item Let $\theta_0 = 4\alpha/\epsilon$ and $\theta_1 =  c_1\sqrt{n}/\eps$ (for a sufficiently large constant $c_1$).
\label{step:set_params_C5}
\item Repeat the following 
$t = \Theta(1/\eps)$ times:
  \begin{enumerate}
     \item Select an edge $e$ by calling the procedure \textsf{Select-an-edge}($\alpha,\epsilon$) (as provided in Section~\ref{subsec:C4-ub}).  If it
     does not return an edge, then continue to the next iteration.
     \item Select an endpoint $v$ of $e$ by flipping a fair coin.
     \item \label{step:random-neighbors-C5} If $d(v) \leq \theta_1$, then select 
     $s_1= \Theta(\sqrt{d(v)/\eps})$ 
     random neighbors of $v$, and for each neighbor $u$ such that $d(u) \leq \theta_0$, query all the neighbors of $u$, and all the neighbors $u'$ of $u$ for which $d(u) \leq \theta_0$.
     That is, from each of the $s_1$ selected neighbors of $v$, perform a restricted BFS of depth 2, where only vertices with degree at most $\theta_0$ are further explored.
     \item \label{step:random-walk-C5} Otherwise ($d(v) > \theta_1$), perform 
     $s_2 = \Theta(\eps^{-3}\cdot\sqrt{n\log n/\theta_1})$ ~($=\tilde{\Theta}(n^{1/4}/\eps^3)$)
     random walks of length $3$ starting from $v$. 
     \item If a $C_5$ is detected, then return it, `Reject' and terminate.
  \end{enumerate}
  \item Return `Accept'.
\end{enumerate}
Since the algorithm rejects a graph only if it observes a $C_5$, it always accepts $C_5$-free graphs. Hence, consider a graph $G$  that is $\eps$-far from being $C_5$-free. 
Similarly to the analysis of the algorithm for testing $C_4$-freeness, we let $\calC$ be a maximal set of edge-disjoint $C_5$s such that no $C_5$ in this set contains an edge between two vertices with degree greater than $\theta_0$. We then define $\calC_1$ and $\calC_2$ to be the subsets of $\calC$ containing $C_5$s with no edge between two vertices with degree greater than $\theta_0$ and at most one (respectively, exactly two) vertices with degree greater then $\theta_1$. Here too, either $|\calC_1|\geq \eps m/16$ or $|\calC_2|\geq \eps m/16$.

If $|\calC_1|\geq \eps m/16$, then by applying a coloring process as in the proof of Theorem~\ref{thm:C4-ub}, we obtain a set $\calC'_1\subseteq \calC_1$
in which every vertex $v$ participating in some $C_5$ in $\calC'_1$ actually participates in  $\Omega(\eps d(v))$ such $C_5$s. Since every $5$-cycle $\rho\in \calC'_1$ contains at most one vertex of degree greater than $\theta_1$, we can label the vertices in $\rho$ by $v_i(\rho)$, $i=0,1,\dots,4$ where $v_2(\rho)$ is the highest degree vertex, such that the following holds. First, $v_0(\rho) \leq \theta_1$, and second, $v_1(\rho)$, $v_3(\rho)$ and $v_4(\rho)$ all have degree at most $\theta_0$. This implies that if the algorithm selects a vertex $v \in V_0(\calC_0') = \bigcup_{\rho\in \calC'_1} \{v_0(\rho)\}$ (which occurs with high constant probability), then with high constant probability the algorithm will detect a $C_5$ 
in Step~\ref{step:random-neighbors-C5}.

If $|\calC_2| \geq \eps m/4$, then we apply a similar coloring process to obtain a set $\calC'_1\subseteq \calC_2$
in which every vertex $v$ participating in some $C_5$ in $\calC'_2$ actually participates in  $\Omega(\eps d(v))$ such $C_5$s. Since every $5$-cycle $\rho\in \calC'_2$ contains two vertices, $v$ and $v'$ of degree greater than $\theta_1$, there is one length-2 path between $v$ and $v'$ that passes through a vertex with degree at most $\theta_0$, and one length-3 path that passes through two vertices with degree at most $\theta_0$. An analysis similar to the given in the proof of Claim~\ref{clm:C4-ub2} shows that in this case (that $|\calC_2| \geq \eps m/4$), with high constant probability, a $C_5$ will be detected in Step~\ref{step:random-walk-C5} of the algorithm.
\ForFuture{T; elaborate a bit more}

\medskip\noindent
We now turn to the lower bound.
\begin{theorem} \label{thm:C5-const-alpha-lb}
Testing $C_5$-freeness in constant-arboricity graphs over $n$ vertices requires $\Omega(n^{1/4})$ queries for constant $\epsilon$.
\end{theorem}

The high-level idea of the proof of Theorem~\ref{thm:C5-const-alpha-lb} is very similar to that of Theorem~\ref{thm:C4-const-alpha-lb}. The main difference is that in both $G_0$ and $G_1$ (based on which the distributions $\calD_0$ and $\calD_1$ are defined), 
some of the vertices in $X_b$ (for $b\in \{0,1\}$) are connected by an edge (so that the graphs are not bipartite),
and for each such edge $(v_1,v_2)$, there are two vertices $u_1$ and $u_2$ in $Y_b$, such that $u_1$ neighbors $v_1$ and $u_2$ neighbors $v_2$.

In $G_0$, every pair of vertices $u_1,u_2$ in $Y_0$ either neighbor a unique vertex $v_0$ in $X_0$ or there is a unique edge $(v_1,v_2)$ between a pair of vertices in $X_0$ such that $u_1$ neighbors $v_1$ and $u_2$ neighbors $v_2$.
On the other hand, in $G_1$, edge-disjoint $C_5$s are created by having the same pair $u_1,u_2$ in $Y_1$ neighbor both a single vertex $v_0$ and the endpoints of an edge $(v_1,v_2)$ as described above, so that $(v_0,u_1,v_1,v_2,u_2)$ is a $C_5$.

\section{An upper bound for $C_6$}\label{sec:C6-ub}
In this section we prove Theorem~\ref{thm:C6-ub}, which is restated next.

\begin{theorem}\label{thm:C6-ub}
There exists a one-sided error algorithm for testing $C_6$-freeness in graphs of constant arboricity  whose query complexity and running time are $\tilde{O}(n^{1/2}\cdot {\rm poly}(1/\eps))$.
\end{theorem}

The algorithm referred to in Theorem~\ref{thm:C6-ub} is presented next.

\medskip\noindent
\textsf{Test-$C_6$-freeness}($n,\epsilon,\alpha$)\label{alg:C6}
\begin{enumerate}
\item Let $\theta_0 = 4 \alpha/\eps$, and $\theta_1 = c_1\cdot n^{1/2}\log^2 n/\eps^2$ 
(for a sufficiently large constant $c_1$).
\item Repeat the following $t= \poly(\log n/\eps)$
times: 
  \begin{enumerate}
     \item Select a vertex $v$ uniformly at random and query its degree.  If $d(v) > \theta_0$, then continue to the next iteration.
     \item Perform a \emph{restricted} BFS starting from $v$ to depth $4$, 
     where the BFS is restricted in the following sense.
         \begin{enumerate}
           \item Whenever a vertex $u$ is reached such that $d(u) \leq \theta_0$, all its neighbors are queried.
           \item Whenever a vertex $u$ is reached such that $d(u) > \theta_0$ and $u$ is reached from a vertex $u'$ such that $d(u') \leq \theta_0$, there are two sub-cases. If $d(u) \leq \theta_1$, then  all of $u$'s neighbors are queried. Otherwise,  $\theta_1$ neighbors of $u$ are selected uniformly at random. \label{step:C6_rand_nbr}
           \item Whenever a vertex $u$ is reached from a vertex $u'$ such that both $d(u)>\theta_0$ and $d(u') > \theta_0$, the BFS does not continue from $u$.
         \end{enumerate}
     \item If a $C_6$ is detected, then return `Reject' (and terminate).
    \end{enumerate}
  \item Return `Accept'.
\end{enumerate}

Since the algorithm rejects only when it detects a $C_6$, it always accepts when the graph $G$ is $C_6$-free, and hence, from this point on, we focus on the case that $G$ is $\eps$-far from $C_6$-free.
Let $\calC$ be a maximal set of edge-disjoint $C_6$s in $G$ where for every $C_6$ in $\calC$, there are no edges in which both endpoints have degree greater than $\theta_0$
(where $\theta_0 = 4\alpha/\eps$, as defined in the algorithm).
Since the arboricity of $G$ is at most $\alpha$, 
we have that $|\calC| \geq \eps m/12$. 

Let $L$ denote the set of vertices with degree at most $\theta_0$, let $M$ denote the vertices with degree greater than $\theta_0$, and let $H$ denote those with degree greater than $\theta_1$ (where $L$ stands for \emph{low}, $M$ for \emph{medium}, and $H$ for \emph{high}).
 We partition the $C_6$s in $\calC$ into three subsets:  $\calC_1$ contains those with at most one vertex in $M$,  and for $b\in \{2,3\}$, $\calC_b$ contains those with exactly $b$ vertices in $M$. For each $b\in \{1,2,3\} $, we consider the case that $|\calC_b| \geq \eps m/36$, where for at least one of the three this lower bound should hold. 

\begin{claim}\label{clm:C6-1}
If $|\calC_1| \geq \eps m/36$, then the algorithm will detect a $C_6$ with probability at least $2/3$.
\end{claim}

\begin{proof}
Consider the following set of \emph{useful} vertices $U$. For each $\rho\in \calC_1$, if all vertices on $\rho$ belong to $L$, then they all belong to $U$. If there is a vertex $u$ on $\rho$ that belongs to $M$, then the vertex $v\in L$ at distance $3$ from $u$ (on $\rho$), belongs to $U$ (that is, the vertex $v$  that is opposite to $u$ on $\rho$). Since $|\calC_1| \geq \eps m/36$, and each vertex in $L$ belongs to at most $\theta_0/2$ edge-disjoint $C_6$s in $\calC_1$, we have that $|U| = \Omega(\eps n)$. Therefore, with probability at least $2/3$, some vertex $v\in U$ is selected in at least one of the iterations of the algorithm. Conditioned on this event, the restricted BFS starting from $v$ will detect a $C_6$.
\end{proof}

\begin{claim}\label{clm:C6-2}
If $|\calC_2| \geq \eps m/36$, then the algorithm will detect a $C_6$ with probability at least $2/3$.
\end{claim}

\begin{proof}
Recall that $\calC_2$ is the set of $C_6$s in $\calC$ with exactly two vertices in $M$, i.e., that have degree greater than $\theta_0$. We consider several sub-cases depending on the number of  these vertices that are also in $H$. 
Let  $\calC_{2,1}$ consists of those $C_6$s in $\calC_2$ that contain at most one vertex in $H$, and $\calC_{2,2}$ those that contain two vertices in $H$, and note that $\calC_2=\calC_{2,1}\cupdot \calC_{2,2}$ so that at least one of these subsets is of size at least $|\calC_2|/2$. 

If $|\calC_{2,1}| \geq |\calC_2|/2$, then let $U$ denote those vertices in $L$ that reside on $C_6$s in $\calC_{2,1}$ that either do not have any vertex in $H$, or are at distance at least two from the single vertex in $H$ on the $C_6$. Similarly to the proof of Claim~\ref{clm:C6-1}, the algorithm selects a vertex from $U$ with probability at least $2/3$, and conditioned on this event, a $C_6$ is detected. 

We now turn to the case that $|\calC_{2,2}| \geq |\calC_2|/2$. We further partition the $C_6$s in $\calC_{2,2}$ according to the distance on the $C_6$ between the two vertices in $H$, which is either $2$ or $3$, and denote the two subsets by $\calC_{2,2,2}$ and $\calC_{2,2,3}$, respectively. In what follows we analyze the case that $|\calC_{2,2,2}| \geq |\calC_{2,2}|/2$.
The case that $|\calC_{2,2,3}| \geq |\calC_{2,2}|/2$, is analyzed similarly.


For each pair of vertices $u,u' \in H$, let $\calC(u,u')$ denote the subset of $C_6$s in $\calC_{2,2,2}$ on which $u$ and $u'$ reside. 
We shall say that the set $\calC(u,u')$ is \emph{useful} if 
$|\calC(u,u')| \geq \frac{(\eps/600)\cdot \max\{d(u),d(u')\}}{|H|}$, otherwise it is \emph{un-useful}.
By this definition, the number of  $C_6$s 
in $\calC_{2,2,2}$ that belong to un-useful sets is at most 
\begin{equation}
    \sum_{u\in H}\sum_{u'\in H:d(u')\leq d(u)}\frac{(\eps/600)\cdot d(u)}{|H|}\leq 
    \eps m/300 \;.
\end{equation}
Let $\calC'_{2,2,2}$ denote the subset of $C_6$s in $\calC_{2,2,2}$ that belong to useful sets, so that
$|\calC'_{2,2,2}| \geq \eps m/300$.

The algorithm ``tries" to detect $C_6$s of this form by starting from a vertex $v$ that is at distance $2$ from both $u$ and $u'$ on some $C_6$ in $\calC(u,u')$, and 
obtaining a path of length 2 and a path of length 4 that collide on either $u$ or $u'$. Assume w.l.o.g. that $d(u')\leq d(u)$. 
Recall that except for $u$ and $u'$, all other vertices on the $C_6$s in $\calC(u,u')$ are in $L$. Therefore, a restricted BFS starting from $v$ will surely reach both $u$ and $u'$ at depth 2. Then, for the algorithm to reach $u$ from  $u'$ in two more steps, it suffices that one of their common neighbors in $L$ is selected in Step~\ref{step:C6_rand_nbr} among the $\theta_1$ sampled random neighbors of $u'$. In such a case the restricted BFS will surely reach $u$  via $u'$,   and thus obtain a $C_6$.
Since the number of common neighbors of $u$ and $u'$ is at last $|\calC(u,u')|$,
the probability of this event is at least 
$$1-\left(1-\frac{(\eps/600)\cdot \max\{d(u),d(u')\}}{|H|\cdot \min\{d(u),d(u')\}}\right)^{\theta_1}\;.$$
Since $|H| \leq 2m/\theta_1$, by the setting of $\theta_1$ (for a sufficiently large constant $c_1$ and recalling that the graph has constant arboricity, so that $m=O(n)$), this probability is at least $9/10$. Combining this with the probability of the selection of some vertex $v$ in $L$ as defined above (which occurs with probability $\Omega(\eps^2)$), the claim follows. 
\end{proof}

\begin{claim}\label{clm:C6-3}
If $|\calC_3| \geq \eps m/36$, then the algorithm will detect a $C_6$ with probability at least $2/3$.
\end{claim}


\begin{proof}
Recall that by the definition of $\calC_3$, for each $C_6$ in $\calC_3$ there are three vertices that belong to $M$, and three that belong to $L$, where each vertex in $M$ neighbors two vertices in $L$.
We start by selecting a subset of $\calC_3$, denoted $\calC_3'$, such that each vertex in $L$ belongs to at most a single $C_6$ in $\calC_3'$. Since each vertex in $L$ has degree at most  $\theta_0 = O(1/\eps)$, if we select $\calC_3'$ in a simple iterative greedy manner, then we get that
$|\calC_3'| \geq \eps_3 m$, where $\eps_3 = \Omega(\eps^2)$. Let $M'\subseteq M$ be the subset of vertices in $M$ that reside on $C_6$s in $\calC_3'$ and let $L'\subseteq L$ be the corresponding subset of vertices in  $L$. Let the edges on the
$C_6$s in $\calC_3'$ be colored green in $G$.

We next define an auxiliary multi-graph $G'$ 
    where (some of the) triangles in $G'$ correspond to $C_6$s in $G$. 
For each vertex in $M'$  we have a vertex in $G'$. In addition, there is a special \emph{ground} vertex, which we denote by $v_g$. For each pair of vertices, $u_1,u_2\in M'$ and for each length-2 green path $(u_1,v,u_2)$ (where $v\in L'$), there is an edge in $G'$. We think of this edge as being colored blue in $G'$ and labeled by the vertex $v$ in $L'$.
In addition, for each  $u\in M'$, and each non-green incident edge that it has in $G$, there is an (``uncolored'') edge between $u$ and $v_g$.
By this construction, there are $|\calC_3'|$ blue edge-disjoint triangles in $G'$ (and the degree of every vertex $u\in M'$ in $G'$ is the same as its degree in $G$).

\begin{claim}\label{clm:tri-G-prime}
Let $G'$ be a multigraph over $m' \leq m$ edges, with $m'_b$ blue edges, and 
at least $\eps_3 m$ edge-disjoint blue triangles. 
Let $D$ be any distribution over edges in $G'$ such that for each blue edge in $G'$, the probability that it is selected according to $D$ is 
$\Omega(1/m'_b)$.
Consider the following process.
First we select a blue edge $(u,u')$ according to $D$. 
Next, for each of $u$ and $u'$, if its degree 
is at most $\theta_1$, then we query all its neighbors in $G'$, and otherwise
we select $\theta_1$  random neighbors in $G'$.
 The probability that the process obtains a triangle $(u,u',u'')$ in $G'$ where $u''\in M'$ is 
$\Omega(\eps_3/\log^4 n)$.
\end{claim}

\begin{proof}
Let $\calT$ denote a maximal set of edge-disjoint triangles in $G'$ whose edges are all colored blue, so that $|\calT| \geq \eps_3 m$. Let $\calT_1$ denote the subset of triangles in $\calT$ that contain at most one vertex in $H$, and let $\calT_2$ denote those that contain at least two vertices in $H$. If $|\calT_1| \geq |\calT|/2$, then we are done:  conditioned on selecting an edge between two vertices in $M'\setminus H$ that belongs to a triangle in $\calT_1$, the procedure will detect this triangle, as it queries all neighbors of the selected edge's endpoints, and the probability of selecting such an edge is 
$\Omega(\eps_3)$.

From this point on we address the case that $|\calT_2| \geq |\calT|/2$. 
For a vertex $u\in M'$ we use $d(u)$ to denote its degree in $G'$. 
For each pair of vertices $u_1,u_2\in M'$, let $E(u_1,u_2)$ denote the set of edges between $u_1$ and $u_2$ that reside on triangles in $\calT_2$ (so that in particular, they are all blue), and let  $e(u_1,u_2) = |E(u_1,u_2)|$. 
Suppose the procedure described in the claim selects an edge $(a,b)\in E(u_1,u_2)$ where $a,b\in H$. Recall that in such a case, $\theta_1$ random neighbors of $a$ and $\theta_1$ random neighbors of $b$ are sampled.
For $i \in [\theta_1]$ and $j\in [\theta_1]$, let $\chi_{i,j}(a,b)$ be a Bernoulli random variable that is $1$ if and only if the $i$th randomly selected neighbor of $a$ equals the $j$th randomly selected neighbor of $b$, where this neighbor is not $v_g$ (thus detecting a blue triangle).
Note that by this definition, the expected number of triangles detected (conditioned on  selecting $(a,b)$) is
$\sum_{i,j\in [\theta_1]}\chi_{i,j}(a,b)$,
where
\begin{equation}\label{eq:a-b-c-1}
    \Pr[\chi_{i,j}(a,b)=1]= \sum_{c\in M}\frac{e(a,c)}{d(a)}\cdot \frac{e(b,c)}{d(b)}\;.
\end{equation}
A simple but important observation is that, since each of the $e(a,b)$ edges in $E(a,b)$ participates in some triangle $(a,b,c)$ in $\calT_2$ where these triangles are edge-disjoint, $e(a,b)$ is upper bounded by
$\sum_{c \in M} \min\{e(a,c),e(b,c)\}$, which in turn is upper bounded by
$\sum_{c\in M} e(a,c)\cdot e(b,c)$. 
That is, 
$$\sum_{c\in M} e(a,c)\cdot e(b,c)\geq \sum_{c \in M} \min\{e(a,c),e(b,c)\} \geq e(a,b)\;.$$
Therefore,
\begin{equation}\label{eq:a-b-c-2}
    \Pr[\chi_{i,j}(a,b)=1] \geq \sum_{c\in M}\frac{e(a,b)}{d(a)\cdot d(b)}\;.
\end{equation}
We next show that there exists a relatively large set of edges between  pairs $(a,b)$ where $a,b \in H$, for which
 $\Pr[\chi_{i,j}(a,b)=1] = \tilde{\Omega}(1/n)$. We shall refer to such pairs $a,b$  as \emph{useful} pairs. 
This will imply  that the expected number of triangles detected when selecting an edge between one of these pairs is $\tilde{\Omega}(\theta_1^2/n)$. 
By the setting of $\theta_1$, this is greater than 1. 
Our definition of  this set of useful pairs  will also ensure that the variance is bounded, allowing us to deduce that a triangle is detected with sufficiently high probability. Details follow.

We start by partitioning the set $\calT_2$ into subsets as follows. First, for each triangle $\Delta \in \calT_2$, we 
denote it by $(a_\Delta, b_\Delta, c_\Delta)$, where, without loss of generality, $a_\Delta, b_\Delta\in H$. For every four integers $x_{\tiny A},x_{\tiny B},y_{\tiny A,C},y_{\tiny B,C}$,  the subset $\calT_2(x_{\tiny A},x_{\tiny B},y_{\tiny A,C},y_{\tiny B,C})$ consists of all triangles $\Delta \in \calT_2$ for which the following holds.
$d(a_\Delta) \in [2^{x_{\tiny A}-1}\cdot \theta_1 , 2^{x_{\tiny A}}\cdot \theta_1)$, $d(b_\Delta) \in [2^{x_{\tiny B}-1}\cdot \theta_1,  2^{x_{\tiny B}}\cdot \theta_1)$, $e(a_\Delta,c_\Delta) \in [2^{y_{\tiny A,C}-1}, 2^{y_{\tiny A,C}})$, and $e(b_\Delta,c_\Delta) \in [ 2^{y_{\tiny B,C}-1},  2^{y_{\tiny B,C}})$.
Let $\calT^*_2 = \calT_2(x^*_{\tiny A},x^*_{\tiny B},y^*_{\tiny A,C},y^*_{\tiny B,C})$ be the largest subset, so that $|\calT^*_2| \geq \eps_3 m/\log^4 n$. 
Let $A^* \eqdef \{a_\Delta: \Delta \in \calT^*_2\}$ and $B^* \eqdef \{b_\Delta:\Delta \in \calT^*_2\}$. 
For each pair of vertices $a\in A^*$ and $b\in B^*$, let $E^*(a,b)$ be the subset of
edges between $a$ and $b$ such that there exists $\Delta\in \calT^*_2$ for which
$a = a_\Delta$ and $b = b_\Delta$, let $e^*(a,b) = |E^*(a,b)|$
and let $C^*_{a,b} \eqdef \{c_\Delta\,:\, a = a_\Delta, b=b_\Delta, \Delta \in \calT^*_2\}$. 
We have:
\begin{equation}\label{eq:sum-e-star}
s^* \eqdef \sum_{a\in A^*, b\in B^*} e^*(a,b) \geq \frac{\eps_3 m}{\log^4 n}\;,
\end{equation}
and for every $a \in A^*$ and $b\in B^*$, similarly to what was argued before regarding $e(a,b)$,
\begin{equation}\label{eq:e-star-ub}
\sum_{c\in C^*(a,b)} e(a,c)\cdot e(b,c) \geq e^*(a,b)\;.
\end{equation}
Let 
\begin{equation}\label{eq:def-P-star}
P^{*}\eqdef \left\{(a,b)\,:\, a\in A^{*},b\in B^{*}, e^*(a,b) \geq \frac{s^*}{2\cdot |A^{*}|\cdot|B^{*}|} \right\}\;.
\end{equation}
Since 
\begin{equation}\sum_{(a,b)\in A^*\times B^* \setminus P^{*}} e^*(a,b) \leq |A^*|\cdot |B^*| \cdot \frac{s^*}{2\cdot |A^{*}|\cdot|B^{*}|} = s^*/2\;,
\end{equation}
by Equation~\eqref{eq:sum-e-star},
\begin{equation}\label{eq:sum-P-star}
\sum_{(a,b)\in P^{*}} e^*(a,b) \geq \frac{\eps_3 m}{2\log^4 n}\;.
\end{equation}

By Equation~\eqref{eq:sum-P-star} and the premise of the claim regarding $D$, if we select an edge according to $D$, then the probability that we obtain an edge in $E^*(a,b)$ for $(a,b)\in P^{*}$ is 
$\Omega(\eps_3/\log^4 n)$.
Fixing any such pair $(a,b)$, we slightly modify the definition of the random variables $\chi_{i,j}(a,b)$ so that $\chi_{i,j}(a,b)=1$ if and only if the $i$th random sampled neighbor of $a$ and the $j$th random sampled neighbor of $b$ are both a common neighbor $c \in C^*(a,b)$ (the original definition allowed any common neighbor $c$ (other than $v_g$)). 
We next upper bound the probability that $\sum_{i,j\in [\theta_1]}\chi_{i,j}(a,b) =0$. 
Let 
\begin{equation}\label{eq:mu-def}
\mu(a,b)\eqdef \sum_{c\in C^*(a,b)}\frac{e(a,c)}{d(a)}\cdot \frac{e(b,c)}{d(b)}\;,
\end{equation}
so that $\Ex[\chi_{i,j}(a,b)] = \Pr[\chi_{i,j}(a,b)=1] = \mu(a,b)$.
Since $(a,b)\in P^*$,
by Equations~\eqref{eq:sum-e-star}--\eqref{eq:def-P-star},
\begin{equation}\label{eq:mu-lb}
\mu(a,b) \geq \frac{e^*(a,b)}{d(a)\cdot d(b)} \geq \frac{\eps_3 m}{2\log^4 n \cdot d(a)\cdot d(b)\cdot |A^*| \cdot |B^*|} \geq \frac{\eps_3 }{8 \log^4 n \cdot m}  \;,
\end{equation}
where we have used the fact that $d(a)\cdot |A^{*}| \leq 2m' \leq 2m$ and $d(b)\cdot |B^{*}| \leq 2m' \leq 2m$. 
From this point on we shall use the shorthand $\chi_{i,j}$ for $\chi_{i,j}(a,b)$ and $\mu$ for $\mu(a,b)$.

At this point we would have liked to apply Corollary~\ref{cor:cheb1} to upper bound the probability that
$\sum_{i,j\in [\theta_1]} \chi_{i,j}=0$, but it does not exactly meet our needs, and hence we give a direct proof.
First we observe that 
\begin{equation}
  \Ex\left[\sum_{i,j\in [\theta_1]}\chi_{i,j}\right] = \theta_1^2\cdot \mu\;.  
\end{equation}
By applying Claim~\ref{clm:var1} (see Appendix~\ref{app:prob}), with $s_1=s_2=\theta_1$ and using the notation $\mu_{1,2} = \Pr[\chi_{i,j_1}=\chi_{i,j_2}=1]$ (for every $i,j_1,j_1\in [\theta_1]$) and
$\mu_{2,1} = \Pr[\chi_{i_1,j} = \chi_{i_2,j} = 1]$ (for every $i_1,i_2,j\in [\theta_1]$) from that claim, we have that
\ForFuture{T: add elaboration}
\begin{equation}\label{eq:var-sum-chis}
    \Var\left[\sum_{i,j\in [\theta_1]}\chi_{i,j}\right] \leq \theta_1^2\cdot \mu + \theta_1^3 \cdot \mu_{1,2} + \theta_1^3 \cdot \mu_{2,1} + {\theta_1\choose 2}^2\cdot \mu^2 - (\theta_1^2\cdot \mu)^2\;.
\end{equation}
By Chebishev's inequality and the above two equations,
\begin{equation}\label{eq:C6-cheb2}
\Pr\left[\sum_{i,j\in [\theta_1]}\chi_{i,j}=0\right] \leq \frac{\Var\left[\sum_{i,j\in [\theta_1]}\chi_{i,j}\right]}{\left(\Ex\left[\sum_{i,j\in [\theta_1]}\chi_{i,j}\right]\right)^2}
\leq  \frac{1}{\theta_1^2\mu} + \frac{\mu_{1,2}}{\theta_1\cdot \mu^2} + \frac{\mu_{2,1}}{\theta_1\cdot \mu^2} \;.
\end{equation}
We next upper bound each of the three terms on the right-hand-side of Equation~\eqref{eq:C6-cheb2}.

By Equation~\eqref{eq:mu-lb} and the setting of $\theta_1=c_1\cdot n^{1/2}\log^2 n/\eps^2$, the first term contributes at most $1/6$ (for a sufficiently large constant $c_1$). 

Turning to the second term, 
\begin{equation}
\mu_{1,2} = 
\Pr[\chi_{i,j_1}=\chi_{i,j_2}=1] 
  = \sum_{c\in C^*(a,b)} \frac{e(a,c)\cdot (e(b,c))^2}{d(a)\cdot (d(b))^2}\;.
\end{equation}
By the definition of $\mu$ in Equation~\eqref{eq:mu-def}
and again using Equations~\eqref{eq:sum-e-star}--\eqref{eq:def-P-star}, the second term is upper bounded by
\begin{eqnarray}
\frac{1}{\theta_1}\cdot \frac{d(a)\cdot \sum_{c\in C^*(a,b)} e(a,c)\cdot (e(b,c))^2 }
                           {\left( \sum_{c\in C^*(a,b)} e(a,c)\cdot e(b,c)\right)^2}  &\leq&
\frac{1}{\theta_1}\cdot \frac{d(a) \cdot |C^*(a,b)| \cdot 2^{y^*_{\tiny A,C}}\cdot 2^{2y^*_{\tiny B,C}}} 
                           {|C^*(a,b)|^2 \cdot 2^{2(y^*_{\tiny A,C}-1)}\cdot 2^{2(y^*_{\tiny B,C}-1)}} \\
 &=&      
 \frac{1}{\theta_1}\cdot \frac{ 16d(a) }
                           {|C^*(a,b)| \cdot 2^{y^*_{\tiny A,C}}} \\   
 &\leq&  \frac{1}{\theta_1}\cdot \frac{16 d(a) }{e^*(a,b)}       \\
 &\leq&    \frac{1}{\theta_1}\cdot \frac{8 \log^4 n \cdot d(a) \cdot |A^{*}|\cdot |B^{*}|}{\eps_3 m }   \\
 &\leq& \frac{16 \log^4 n \cdot |B^{*}|}{\eps_3 \theta_1}  \;.
\end{eqnarray}
Since $|B^*| \leq 2m/\theta_1$ (as $B^* \subseteq H$), by the setting of $\theta_1$, the contribution of this term is at most $1/6$ as well.

For the third term we have that 
\begin{equation}\label{eq:cheb-C6-case3}
\mu_{2,1} = 
  \Pr[\chi_{i_1,j_1}=\chi_{i_1,j_2}] = 
   \sum_{c\in C^*(a,b)} \frac{(e(a,c))^2\cdot e(b,c)}{d(a)\cdot d(b)}\;.
\end{equation}
By essentially the same argument as the one bounding the second term (replacing the roles of $a$ and $b$ ($A^*$ and $B^*$), this term contributes at most $1/6$ as well. 

We have thus shown that, conditioned on selecting an edge between a pair $(a,b) \in P^*$, the probability that we obtain a triangle is at least $1/2$, and Claim~\ref{clm:tri-G-prime} follows.
\end{proof}

\medskip
In order to complete the proof of
Claim~\ref{clm:C6-3} by applying Claim~\ref{clm:tri-G-prime}, we make the following observations. First, by the construction of $G'$, each vertex in $L'$  corresponds to exactly one blue edge in $G'$. In particular, if  we select a vertex in $L'$ and query all its neighbors (as done in the algorithm), then we effectively obtain an edge in $G'$. Since $|\calC_3'|= \Omega(\eps^2 m)$, we have that $|L'| = \Omega(\eps^2 n)$.
Therefore, the probability in each iteration that the algorithm selects a vertex in $L'$, and hence an edge in $G'$, is $\Omega(\eps^2)$. 

Next, consider each of the endpoints $u$ of such an edge. If $d(u)\leq \theta_1$, then our algorithm queries all the neighbors of $u$, and for each neighbor $v$ of $u$ such that $d(v) \leq \theta_0$, the algorithm queries all the neighbors of $v$. This implies that  the algorithm reaches all of the  neighbors $u'\in M'$ of $u$ in $G'$ (as in the corresponding case described in Claim~\ref{clm:tri-G-prime}). Similarly, if $d(u)> \theta_1$, then our algorithms samples $\theta_1$ random neighbors of $u$, and for each sampled neighbor $v$ such that $d(v) \leq \theta_0$, it queries all the neighbors of $v$. This implies that for every $u'\in M'$, the probability that it is reached from $u$ by such a length-2 path in $G$, is at least as large as the probability that it is reached in one random step from $u$ in $G'$.
Since blue triangles in $G'$ correspond to $C_6$s in $G$, Claim~\ref{clm:C6-3} follows.
\end{proof}

\medskip
The proof of Theorem~\ref{thm:C6-ub} follows directly from Claims~\ref{clm:C6-1}--\ref{clm:C6-3} (together with the description of the algorithm for obtaining the upper bound on the complexity of the algorithm).

\section{A lower bound of $\Omega(n^{1/3})$ for all constant-length cycles }\label{sec:Ck-lb}

For the ease of readability, we first restate Theorem~\ref{thm:lb_ck_const_alpha}

\medskip
\noindent{\textbf{Theorem~\ref{thm:lb_ck_const_alpha}}}~
\emph{\GenLB}

\medskip

As noted in the introduction, the lower bound stated in Theorem~\ref{thm:lb_ck_const_alpha} also applies to graphs with non-constant arboricity by adding a $C_k$-free subgraph with  higher arboricity. For an odd $k$, it suffices to add a dense bipartite graph, and for even $k$, by the Erd\H{o}s girth conjecture~\cite{E63}, one can add a subgraph with arboricity $n^{2/k}$.

\ForFuture{I still feel a bit weird about this, so just leaving a comment}

To prove the theorem we shall reduce (in a non black-box manner) from the following result:

\begin{theorem}[\protect{\cite[Lemma 2]{AKKR08}}]\label{thm:triangle_freeness}
    Every one-sided error algorithm for testing triangle-freeness in graphs with $\trn$ vertices and average degree $d$ must
perform $\Omega(d,\trn/d)$ queries. This lower bound holds even when 
the maximum degree is $O(d)$.
\end{theorem}

\begin{lemma}[Implicit in the proof of Lemma 2 in \cite{AKKR08}]\label{lem:tri_free_hard_family}
There exists a family of (multi-)graphs $\mG_{\trn}$ such that 
the following holds.
\begin{itemize}
    \item The graphs  are tripartite and have $\trn$ vertices in each part
    \item The graphs are $d$-regular for $d=\Theta(\sqrt{\trn})$, 
    \item All but at most a small constant fraction of  graphs in the family  are $\Omega(1)$-far from triangle-free. 
\end{itemize}

Every algorithm that uses neighbor
and/or vertex-pair queries on a graph selected uniformly at random from  $\mG_{\trn}$ must perform $\Omega(\sqrt{\trn})$ queries before it 
views a triangle with sufficiently high constant probability.  
Furthermore, this holds even when the algorithm knows in advance which vertex belongs to which of the three parts.
\end{lemma}


\begin{proofof}{Theorem~\ref{thm:lb_ck_const_alpha}}
To prove the theorem, assume towards a contradiction that there exists a one-sided error algorithm $\mA$ for testing $C_k$-freeness using at most 
$n^{1/3}/c$ queries  (for some constant $c$)
in graphs with $n$ vertices and arboricity $2$.
We shall prove that  there exists an algorithm $\trA$ that 
finds, with high constant probability, a triangle in graphs selected uniformly from $\mG_{\trn}$ using at most $\sqrt{\trn}/c'$ queries 
 for some constant $c'$, thus reaching a contradiction to Lemma~\ref{lem:tri_free_hard_family}.

Algorithm $\trA$ will operate as follows. Given query access to a graph $\trG\in \mG_{\trn}$, algorithm $\trA$ (implicitly) defines a graph $G$ and invokes algorithm $\mA$ on it, where each query to $G$ is answered using a constant number of queries to $\trG$ (in expectation). 
Once $\mA$ terminates its execution on $G$, 
if it found  a $C_k$ in $G$, 
then $\trA$ returns a corresponding triangle in $\trG$, 
where there is a small  probability ($o(1)$) that $\trA$ terminates early (before $\mA$ makes its decision). 

We shall show that 
if $\trG$ is $\treps$-far from triangle-free, then $G$ is $\eps=\Omega(\treps)$-far from $C_k$-free. In such a case (when executed with a distance parameter $\eps$),  $\mA$ must reject $G$ with high constant probability. Since $\mA$ has one-sided error, when it rejects, it must have evidence in the form of a $C_k$.\footnote{This statement holds for the graphs constructed, and assuming the number of queries performed by $\mA$ is indeed at most $n^{1/3}/c$. The reason is that if after performing these many queries the algorithm did not detect a $C_k$, then the algorithm's  ``knowledge graph'' (which includes all queried edges as well as degrees) can be completed to a graph that is $C_k$-free. \ForFuture{work on phrasing}}

For every graph $\trG$ in $\mG_{\trn}$,   we consider an orientation of its edges so that for  edge $(u,v)$  if $id(u)< id(v)$, 
then the edge is oriented from $u$ to $v$.  For ease of presentation, assume for now that $k$ is divisible by $3$.
The graph $G$ is constructed in two steps where the first is deterministic and the second is randomized. First a graph $G^+$ is obtained by replacing each oriented edge $e=(u,u')$ in $\trG$ with a path of length $k/3$, where the original endpoints are kept in $G^+$, and $\ell=k/3-1$ new vertices $v_{e}^1, \ldots, v_{e}^{\ell}$ are added along the path (where $u=v_{e}^0$ and $u'=v_e^{\ell+1}$).
We refer to the set of vertices in $G^+$ that originated from endpoints of edges in $\trG$ as ``original vertices'', and to the set of new vertices as ``path vertices''.
See Figure~\ref{fig:gen-lb} for an illustration.
(Note that even if the graph $\trG$ has parallel edge, since every edge in $\trG$ is replaced by a distinct path in $G^+$, the resulting graph $G^+$ has no parallel edges.)

In the second step, the graph $G$ is obtained by applying a random permutation on the ids of the vertices of $G^+$, where the permutation is applied separately to the ids of the original vertices and to the ids of the path vertices. That is, the algorithm $\trA$ knows at the beginning of its execution, what is the set of ids of the original vertices and  of path vertices. If $k$ is not divisible by $3$, then  to obtain $G^+$ from $\trG$, 
a similar process to the above is done with either $\lceil k/3\rceil, \lceil k/3\rceil, \lfloor k/3\rfloor$ or $\lceil k/3\rceil, \lfloor k/3\rfloor, \lfloor k/3\rfloor$, where all edges between the same two parts in the original graph $\trG$, are replaced by a path of the same length in $G^+$.

Recall that all graphs $\trG$ in $\mG_{\trn}$ have $\trm=\Theta(\trn\cdot d)=\Theta(\trn^{3/2})$ edges so that $G$ has $n=\Theta (k\cdot \trm)=\Theta(\trn^{3/2})$ vertices and  $m=\Theta(\trn^{3/2})$ edges (recall that $k$ is a constant). Also note that $G$ has arboricity $2$, as for any subgraph  of $G$, its average degree is at most $2$.  

Observe that by the above construction, every triangle in $\trG$ is transformed into a $k$-cycle in $G$, and every $k$-cycle in $G$ originates from a triangle in $\trG$. Therefore, if a $k$-cycle is detected in $G$, then a triangle is detected in $\trG$.
We also claim that the distance is preserved (up to a constant factor), i.e., that $dist(G, C_k\mbox{-}free)= \Omega(dist(\trG, C_3\mbox{-}free))$. 
To verify this, let $R$ be a 
minimum-size set of edges so that if we remove $R$ from $G$, then it becomes $C_k$-free. Map each edge $e\in R$ that resides on some path in $G$ to the edge $\tre\in E(\trG)$ that this path replaced, and let $\trR$ be the set of edges that $R$ maps to. Then $|\trR|= |R|$, and  removing $\trR$ from $\trG$ will make $\trG$ triangle-free.\footnote{Assume the contrary, i.e., that after removing $\trR$ from $\trG$, there exists a triangle in $\trG$. Then the three paths in $G$ that replaced this triangle are disjoint from $R$, as otherwise this path would not have survived. Therefore, $G$ after removing $R$ is not $C_k$-free which is a contradiction to the definition of $R$.} Therefore,

\begin{equation} dist(\trG, C_3\mbox{-}free)\leq \frac{|\trR|}{\trm}= \frac{|R|}{\trm}\leq \frac{|R|}{m\cdot \lfloor 3/k\rfloor}=\Omega(dist(G,C_k\mbox{-}free))\;,
\end{equation}
as claimed.

Hence, given access to $\trG$ in $\mG_{\trn}$,  we shall invoke $\mA$ on $G$.
If $\trG$ is $\treps$-far from being triangle-free (for some constant $\treps$, as almost all graphs in $\mG_{\trn}$ are), then $G$ is $\treps/c'$-far from being $C_k$-free for some constant $c'$. Therefore, by the assumption on $\mA$, we can, with high constant probability, using 
at most $n^{1/3}/c \leq \sqrt{\trn}/c'$ 
queries on $G$, 
obtain a triangle in $\trG$.
Thus we get a contradiction to Lemma~\ref{lem:tri_free_hard_family}. 

It remains to show that  the invocation of $\mA$ on $G$ can be simulated by giving $\trA$ query access to $\trG$, with the same order of query complexity. 

To answer queries to $G$ using queries to $\trG$, $\trA$ constructs uniform permutations $\sigma_1\in\Pi_{\trn}$ and $\sigma_2 \in \Pi_{n-\trn}$ ``on the fly", so that  $\sigma_1$ maps original vertices in $G$ to original vertices in $G^+$ and $\sigma_2$ maps path vertices in $G$ to path vertices in $G^+$.  

Let $I_{1,t}\subset[\trn]$ and $I_{2,t}\subset [\trn+1, n]$ be the set of indices,  for which $\sigma_1$ and $\sigma_2$, respectively,  were not yet decided on by the $t$th query.  Further let  $\sigma(I_{1,t})$ and $\sigma(I_{2,t})$ be the set of indices in $[\trn]$ and $[\trn+1, n]$, respectively, that were not yet mapped to by $\sigma$ by the $t$th query.
For any $t\geq 0$,  the $t+1$ query is   answered by $\mA$ as follows.

\paragraph{Degree queries $deg(u)$:}

Recall that every $\trG\in \mG_{\trn}$ is $d$-regular, so that all the original vertices in $G$ have degree $d$. Also, by construction, all path vertices in $G$ have degree exactly $2$. Finally, by construction, the set of ids of each group of vertices is known in advance. Hence, we can assume without loss of generality that no degree queries are performed.



\paragraph{Neighbor queries $nbr(u,p)$:} 
The query is answered differently depending on whether $u\in [\trn]$ or $u\in [\trn+1,n]$.
If $u\in [\trn]$, then $\trA$ first selects $\sigma_1(u)$ by choosing an index in $[\trn]\setminus \sigma_1(I_t)$ u.a.r.. Then $\trA$ performs a neighbor query on the original vertex $\sigma(u)$ in $\trG$. That is, $\trA$ performs the query $nbr(\sigma(u), p)$ on $\trG$. 
    Let $\tre=(\sigma(u),v)$ denote the returned edge. 
    If $v$ was not previously discovered (i.e., was not previously queried or was returned as an answer to a previous query), then  
    $\trA$ decides on its id in $G$ , i.e., $\sigma^{-1}_1(v)$, by choosing uniformly from the indices in $[\trn]\setminus I_{1,t}$. 
    $\trA$ also performs a \emph{path labeling  process} on the path that replaces the edge $\tre$ in $G$:
    Let $v_{\tre}^1, \ldots, v_{\tre}^{\ell}$ be the path vertices in $G^+$ that 
    on the path that replaced the edge $\tre$ in $\trG$ (where $\sigma(u)=v_{\tre}^0$ and $v=v_{\tre}^{\ell+1}$).
    $\trA$  labels the path vertices  by choosing for each $v_{\tre}^r$ a value $\sigma_2^{-1}(v_{\tre}^r)$ u.a.r. in $[\trn+1, n]\setminus I_{2,t}$.
    Finally, $\trA$ answers the query with all the vertices of the path and their ids. 
    We refer to this as a ``path answer".

    We turn to the case that the neighbor query is performed on a path vertex. That is, $u\in [\trn+1, n]$.  
    In this case we would like to map $u\in G$ to an edge $\tre\in \trG$.  
    Recall that the id $u$ not only specifies that $u$ is a path vertex, but also to which two parts its path belongs to. Assume without loss of generality that $u$ belongs to the set of path vertices that reside on paths between parts $1$ and $2$. In order to map $u$ to an edge between parts $P_1,P_2$, we need to first sample an edge $e$ between $P_1$ and $P_2$ uniformly at random  (we shortly explain how).  If the edge $\tre\in \trG$ has already been  discovered, then $\trA$ terminates and returns that $\trG$ is triangle free. This only happens with negligible probability $O(Q/m)=o(1)$, where $Q$ is the number of queries performed by $\mA$. Otherwise, $\trA$  draws   a uniform index $r\in [\ell]$,  
  and lets $\sigma(u)=v_{\tre}^{r}$. Then 
    $\trA$ performs a path labeling process on the remaining path vertices and returns a path answer for $\tre$.
    
    To sample a uniform edge in $\trG$ between two specific parts, say $P_1$ and $P_2$,  $\trA$ samples $w$ in the set of indices of $P_1$ and $r'\in [d]$ u.a.r. and performs  neighbor queries $nbr(w,r')$ on $\trG$ until the returned neighbor belongs to the desired part, in this case $P_2$. 
    Since $\trG$ is $d$-regular, this results in a uniformly distributed edge $\tre$ in $E(P_1,P_2)$. Note that since every vertex $v\in G$ has exactly half of its neighbors in each of the other two parts, the above process returns an edge in the desired part with probability $1/2$. Hence, the expected number of queries per one query simulation  is constant.

    \paragraph{Pair queries $pair(u_1,u_2)$:} 
    If the two queried vertices are original vertices in $\trG$ then the answer to the query is no. If at least one of them is a path vertex, then $\mA$ maps it to an edge, as described above for neighbor queries, and answers accordingly.

\bigskip
Finally, as each query to $G$ can be simulated in a constant number of queries to $\trG$ in expectation, it follows that with high probability $1-o(1)$ (over the randomness of $\trA$), the total number of queries does not exceed that of $\mA$ by more than a constant factor, 
and otherwise $\trA$ may terminate.
  This concludes the proof.
\end{proofof}


\section{A general upper bound for testing $C_k$-freeness (and $F$-freeness in general)}\label{sec:Ck}


\newcommand{\VC}{\mathcal{VC}}

In this we section describe and analyze our algorithms for testing
$C_k$-freeness, for even and odd $k$, and more generally for $F$-freeness, for any fixed subgraph $F$. 
We shall use $k$ to denote the number of vertices in $F$ (which is consistent with the special case that $F=C_k$).
In order to present the result for $F$-freeness, we introduce a measure that extends the notion of the size of a minimum vertex cover.

\begin{definition}\label{def:ell}
    For a graph $F = (V_F, E_F)$ let $\VC(F)$ denote the set of all vertex covers of $F$. For a vertex cover $Z$ of $F$ we denote by $\VC'(Z)$ the set of vertex covers of $F$ that are subsets of $Z$.
    We define $\ell(F) = \max_{Z \in \VC(F)} \left\{min_{B \in \VC'(Z)} \left(|B|\right)\right\}$.
\end{definition}
Observe that by Definition~\ref{def:ell}, we have that $\ell(F)$ is lower bounded by the size of a minimum vertex cover of $F$ and is upper bounded by $k = |V_F|$.

Our algorithm for testing $F$-freeness appears next. The algorithm receives the number of edges, $m$, as a parameter. We note that $m$ can be an upper bound on the number of edges of the graph. The algorithm also receives $\alpha$ as a parameter. As noted before, if $m$ is given (and not just an upper bound) then we can estimate the effective arboricity instead of receiving $\alpha$ as a parameter. Another option is to receive only $\alpha$ (without $m$) and use $\alpha n$ as an upper bound on $m$.   

\medskip
    \textsf{Test-subgraph-freeness}($F,n, m, \alpha, \epsilon$)\label{alg:Ffree}
\begin{enumerate}
\item Set $\theta_0 = 4\alpha/\epsilon$.
\item Sample u.a.r. $s = $$\Theta\left(k^{2+1/\ell(F)} \cdot m \cdot \left(\frac{\alpha}{m }\right)^{1/\ell(F)} \cdot \left(\frac{1}{\epsilon}\right)^{1+2/\ell(F)}\right)$ vertices from $G$ (where $\ell(F)$ is as defined in Definition~\ref{def:ell}).
\item Query all the neighbors of every vertex $v$ in the sample such that $d(v) \leq \theta_0$ .
\item Reject if and only there is a witness for $F$ in the resulting subgraph.
\end{enumerate}

\medskip
In order to analyze the algorithm Test-subgraph-freeness, we introduce some definitions, and 
prove a claim regarding sampling of tuples.

For the threshold $\theta_0$ as defined in the algorithm, let $L$ denote the subset of vertices in $G$ whose degree is at most $\theta_0$, and let $H$ denote the subset of vertices whose degree is greater than $\theta_0$. 
We refer to the former as \emph{light} vertices and the latter as \emph{heavy} vertices. 
We say that a subset of subgraphs of $G$ (in particular, that are isomorphic of $F$), are 
{\em light-vertex-disjoint} if they do not share any light vertex.
Let $G_{\leq \theta_0}$ denote the subgraph of $G$ that contains all vertices of $G$ but only edges that have at least one light endpoint.

\begin{claim}\label{clm:ver-num}
Let $X$ be a set of elements 
and let $\mathcal{T}=\{(x^j_1, \ldots, x^j_\ell)\}_{j\in [|\calT|]}$ be a set of $\ell$-tuples of elements of $X$ such that each $x\in X$ appears at most once in some $\rho\in \mathcal{T}$ (namely, the $\ell$-tuples are disjoint and each tuple contains distinct elements).
Let $S = \{u_1, \ldots u_s\}$ be a multi-set of $s$ elements chosen independently, uniformly, at random from $X$. 
If $s \geq 16 \ell |X| / (|\mathcal{T}|^{1/\ell})$, then with high constant probability over the choice of $S$, it contains $\ell$ elements that belong to a common tuple in $\calT$.
\end{claim}

\begin{proof}
We say that an $\ell$-tuple of indices $(i_1, \ldots i_\ell)$ where $i_j \in [s]$ for each $j\in [\ell]$ is \emph{good} with respect to $S$ and $\mathcal{T}$ if $(u_{i_1}, \ldots u_{i_\ell})$ belongs to $\mathcal{T}$.
Let $r = {s \choose \ell } \cdot \ell!$ denote the number of different $\ell$-tuples of indices in $[s]$.  
Consider any fixed ordering of these tuples, and let $\chi_i$ 
be the indicator variable for the event that the $i$-th tuple of indices is good with respect to $S$ and $\mathcal{T}$.
We next upper bound the probability that
$\sum_{i \in [r]} \chi_{i}=0$.
Let $\mu = \Ex[\chi_i]$ and $\bar{\chi}_{i} = \chi_{i} - \mu$, so that $\Ex[\bar{\chi}_{i}] = 0$.
By Chebyshev's inequality,
\begin{equation}\label{eq:cheb-Ck}
\Pr\left[\sum_{i\in [r]}\chi_{i}=0\right] \leq \frac{\Var\left[\sum_{i\in [r]}\chi_{i}\right]}{\left(\Ex\left[\sum_{i\in [r]}\chi_{i}\right]\right)^2}
= \frac{\sum_{i,j} \Ex[\bar{\chi}_{i}\cdot \bar{\chi}_{j}]}{r^2\cdot \mu^2} \;.
\end{equation}

We first calculate:

\begin{equation}
\Ex\left[\sum_{i\in [r]}\chi_{i}\right] = r \cdot \mu = {s \choose \ell } \cdot \ell! \cdot |\mathcal{T}| \cdot \left(\frac{1}{|X|}\right)^{\ell}
= \frac{s!}{(s-\ell)!} \cdot |\calT| \cdot \left(\frac{1}{|X|}\right)^{\ell} \geq (s/2)^\ell \cdot |\calT| \cdot \left(\frac{1}{|X|}\right)^{\ell}  \geq (8\ell)^\ell\;.
\end{equation}

We next break the sum $\sum_{i,j} \Ex[\bar{\chi}_{i}\cdot \bar{\chi}_{j}]$ into several sub-sums as follows.
We say that $(i,j) \in [r]\times [r]$ is in $R_t$ if the number of elements (namely, indexes of samples) that belong both to the $i$-th tuple and the $j$-th tuple is exactly $t$ (namely the intersection of the tuples is of size $t$ when they are viewed as sets).

If $(i,j) \in R_0$ then $\chi_{i}$ and $\chi_{j}$ are independent and so
\begin{equation}\label{eq:cheb-Ck-case1}
\Ex[\bar{\chi}_{i}\cdot \bar{\chi}_{j}] =  \Ex[\bar{\chi}_{i}]\cdot \Ex[\bar{\chi}_{j}] = 0\;.
\end{equation}
Hence, the pairs in $R_0$ do not contribute anything to the right-hand side of Equation~\eqref{eq:cheb-Ck}.

If $i = j$ then
\begin{equation}\label{eq:cheb-Ck-case2}
\Ex[\bar{\chi}_{i}\cdot \bar{\chi}_{j}] = \Ex[(\bar{\chi}_{i})^2]
\leq \Ex[(\chi_{i})^2] = \Ex[\chi_{i}] = \mu \;,
\end{equation}
where the second inequality follows from the definition of $\bar{\chi}_{i}$.

Given that $(i,j) \in R_t$ it holds that

\begin{equation}\label{eq:cheb-Ck-case3}
\Ex[\bar{\chi}_{i}\cdot \bar{\chi}_{j}]
\leq \Ex[\chi_{i} \chi_{j}] \leq  |\calT| \cdot \left(\frac{1}{|X|}\right)^{2\ell -t} \;.
\end{equation}
Since there are at most ${\ell \choose t} \cdot s^{2\ell -t}$ pairs in $R_t$ we obtain that
\begin{eqnarray}
\sum_{(i, j)\in R_t}\Ex[\bar{\chi}_{i}\cdot \bar{\chi}_{j}]
&\leq&  {\ell \choose t} \cdot s^{2\ell -t} \cdot |\calT|\cdot   \left(\frac{1}{|X|}\right)^{2\ell -t} \leq \ell^t \cdot |\calT| \cdot \left(\frac{s}{|X|}\right)^{2\ell -t} \;\leq\;  \ell^\ell \cdot |\calT| \cdot   \left(\frac{s}{|X|}\right)^\ell \nonumber\\
&=&  \left( \frac{|\calT|^{1/\ell} \cdot \ell \cdot s}{|X|}\right)^\ell \;\leq\; (16\ell^2)^\ell\;,
\label{eq:cheb-Ck-case4}
\end{eqnarray}

Thus,

\begin{equation}
\frac{\sum_{i,j} \Ex[\bar{\chi}_{i}\cdot \bar{\chi}_{j}]}{r^2\cdot \mu^2}
\leq  \frac{1}{r \mu} + \frac{\ell (16\ell^2)^\ell}{(r \mu)^2} \leq
\frac{1}{(8\ell)^\ell} + \frac{\ell (16\ell^2)^\ell}{(8\ell)^{2\ell}} \leq \frac{1}{8} +  \frac{\ell}{2^{2\ell}} \leq \frac{3}{8}\;,
\end{equation}
which concludes the proof of the claim.  
\end{proof}


\begin{theorem}\label{thm:Gen-F-ub}
The algorithm \TestFfree\ is a one-sided error tester for $F$-freeness whose query complexity is $$O\left(k^{2+1/\ell(F)} \cdot m \cdot \left(\frac{\alpha}{m }\right)^{1/\ell(F)} \cdot \left(\frac{1}{\epsilon}\right)^{1+2/\ell(F)}\right)\;.$$
\end{theorem}

\begin{proof}
The completeness of the tester follows by construction.
We next prove its soundness (where we start similarly to previous proofs). Assume $G$ is $\epsilon$-far from being $F$-free.
By the setting of $\theta_0 = 4\alpha/\eps$, the graph $G_{\leq \theta_0}$ is $\eps/2$-far from being $F$-free. By the definition of $G_{\leq \theta_0}$, every copy of $F$ in $G_{\leq \theta_0}$ (i.e.,  subgraph of $G_{\leq \theta_0}$ that is isomorphic to $F$) only contains edges that are incident to light vertices.
Consider an iterative process, that constructs a set $\mathcal{F}$ of light-vertex-disjoint copies of $F$ in $G_{\leq \theta_0}$ in the following greedy manner. In each iteration it selects a new copy of $F$ in $G_{\leq \theta_0}$ that does not share any light vertices with previously selected copies. 
It follows that $|\mathcal{F}| \geq \eps m/ (2\theta_0 k)$ (to see this consider a greedy process which removes all the edges that are incident to the light vertices of previously selected copies). 

From each one of these copies we pick a subset of light vertices that covers the edges of this copy (since the copy is in $G_{\leq \theta_0}$ there exists a vertex cover which is composed of only light vertices).
Moreover, by the definition of $\ell(F)$, it suffices to pick $\ell(F)$ light vertices for each copy.
By ordering these vertices (arbitrarily) we obtain an $\ell(F)$-tuple of light vertices for each one of these copies.
Let $\calT$ to be the set of $\ell(F)$-tuples corresponding to these copies. According to Claim~\ref{clm:ver-num}, if we set $s$ to be at least $16 \ell(F) n / ((\epsilon m/ (2\theta_0 k))^{1/\ell(F)})$ vertices, then w.h.c.p. there exists a copy of $F$ in $G$ such that we hit all its light vertices. Conditioned on this event, since the algorithm queries all the neighbors of all light vertices in the sample, the copy will be revealed.
By Claim~\ref{clm:average}, w.h.p., the number of queries the algorithm needs to preform is bounded by $s \cdot 2\bar{d} = O\left(k^{2+1/\ell(F)} \cdot m \cdot \left(\frac{\alpha}{m }\right)^{1/\ell(F)} \cdot \left(\frac{1}{\epsilon}\right)^{1+2/\ell(F)}\right)$. If the number of queries exceeds this bound (and the algorithm did not find a witness for rejection) the algorithm can simply accept. 
\end{proof}


\begin{corollary}\label{cor:Ck-ub}
There exists a one-sided error tester for $C_k$-freeness whose query complexity is
$$O\left(k^{2+(2/k))} \cdot m \cdot \left(\frac{\alpha}{m }\right)^{2/k} \cdot \left(\frac{1}{\epsilon}\right)^{1+(4/k)}\right)$$
for even $k$, and
$$O\left(k^{2+2/(k+1)} \cdot m \cdot \left(\frac{\alpha}{m }\right)^{2/(k+1)} \cdot \left(\frac{1}{\epsilon}\right)^{1+4/(k+1))}\right)$$
for odd $k$.
\end{corollary}

\bigskip
We next turn to describe and analyze an algorithm for testing $C_k$-freeness when $k$ is odd, where the algorithm improves on the upper bound stated in Corollary~\ref{cor:Ck-ub} (when $k$ is odd) for a certain range of values of $\alpha$. 

\medskip
We first slightly modify the algorithm Select-an-edge from Section~\ref{subsec:C4-ub} to obtain an algorithm that samples a uniform edge from $G_{\leq \theta_0}$. As shown in~\cite{Reut}, the expected running time of this algorithm is $O(\theta_0 n/ m)$.

\medskip
\textsf{Sample a uniform edge from $G_{\leq\theta_0}$}
\begin{itemize}
\item Repeat until an edge is returned:
\begin{enumerate}
\item Sample u.a.r. a vertex $v$ from $V(G)$.
\item Pick $j\in [\theta_0]$ and toss a fair coin $b\in \{\rm{Heads}, \rm{Tails}\}$.
\item If $v\in L$ and $v$ has a $j$-th neighbour, $u$ then:
\begin{enumerate}
\item If $u \in H$, then return $\{u,v\}$.
\item If $u \in L$ and $b=\rm{Heads}$, then return $\{u,v\}$.
\end{enumerate}
\end{enumerate}
\end{itemize}

\begin{claim}\label{clm:sampleedges}
Let $E'$ be a set of $\beta m$ edges in $G_{\leq \theta_0}$, where $\beta \in (0, 1]$ and let $m'$ denote the number of edges in $G_{\leq \theta_0}$. If we sample $\log n/\beta  \leq y \leq m'/2$ edges uniformly from $G_{\leq \theta_0}$, then with high probability we sample at least $c \beta y$ different edges from $E'$, where $c$ is an absolute constant.
\end{claim}

\begin{proof}
    Let $X_i$ denote the indicator variable for the event that the $i$-th sample is an edge from $E'$. Since the edges are sampled uniformly from $G_{\leq \theta_0}$, $\Ex(X_i) = \beta$.
    Thus, by multiplicative Chernoff's bound, with high probability $\sum_i X_i = b \beta y$, where $b$ is a constant.
    Let $Y_i$ be the indicator variable for the event that the $i$-th sample is a new edge from $E'$ (with respect to the $(i-1)$ first samples).
    We consider, without loss of generality, the sampling process as if it first decides the values of $X_i$ for each sample $i$ and then samples the edges one by one accordingly.
    Consider the $j$-th sample.
    Given the values of $X_i$ for every $i$ and the values of the samples for each $i < j$, the probability that the $j$-th sample is a new edge, conditioned on the event that $X_j = 1$ is constant (since $y \leq x/2)$. Therefore, with high probability we sample $c \beta y$ different edges from $E'$ for some absolute constant $c$.
\end{proof}

\bigskip
\textsf{Test-$C_k$-freeness when $k$ is odd}($n, m, \alpha, \epsilon$)
\begin{enumerate}
\item Set $\theta_0 = 4\alpha/\epsilon$.
\item Sample u.a.r. $s_1 = \Theta \left(k \cdot m^{1-\frac{2}{k-1}} \cdot (1/\alpha)^{1-\frac{4}{k-1}} \cdot \epsilon^{1-\frac{6}{k-1}}\right)$ edges from $G_{\leq \theta_0}$ (as described above).
For each edge in the sample,  reveal the neighborhood of both endpoints if both are light.
\item If $k > 3$, then sample u.a.r. $s_2 = O\left(k \cdot n \cdot (\alpha^2/m)^{\frac{2}{k-1}} \cdot \left( \frac{1}{\epsilon}\right)^{\frac{6}{k-1}} \right)$ vertices and reveal the neighborhood of all the light vertices in the sample.
\item Reject iff there is a witness for $C_k$ in the resulting subgraph.
\end{enumerate}

The following bound is an improved bound for testing $C_k$-freeness for odd $k$ whenever $m = \Omega(\alpha^{(k+3)/2})$ and in particular when $\alpha$ is a constant.
\begin{theorem}\label{thm:Ck-ub-odd-k}
There exists a one-sided error tester for $C_k$-freeness for odd $k$ whose query complexity is
$$O\left(k \cdot m \cdot (\alpha^2/m)^{\frac{2}{k-1}} \cdot \left( \frac{1}{\epsilon}\right)^{\frac{6}{k-1}} \right)$$
\end{theorem}

\begin{proof}
The completeness of the tester follows by construction. 
We next prove its soundness. Assume $G$ is $\epsilon$-far from being $C_k$-free.
As in the proof of Theorem~\ref{thm:Gen-F-ub}, there exists a set $\calC$ of light-vertex-disjoint copies of $C_k$ in $G_{\leq \theta_0}$ such that $|\calC|\geq \epsilon m/ (2\theta_0 k)$.
For each one of the copies in $\calC$ we pick a subset of light vertices that covers the edges of this copy (since the copy is in $G_{\leq \theta_0}$ there exists a vertex cover which is composed of only light vertices).
By the structure of $C_k$, it suffices to pick $\lceil k/2 \rceil$ light vertices from each copy.
Moreover, exactly one pair of these vertices are on the same edge of the corresponding copy $C_k$.
By excluding this pair and ordering the rest of these vertices (arbitrarily) , we obtain an $(k-3)/2)$-tuple of light vertices for each one of these copies.

Let $\calT$ denote the set of these tuples.
We also consider the set of pairs of vertices we excluded from each copy.
Each such pair is an edge of a $C_k$ in $\calC$.
Denote this set of edges by $Y$, so that
$|Y| \geq \epsilon m/ (2\theta_0 k)$.
By Claim~\ref{clm:sampleedges}, w.h.p., we sample at least $c \epsilon s_1/ (2\theta_0 k) $ edges from $Y$ (where $c$ is a constant).
Let $\calT' \subseteq \calT$ be the set of tuples corresponding to copies for which we sampled an edge from $Y$.
By the above, w.h.p., $|\calT'| \geq c \epsilon s_1/ (2\theta_0 k) $.
Conditioning on this event, according to Claim~\ref{clm:ver-num}, if $s_2$ is at least $16 k n / (c \epsilon s_1/ (2\theta_0 k))^{2/(k-3)}$, then w.h.c.p. the algorithm will find a copy of $C_k$.
By Claim~\ref{clm:average} the expected query complexity of the algorithm is bounded by $O(s_1 \cdot \theta_0 + s_2 \cdot \bar{d}) = O\left(k \cdot m^{1- \frac{2}{k-1}} \cdot \alpha^{\frac{4}{k-1}} \cdot \left( \frac{1}{\epsilon}\right)^{\frac{6}{k-1}} \right)$. Thus, by Markov's inequality, if the number of queries exceeds this bound (and the algorithm did not find a witness for rejection) the algorithm can simply accept. 
\end{proof}

\begin{claim}\label{clm:average}
	Let $\bar{d}$ be the average degree in $G$, and 
	let $S=\{v_1, \ldots, v_s\}$ be a uniform sample of size at least $s=(\theta_0/\bar{d})\cdot 12\ln(1/\delta)$ vertices (with repetitions).
	Let $\chi_i=d(v_i)$ if $d(v_i)\leq \theta_0$ and $\chi_i=0$ otherwise.
	Then with probability at least $1-\delta$, $\frac{1}{s}\sum_{i=1}^s\chi_i\leq 2\bar{d}$.
\end{claim}

\begin{proof}
	It holds that $\Ex[\chi_i] \leq \frac{1}{n} \sum_{v \in V} d(v) = \bar{d}$.
	For any $\epsilon < \frac{1}{2}$, 
	$\Ex[\chi_i]\geq
	\frac{1}{n}\sum_{v : d(v) \leq t} d(v) 
	\geq \frac{1}{n}\cdot 2(1-\epsilon)m
	\geq \bar{d}$.
	
	Also, for every $i\in [s]$, $\chi_i\in [0,\theta_0]$ for every $i$. 
	Let $x_i=\chi/\theta_0$ and $X=\sum_{i=1}^s x_i$.
	Then the $x_i$'s are bounded independent   random variables in $[0,1]$ with $\Ex[X]\leq s\cdot \bar{d}/\theta_0$. Therefore, by  a coupling argument and the multiplicative Chernoff's bound
	\[
	\Pr\left[X > 2s\cdot \frac{\bar{d}}{\theta_0}\right] \leq  \exp\left( -s\cdot \bar{d}/3\theta_0 \right) < \delta.
	\]
\end{proof}

\section{Some probabilistic claims}\label{app:prob}


\begin{claim}\label{clm:var1}
For integers $s_1$ and $s_2$  let
$\{\chi_{i,j}\}_{(i,j)\in [s_1]\times [s_2]}$ be Bernoulli random variables for which the following holds.
\begin{enumerate}
\item  $\Pr[\chi_{i,j} = 1] = \mu$ for every $(i,j)\in [s_1]\times [s_2]$.
\item For every $i_1,i_2 \in [s_1]$, $j_1,j_2\in [s_2]$ such that $i_1\neq i_2$ and $j_1\neq j_2$,
the random variables $\chi_{i_1,j_2}$ and $\chi_{i_2,j_2}$ are independent.
\item For every $i\in [s_1]$, $j_1,j_2\in [s_2]$ such that  $j_1\neq j_2$,
$\Pr[\chi_{i,j_1}=\chi_{i,j_2} = 1] = \mu_{1,2}$, and for every $i_1,i_2 \in [s_1]$ and $j \in [s_2]$ such that $i_1\neq i_2$,  $\Pr[\chi_{i_1,j}=\chi_{i_2,j} = 1] = \mu_{2,1}$.
\end{enumerate}
Then
$$\Var\left[\sum_{i\in [s_1],j\in [s_2]}\chi_{i,j}\right] \leq s_1\cdot s_2 \cdot \mu + s_1\cdot {s_2\choose 2}\cdot \mu_{1,2} + {s_1\choose 2} \cdot s_2\cdot \mu_{2,1}\;.$$
\end{claim}

\begin{proof}
For each pair $(i,j)\in [s_1]\times [s_2]$,
let $\bar{\chi}_{i,j} = \chi_{i,j} - \mu$, so that $\bar{\chi}_{i,j} \leq \chi_{i,j}$, and
$\Ex[\bar{\chi}_{i,j}] = 0$. By the definition of the variance,
\begin{equation}\label{eq:var1}
\Var\left[\sum_{i\in [s_1],j\in [s_2]}\chi_{i,j}\right] =
  \sum_{(i_1,j_2),(i_2,j_2)\in [s_1]\times[s_2]} \Ex[\bar{\chi}_{i_1,j_1}\cdot \bar{\chi}_{i_2,j_2}]\;.
\end{equation}

We upper bound the right-hand-side of Equation~\eqref{eq:var1} by breaking
 the sum $\sum_{i_1,j_2,i_2,j_2} \Ex[\bar{\chi}_{i_1,j_1}\cdot \bar{\chi}_{i_2,j_2}]$ into four sub-sums.

 \paragraph{The first} is over $(i_1,j_1), (i_2,j_2)\in [s_1]\times [s_2]$ such that all four indices are distinct. By the second condition in the claim, for such quadruples of indices, $\chi_{i_1,j_1}$ is independent of $\chi_{i_2,j_2}$ and so
\begin{equation}\label{eq:cheb-case1}
\Ex[\bar{\chi}_{i_1,j_1}\cdot \bar{\chi}_{i_2,j_2}] =  \Ex[\bar{\chi}_{i_1,j_1}]\cdot \Ex[\bar{\chi}_{i_2,j_2}] = 0\;.
\end{equation}

\paragraph{The second} is over $i_1 = i_2$ and $j_1 = j_2$, so that
\begin{equation}\label{eq:cheb-case2}
\Ex[\bar{\chi}_{i_1,j_1}\cdot \bar{\chi}_{i_2,j_2}] = \Ex[(\bar{\chi}_{i_1,j_1})^2]
\leq \Ex[(\chi_{i_1,j_1})^2] = \Ex[\chi_{i_1,j_1}] = \mu \;,
\end{equation}
where the second inequality follows from the definition of $\bar{\chi}_{i,j}$.
There are $s_1\times s_2$ such quadruples.

\paragraph{The third} is over $i_1 = i_2$ and $j_1 \neq j_2$, in which case we have
\begin{equation}\label{eq:cheb-case3}
\Ex[\bar{\chi}_{i_1,j_1}\cdot \bar{\chi}_{i_1,j_2}]
\leq \Ex[\chi_{i_1,j_1}\cdot \chi_{i_1,j_2}] = \Pr[\chi_{i_1,j_1}=\chi_{i_1,j_2}=1]
  = \mu_{1,2}\;.
\end{equation}
There are $s_1\times {s_2\choose 2}$ such pairs.

\paragraph{The fourth} is over $i_1 \neq i_2$ and $j_1 = j_2$, where similarly to the third case we get
\begin{equation}\label{eq:cheb-case4}
\Ex[\bar{\chi}_{i_1,j_1}\cdot \bar{\chi}_{i_2,j_2}]  
\leq 
\mu_{2,1}\;
\end{equation}
where there are ${s_1\choose 2}\cdot s_2$ such pairs.

The claim follows by summing all four sub-sums.
\end{proof}

The next claim is very similar to Claim~\ref{clm:var1}, but the setting is slightly different, and will be used in our analysis several time.
In what follows, for an integer $s$ we use $\Phi(s)$ to denote the set of pairs $i,j\in [s]$ such that $i<j$.
\begin{claim}\label{clm:var2}
For an integer $s$l let $\{\chi_{i,j}\}_{(i,j)\in \Phi(s)}$ be Bernoulli random variables for which the following holds.
\begin{enumerate}
\item  $\Pr[\chi_{i,j} = 1] = \mu$ for every $(i,j)\in \Phi(s)$.
\item For every $i_1,i_2 \in [s_1]$, $j_1,j_2\in [s_2]$ such that $i_1\neq i_2$ and $j_1\neq j_2$,
the random variables $\chi_{i_1,j_2}$ and $\chi_{i_2,j_2}$ are independent.
\item For every $(i_1,j_1)\in \Phi(s)$ and $(i_2,j_2)\in \Phi(s)$  such that exactly two of the four indices are the same,
$\Pr[\chi_{i_1,j_1}=\chi_{i_2,j_2} = 1] = \mu_{1,2}$.
\end{enumerate}
Then $\Var\left[\sum_{(i,j)\in \Phi(s)}\chi_{i,j}\right] \leq {s\choose 2}\cdot \mu + {s\choose 3}\mu_{1,2}$.
\end{claim}
The proof of Claim~\ref{clm:var2} is very similar to the proof of Claim~\ref{clm:var1}, and is hence omitted. As a corollary of this claim we get:

\begin{corollary}\label{cor:cheb1}
For an integer $s$
let $\{\chi_{i,j}\}_{(i,j)\in \Phi(s)}$ be Bernoulli random variables where
 $\Pr[\chi_{i,j} = 1] = \mu$ for every $(i,j)\in \Phi(s)$. Suppose that the following conditions hold
 for some $c_1>0$ and $c_2 > 4$.
\begin{enumerate}
\item For every $(i_1,j_1)\in \Phi(s)$ and $(i_2,j_2)\in \Phi(s)$ such that the four indices are distinct, $\chi_{i_1,j_1}$
and $\chi_{i_2,j_2}$ are independent.
\item For every $(i_1,j_1)\in \Phi(s)$ and $(i_2,j_2)\in \Phi(s)$  such that exactly two of the four indices are the same,
$\Pr[\chi_{i_1,j_1}=\chi_{i_2,j_2} = 1] \leq c_1\cdot \mu^{3/2}$.
\item 
   $s \geq c_2/\sqrt{\mu}$\;.
\end{enumerate}
Then $\Pr\left[ \sum_{(i,j)\in \Phi(s)} \chi_{i,j} = 0\right] \leq \frac{1+c_1}{c_2}$.
\end{corollary}

\begin{proof}
By Chebishev's inequality,
\begin{equation}\label{eq:cheb1}
\Pr\left[\sum_{(i,j)\in \Phi(s)}\chi_{i,j}=0\right] \leq \frac{\Var\left[\sum_{(i,j)\in \Phi(s)} \chi_{i,j}\right]}{\left(\Ex\left[\sum_{(i,j)\in \Phi(s)}\chi_{i,j}\right]\right)^2}\;
\end{equation}
The denominator of the right-hand-side of Equation~\eqref{eq:cheb1} equals
${s\choose 2}^2\cdot \mu^2$.
By Claim~\ref{clm:var2} and the first two conditions of the current claim, the numerator of the right-hand-side of Equation~\eqref{eq:cheb1} is upper bounded by
${s\choose 2}\cdot \mu + {s\choose 3}\cdot c_1\cdot \mu^{3/2}$, so that the claim follows.
\end{proof}

\end{document}